\documentclass[12pt]{article}
\usepackage[utf8]{inputenc}
\usepackage{booktabs}
\usepackage{eqparbox}
\usepackage[dvipsnames]{xcolor}
\usepackage{soul}
\newcommand{\mathcolorbox}[2]{\colorbox{#1}{$\displaystyle #2$}}
\newcommand\MyLeftColumn[1]{\eqmakebox[A][r]{$#1$}}
\usepackage{graphicx}
\usepackage{fullpage} 
\usepackage{subcaption} 
\usepackage{braket}
\usepackage{amsmath}

\usepackage{tikz-cd}
\usepackage{dsfont}
\usepackage{amsfonts}
\usepackage{amsthm}
\usepackage{amssymb,graphicx}
\usepackage{hyphenat}
\usepackage{csquotes}
\usepackage{hyperref}
\numberwithin{equation}{section}
\hypersetup{
    colorlinks=true,
    linkcolor=blue,
    filecolor=blue,      
    urlcolor=blue,
    citecolor=blue,
    linktoc=page,
}
\newcommand{\discolorlinks}[1]{{\hypersetup{hidelinks}#1}}
\usepackage[font={small}]{caption}
\usepackage{float}
\usepackage[parfill]{parskip}
\usepackage{multirow}
\usepackage{braket}
\newcommand{\e}[1]{é}
\newcommand{\JL}[1]{\textcolor{blue}{#1}}

\newcommand{\edit}[1]{\textcolor{green}{#1}}

\begin{document}

\begin{titlepage}
\thispagestyle{empty}
\begin{centering}
	\vspace*{1cm}
	{\scshape\Huge A pedagogical review of gravity as a gauge theory \par}
	\vspace*{1cm}
	{\mdseries\Large Jason Bennett\par}
	\vspace*{1cm}
	{\mdseries Supervised by Professor Eric Bergshoeff, Johannes Lahnsteiner, and Ceyda Şimşek\par}
	
\end{centering}	
\vspace*{2cm}

After showing how Albert Einstein's general relativity (GR) can be viewed as a gauge theory of the Poincaré algebra, we show how Élie Cartan's geometric formulation of Newtonian gravity (Newton-Cartan gravity) can be viewed as a gauge theory of the Bargmann algebra following the construction of [R. Andringa, E. Bergshoeff, S. Panda, and M. de Roo, ``Newtonian Gravity and the Bargmann Algebra," \href{https://iopscience.iop.org/article/10.1088/0264-9381/28/10/105011}{Class. Quant. Grav. 28 (2011) 105011}, \href{https://inspirehep.net/literature/875549}{arXiv:1011.1145 [hep-th]}]. In doing so, we will touch on the following auxiliary topics: the extension of Yang-Mills gauge theory to a more generic formalism of gauge theory, the fiber bundle picture of gauge theory along with the soldering procedure necessary to complete the gravity as a gauge theory picture, the vielbein formalism of GR, Lie algebra procedures such as central extensions and İnönü-Wigner contractions, and the hallmarks of Newtonian gravity which differentiate it from GR. The objective of the present work is to pedagogically fill in the gaps between the above citation and an undergraduate physics and mathematics education. Working knowledge of GR and some familiarity with classical field theory and Lie algebras is assumed.

\vspace*{3cm}
\begin{centering}
{\scshape \large Van Swinderen Institute, University of Groningen \par}
	{\large July 02, 2020 \par}
	
\end{centering}
\end{titlepage}

\pagebreak

\begin{titlepage}
\vspace*{5cm}
\begin{displayquote}
{\large
Physics and geometry are one family.
\newline
Together and holding hands they roam
\newline
\hspace*{20mm} to the limits of outer space.
\newline
Black hole and monopole exhaust
\newline
\hspace*{20mm} the secret of myths;
\newline
Fiber and connections weave to interlace
\newline
\hspace*{20mm} the roseate clouds.
\newline
Evolution equations describe solitons;
\newline
Dual curvatures defines instantons.
\newline
Surprisingly, Math. has earned 
\newline
\hspace*{20mm} its rightful place
\newline
\hspace*{20mm} for man and in the sky;
\newline
Fondling flowers with a smile — just wish 
\newline
\hspace*{20mm} nothing is said!
\begin{center}
\textit{ — Shiing-Shen Chern} \cite{chern}
\end{center}}
\end{displayquote}
\end{titlepage}

\begin{titlepage}
\thispagestyle{empty}
\tableofcontents
\thispagestyle{empty}
%\JL{Johannes's comments}

%CS{Ceyda's comments}

%\edit{Things I need to edit in the future}
\end{titlepage}
\pagebreak

\section*{Acknowledgements}
\addcontentsline{toc}{section}{Acknowledgements}

I have met several versions of Professor Eric Bergshoeff. I met him as a huge name in his field through his INSPIRE-HEP profile while searching for mentors for my Fulbright project. I met him as a supportive sponsor after he responded to my proposal and we began to draft my application. Arriving in Groningen and meeting him in person I have over the course of nine months met the advisor who is deftly instructive in providing suggestions to his advisees; the professor and public speaker with an infectious love for physics; and the warm, amicable, and humorous person who both draws a crowd at social functions and who never failed to evoke at least a few fits of laughter from his advisees during every research meeting. I cannot thank Professor Bergshoeff enough for making this experience possible.

I look up to Professor Bergshoeff's PhDs Johannes Lahnsteiner and Ceyda Şimşek for so many reasons. Mastery of their subject and related fields, the ability to slowly unravel new concepts to students learning the subject for the first time, and a mystic ability to anticipate where their students are struggling are just a few reasons. Thank you both for being who you are — having you both as role models to attempt to emulate as I begin my PhD next year means the world.

The courses I have taken here have been some of the best in my life because of the courses' professors and other students. Thank you Daniël Boer, Simone Biondini, Eric Bergshoeff, Elisabetta Pallante, Johannes Lahnsteiner, Anupam Mazumdar, and my classmates for valuing pedagogy so highly. Thank you Jasper Postema as well for being a helpful partner to learn the vielbein formalism of GR alongside during this work.

Thank you Arunesh Roy and Ruud Peeters for being great office mates, and thank you Ruud Peeters, Pi Haase, Alba Kalaja, Femke Oosterhof, Johannes Lahnsteiner, and Ceyda Şimşek for letting me tag along to lunch with the PhDs to feel older and smarter than I was.

Thank you Iris de Roo-Kwant, Annelien Blanksma, and Hilde van der Meer of the Van Swinderen Institute (VSI), University of Groningen's International Service Desk, and Gemeente Groningen's International Welcome Center North for working for months with me to iron out the practical matters of coming to Groningen. Your assistance made this process incredibly smooth.

Thank you Gideon Vos for passionately explaining advanced topics on our train trips to Delta Holography meetings; Roel Andringa for supplementing his masterpiece of a thesis with some pointers that were very helpful for parts of this work; Diederik Roest for bringing out the best in every speaker by being a brilliantly social physicist and asking great questions; Simone Biondini, Sravan Kumar, Ivan Kolar, and Luca Romano for organizing lunch talks and journal clubs for VSI; and thank you to the organizers of the Delta ITP Holography meetings for creating a great network of science in the Netherlands/Belgium.

Thank you to the Fulbright Scholarship program and the Netherland-America Foundation for providing the financial support and infrastructure necessary for programs like this to exist.

Thank you Kelly Sorensen, Sera Cremonini, Eric Bergshoeff, Tom Carroll, Lew Riley, and Becky Evans for tirelessly working with me to polish, re-polish, ... and re-polish my application to the Fulbright. Thank you Tom Carroll, Sera Cremonini, Erin Blauvelt, Lew Riley, Nicholas Scoville, Christopher Sadowski, Casey Schwarz, and Kassandra Martin-Wells for teaching me the ways of physics, mathematics, research, teaching, writing, and outreach.

Thank you Charley for showing me this country in a way I never expected, by going on unforgettable dates with an amazing girlfriend. Dankje mijn aanmoediger for supporting me at every turn. And thank you to my family for visiting me in my amazing world here in Groningen during my stay and supporting me always.

\pagebreak

\section*{Outline}
\addcontentsline{toc}{section}{Outline}

The objective of the present work is to pedagogically fill in the gaps between an undergraduate physics and mathematics education  and a comprehensive understanding of the gauging procedure that one follows in order to build up a gravitation theory from an algebra. Working knowledge of GR and some familiarity with classical field theory and Lie algebras is assumed.
In Chapter \ref{sec:intro} we introduce the idea of using the tools of mathematics to study symmetries underlying physical systems, and we provide two motivations (which can also be viewed as further directions) for this work. As a warm up to introduce ourselves to gauge theory before diving into the formalism, in Chapter \ref{chap:gauge1} we look at the gauging of a U(1) symmetry. Then in Chapter \ref{chap:diggingDeeper} we explore the simplest non-abelian gauge theory, SU(2) Yang-Mills theory, taking the scenic route by exploring geometric interpretations of the gauge field/connection, the covariant derivative, and the field strength/curvature. As a prerequisite to working with gravity as a gauge theory, in Chapter \ref{sec:manylegs} we hash out the vielbein formalism of general relativity. In Chapter \ref{chap:lgauge} we take a final step in pure gauge theory by considering the pure Lorentz algebra. In Chapter \ref{gauge} we formalize the concepts introduced in previous chapters to describe symmetry transformations in a totally abstract formulation of gauge theory. Chapter \ref{sec:pgauge} marks the distinction between pure gauge theory and gravity as a gauge theory, where we study the issues that arise when naively gauging local spacetime translations. In Chapter \ref{sec:NRlas} we introduce the non-relativistic counterparts to the symmetry algebras of previous chapters. Finally, in Chapter \ref{sec:nrgauge}, after introducing Newtonian gravity as well as the frame-independent geometric formulation of it — Newton-Cartan gravity, we work through the gauging procedure a second time. This time, as opposed to gauging the Poincaré algebra to reproduce GR, we gauge the Bargmann algebra to reproduce Newton-Cartan gravity. After summarizing what we have accomplished in this work in Chapter \ref{sec:conclusion}, we outline some further directions (which also serve as motivations) for this work in Chapter \ref{sec:fd}.

\pagebreak
\section{Introduction}
\label{sec:intro}

Symmetries of a physical system can be catalogued in an algebraic framework. For instance, the symmetry group of spatial rotations is SO(3), the symmetry group of spatial rotations and Lorentz spacetime boosts is the Lorentz group, and adding translations in spacetime to the Lorentz group forms the Poincaré group. Each symmetry transformation in these collections correspond to a particular group element of the symmetry group (called a Lie group — a group with a manifold structure). One can also think about infinitesimally small symmetry transformations. In this case, we can look at a Lie group's associated Lie algebra and study the structure of that. While a Lie algebra may not always capture the global/topological aspects of the Lie group, working with algebras is in most cases totally sufficient to extract physically interesting data. On top of the beauty of studying the mathematical structure underlying physical symmetries, there exists a way to \textit{build up} physical theories directly from the structure of Lie algebras.

In the context of field theory, there exists local (spacetime dependent) symmetry transformations whose structure is defined by a Lie group and corresponding Lie algebra. If a physical theory is invariant under these local symmetry transformations, it is coined, a \textit{gauge} theory. While not realized at the time, Maxwell's theory of electromagnetism is the simplest example of a gauge theory, the \textit{gauge (Lie) group} of the symmetries transformations on the theory is the U(1) circle group \cite{EM}. More complicated gauge theories came into play when Yang and Mills studied the orientation of the isotopic spin \cite{YM}. It turned out that, just as the electromagnetic field necessitated certain invariance properties, the existence of the Yang-Mills field necessitated certain invariances. It was in this background that researchers in the 1950s and 1960s began to consider the relationship between the existence of the gravitational field and Lorentz invariance \cite{gaugegravity}.

Beginning with the Lorentz algebra, Utiyama began work on gauging an algebraic structure to obtain a theory of gravity in 1956 \cite{uti}. Several years later in 1960, Sciama and Kibble extended Utiyama's theory by considering the full Poincaré algebra \cite{Sciama} \cite{Kibble}. Their derivation of Einstein's theory of general relativity marked the beginning of the perspective of gravity as a gauge theory.

There are several motivations for studying gravity as a gauge theory.
\pagebreak
\subsection{Motivation \# 1: Quantum gravity}

The Bronstein cube of gravitation (G), relativity (c), and quantum mechanics ($\hbar$) helps to visualize three possible routes to a theory of quantum gravity (theory of everything) \cite{bronstein}. 

\begin{figure}[H]
\centering
\includegraphics[scale=0.5]{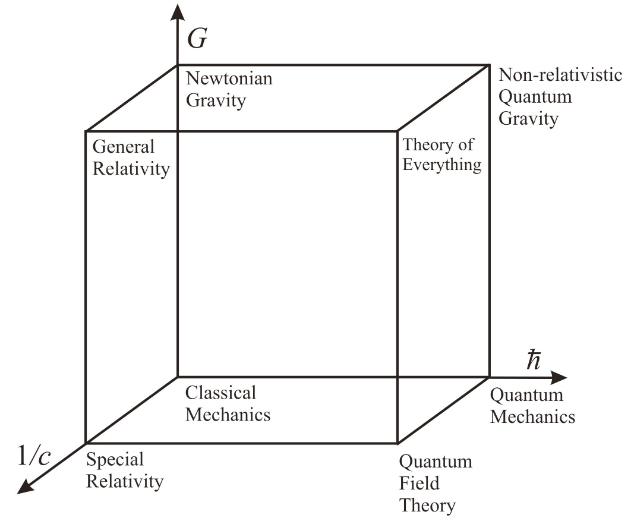}
\captionsetup{format=hang}
\caption{Bronstein cube \cite{cubepic}.}
\label{fig:bronstein}
\end{figure}

Notably, one can try to quantize general relativity (as loop quantum gravity attempts), add gravity to quantum field theory (as string theory attempts), or let velocities approach the speed of light in a theory of non-relativistic (NR) quantum gravity \cite{johns}. One problem in this last NR quantum gravity route is the lack of an understanding of this corner of the Bronstein cube. While general relativity and quantum field theory are two of the most astonishingly successful theories in physics, there does not exist a theory of NR quantum gravity. One route to approach such a theory would be to start at the origin of the Bronstein cube, progress along the G-axis toward a theory of Newtonian gravity, and then along the line parallel to the $\hbar$-axis to our goal. This first step towards a theory of Newtonian gravity is notable in its own right.

Newton developed the first formalization of gravity with his law of universal gravitation \cite{newton}. However, Newtonian gravity as originally formulated by Newton is not a frame-independent theory like general relativity. In developing general relativity, Einstein achieved two things \cite{gr}. First, he recognized that the gravitational force does not act instantaneously like in Newtonian gravity, but instead propagates at the speed of light, by making use of the curvature of spacetime to describe gravity. In this way he made gravity consistent with his theory of special relativity, which states that no information can propagate faster than the speed of light. To describe the curvature of spacetime he needed to use a piece of mathematics, called Riemannian geometry, that was developed in 1854 by Bernhard Riemann and that was not available when Newton formulated his theory. Secondly, Einstein gave a frame-independent formulation of his relativistic gravitational theory. It was only 8 years later that Cartan was able to give a frame-independent formulation of Newtonian gravity, called Newton-Cartan (NC) gravity, using the geometric ideas of Einstein \cite{nc}. It is NC gravity that occupies the ``Newtonian gravity" corner of the Bronstein cube.

The motivation to study gravity as a gauge theory was revamped in 2011 when Professor Bergshoeff et al. discovered a way to mimic the procedure of Utiyama, Sciama, and Kibble to obtain NC gravity as opposed to Einstein's GR \cite{bergshoeff}. Notably, instead of gauging a relativistic symmetry algebra like the Poincaré algebra, they gauged a version of the NR Galilean algebra called the Bargmann algebra. It was not long before this method of gauging algebras provided a route to the NR quantum gravity corner of the cube. The next year, Bergshoeff et al. gauged an extended `stringy' Galilean algebra to obtain a string-theoretical version of NC gravity \cite{stringnc}. In this way, the procedure of gauging Lie algebras has made a significant advancement in the NR quantum gravity corner of the Bronstein cube by introducing new techniques to build NR string theories. The more developed this corner of the cube gets, the closer we are to establishing a bona fide third route to a theory of quantum gravity.  This progress in NR string theory has spurred further developments in the field, with Groningen remaining one of the leading programs at the forefront of these developments \cite{nrs1} \cite{nrs2} \cite{nrs3} \cite{nrs4} \cite{nrs5} \cite{nrs6}.

\pagebreak
\subsection{Motivation \# 2: Non-relativistic holography}

The use of perturbation theory in quantum field theory (QFT) has been wildly successful to the point of developing the most rigorously tested theory of physics, the Standard Model. However, perturbation theory relies on the ability to modify known solutions by introducing small perturbations — thus reaching new solutions. These perturbations are proportional to the interaction strength (between the constituents) of the system. When the system's interaction strength is large, there is no small quantity with which to perturb the known solution by. And so the system is deemed non-perturbative — methods other than perturbation theory are necessary to explore the properties of the system. 

Holography has emerged as an incredibly enticing way to study non-perturbative systems. The holographic principle states that all information about gravity in a given volume of space (often called the bulk) can be viewed as encoded in fewer dimensions by a QFT on the boundary of said volume \cite{'tHooft} \cite{susskind}. The most thoroughly studied case of holography is the anti-de Sitter/conformal field theory correspondence (AdS/CFT) \cite{maldacena}. The conjecture equates gravitational (string) theories in (d+1)-dimensional AdS to d-dimensional conformal QFTs on the boundary of AdS.

\begin{figure}[H]
\centering
\includegraphics[scale=0.5]{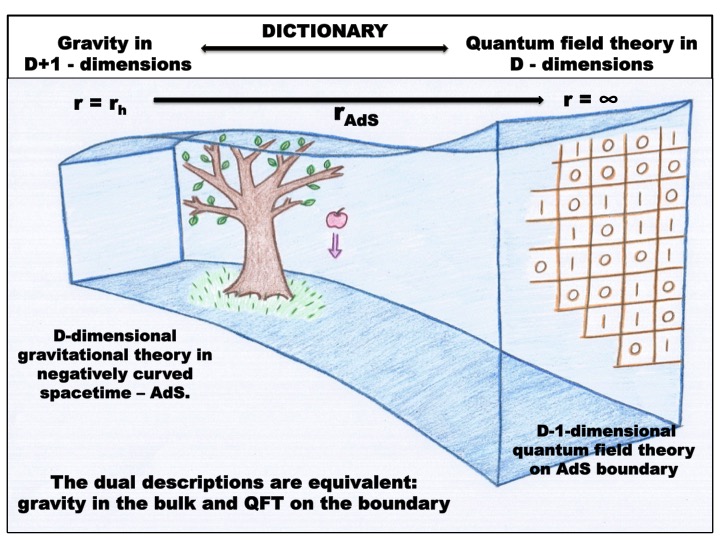}
\captionsetup{format=hang}
\caption{A cartoon depiction of the AdS/CFT correspondence \cite{steven}.}
\label{fig:adscft}
\end{figure}

Holography represents a special kind of duality (statement of equivalence) called a weak-strong duality. This means that when the QFTs on the boundary are strongly interacting (and as discussed above, perturbation theory fails), the gravitational theories in the bulk are necessarily weakly interacting, and perturbation theory can be successfully applied on that side of the duality. In other words, upon encountering a seemingly intractable strongly interacting quantum system, we can translate it to its dual (equivalent) weakly interacting gravitational system, and use perturbation theory on the weakly interacting side of the duality — thus garnering information about a previously non-perturbative original system. 

While holography was originally formulated using relativistic gravity in the volume of the space (the bulk) and relativistic QFTs at the boundary of the volume (the boundary), this need not always be the case. In fact, as holography began to be seen as a tool, it was soon realized that it could be generalized to the NR case to work with the vast majority of QFTs that are NR \cite{soncoldatoms}\cite{mcgreevy}\cite{herzog}\cite{maldacenaprime}. Currently, NR phenomena that have been probed with the aid of non-relativistic holography include: condensed matter systems such as strange metals and high temperature superconductors, ultra cold atomic systems, and quantum critical points.

However, in this picture of NR holography, for an NR QFT with a given symmetry, a dual gravity is proposed which realizes the symmetry as an isometry of its geometry, and then the gravity theory is embedded into an (at the end of the day still) \textit{relativistic} string theory \cite{taylor}. One can naturally envision a more ``pure" version of NR holography where one would use NR string theory in the bulk to describe an NR QFT on the boundary \cite{nrh1} \cite{nrh2} \cite{nrh3} \cite{nrh4}. This new method allows researchers to probe physical systems whose NR symmetries cannot be aptly described by relativistic gravitational theories in the bulk

Here we come to how charting the NR quantum gravity corner of the Bronstein cube can lead to revelations in NR holography. This method of gauging NR algebras to obtain novel NR string theories as Professor Bergshoeff has done opens the door to new opportunities in NR holography because there exists many more algebras with conformal symmetries that could be gauged to obtain useful gravitational theories to add to holography's toolbox, notably the Galilean algebra with conformal symmetries added, the Schrödinger algebra and the Lifshitz algebra. This variation is a result of the non-uniqueness of NR gravity. Unlike the unique relativistic gravity of GR, there exist several distinct NR gravitational theories all with their own properties.

\pagebreak

\section{U(1) gauging procedure}
\label{chap:gauge1}
Consider the following complex field Lagrangian,

\begin{equation}
    \mathcal{L} = \partial_\mu \bar{\phi}\partial^\mu \phi - m^2 \bar{\phi}\phi
\end{equation}

This has a global symmetry, notably that fields can transform like
\begin{eqnarray} 
\notag
\phi(x) &\rightarrow& e^{i\alpha}\phi(x)
\\
\bar{\phi}(x) &\rightarrow& e^{-i\alpha}\bar{\phi}(x)
\end{eqnarray}
and leave the Lagrangian invariant, i.e. the same as it was before the transformation.

However, if we ``gauge'' this symmetry, i.e. make it local, so that the fields transform like,
\begin{eqnarray} 
\notag
\phi(x) &\rightarrow& e^{i\alpha(x)}\phi(x)
\\
\bar{\phi}(x) &\rightarrow& e^{-i\alpha(x)}\bar{\phi}(x)
\end{eqnarray}
then the Lagrangian is no longer invariant.

To see why, consider the quantity, $\partial_\mu (\phi(x))$, we would like this to transform like the fields do, i.e. $\partial_\mu (\phi(x)) \rightarrow e^{i\alpha(x)}\partial_\mu (\phi(x))$, so that the $\partial_\mu \phi(x)\partial^\mu \bar{\phi(x)}$ term in the Lagrangian is invariant. Let's work $\partial_\mu (\phi(x))$ out,

\begin{eqnarray} 
\notag
\partial_\mu\phi(x) &\rightarrow& \partial_\mu(e^{i\alpha(x)}\phi(x))
\\
\notag
&=& \partial_\mu(e^{i\alpha(x)}))\phi(x) +e^{i\alpha(x)}\partial (\phi(x))
\\
\notag
&=& i\partial_\mu(\alpha(x))e^{i\alpha(x)}\phi(x) +e^{i\alpha(x)}\partial (\phi(x))
\\
&\neq& e^{i\alpha(x)}\partial_\mu (\phi(x))
\end{eqnarray}

So let's define a new derivative, which we will call the ``covariant derivative," by $D_\mu=\partial_\mu-iA_\mu(x)$, where $A_\mu(x)$ is a gauge field. So then our new Lagrangian that we claim is invariant is 

\begin{equation}\mathcal{L}' = D_\mu \bar{\phi}D^\mu \phi - m^2 \bar{\phi}\phi\end{equation}

Now let's see whether our new fancy derivative satisfies, $D_\mu(\phi(x)) \rightarrow e^{i\alpha(x)}D_\mu(\phi(x))$,

\begin{eqnarray} 
\notag
D_\mu(\phi(x)) &\rightarrow& D_\mu(e^{i\alpha(x)}\phi(x))
\\
\notag
&=& \partial_\mu(e^{i\alpha(x)}\phi(x)) -i A_\mu(x) e^{i\alpha(x)}\phi(x)
\\
\notag
&=& i\partial_\mu(\alpha(x))e^{i\alpha(x)}\phi(x) +e^{i\alpha(x)}\partial (\phi(x)) -i A_\mu(x) e^{i\alpha(x)}\phi(x)
\\
\notag
&=& e^{i\alpha(x)}(i\partial_\mu(\alpha(x))\phi(x) + \partial (\phi(x)) -i A_\mu(x) \phi(x))
\\
&\neq& e^{i\alpha(x)}D_\mu(\phi(x))
\end{eqnarray}

Something is clearly amiss. There is another field in the mix now, $A_\mu(x)$. If we want the quantity in the parenthesis above to equal $D_\mu(\phi(x))$ maybe we can define how the $A_\mu(x)$ field transforms to make everything work out.

Let's require $A_\mu(x)$ to transform like,

\begin{equation}A_\mu(x) \rightarrow A_\mu(x) + \partial_\mu(\alpha(x))\end{equation}

and see how things work out. 
\begin{eqnarray} 
\notag
D_\mu(\phi(x)) &\rightarrow& D_\mu(e^{i\alpha(x)}\phi(x))
\\
\notag
&=& \partial_\mu(e^{-i\alpha(x)}\phi(x)) -i A_\mu(x) e^{i\alpha(x)}\phi(x)
\\
\notag
&\rightarrow& i\partial_\mu(\alpha(x))e^{i\alpha(x)}\phi(x) +e^{i\alpha(x)}\partial (\phi(x)) -i A_\mu(x) e^{i\alpha(x)}\phi(x) - i\partial_\mu(\alpha(x))e^{i\alpha(x)}\phi(x)
\\
\notag
&=& e^{i\alpha(x)}\partial (\phi(x)) -i A_\mu(x) e^{i\alpha(x)}\phi(x)
\\
\notag
&=& e^{i\alpha(x)}(\partial (\phi(x)) -i A_\mu(x)\phi(x))
\\
&=& e^{i\alpha(x)}D_\mu(\phi(x))
\end{eqnarray}

It should be clear now that the first term in $\mathcal{L}'$ is invariant,

\begin{eqnarray} \notag 
D_\mu(\phi(x)) &\rightarrow& e^{i\alpha(x)}D_\mu(\phi(x))
\\ \notag 
D^\mu(\bar{\phi}(x)) &\rightarrow& e^{-i\alpha(x)}D^\mu(\bar{\phi}(x))
\\ \notag 
D_\mu(\phi(x))D^\mu(\bar{\phi}(x)) &\rightarrow& e^{i\alpha(x)-i\alpha(x)}D_\mu(\phi(x))D^\mu(\bar{\phi}(x))
\\
&=& D_\mu(\phi(x))D^\mu(\bar{\phi}(x))
\end{eqnarray}

To finish up, lets calculate the variance of a few crucial quantities, $\delta \phi(x), \delta D_\mu(\phi(x)),$ and $\delta A_\mu(x).$ 

%\CS{What is the Noether current corresponding to that symmetry? How many conserved quantity does this system have?}

For infinitesimal $\alpha(x)$ and ignoring $\mathcal{O}(\alpha(x))^2$ terms, we have
\begin{eqnarray} \notag
\delta \phi(x) &=& \phi(x)'-\phi(x)
\\ \notag
&=& e^{i\alpha(x)}\phi(x)-\phi(x)
\\ \notag
&\approx& (1 +i\alpha(x) + \mathcal{O}(\alpha(x))^2 + ...)\phi(x)-\phi(x)
\\ \notag
&=& \phi(x) + i\alpha(x)\phi(x) - \phi(x)
\\
&=& i\alpha(x)\phi(x).
\end{eqnarray}

\begin{eqnarray} \notag
\delta D_\mu(\phi(x)) (x) &=& D_\mu(\phi(x))'-D_\mu(\phi(x))
\\ \notag
&=& e^{i\alpha(x)}D_\mu(\phi(x)) - D_\mu(\phi(x))
\\ \notag
&\approx& (1 +i\alpha(x) + \mathcal{O}(\alpha(x))^2 + ...)D_\mu(\phi(x)) - D_\mu(\phi(x))
\\ \notag
&=& D_\mu(\phi(x)) +i\alpha(x)D_\mu(\phi(x))- D_\mu(\phi(x))
\\
&=& i\alpha(x)D_\mu(\phi(x)).
\end{eqnarray}

And finally,

\begin{eqnarray} \notag
\delta A_\mu (x) &=& A_\mu(x)'-A_\mu(x)
\\ \notag
&=& A_\mu(x)+ \partial_\mu(\alpha(x)) -A_\mu(x)
\\ 
&=& \partial_\mu(\alpha(x)).
\end{eqnarray}

Another quantity involving $A_\mu$ we can check is invariant is $\frac{1}{4}F^{\mu\nu}F_{\mu\nu}$ under the transformation of $A_\mu$. We have,

\begin{eqnarray} \notag
F_{\mu\nu} &=& \partial_\mu (A_\nu(x)) - \partial_\nu (A_\mu(x))
\\ \notag
&\rightarrow& \partial_\mu (A_\nu(x)+ \partial_\nu(\alpha(x))) - \partial_\nu (A_\mu(x)+ \partial_\mu(\alpha(x)))
\\ \notag
&=& \partial_\mu (A_\nu(x))+ \partial_\mu\partial_\nu(\alpha(x)) - \partial_\nu (A_\mu(x)) + \partial_\nu\partial_\mu(\alpha(x))
\\ \notag
&=& \partial_\mu (A_\nu(x)) - \partial_\nu (A_\mu(x)) 
\\
&=& F_{\mu\nu} \label{fmunuu1}
\end{eqnarray}

\iffalse
\textbf{Question: Why is $\partial_\mu\partial_\nu(\alpha) = - \partial_\nu\partial_\mu(\alpha)$? In differential forms/vielbein context, maybe that makes sense (basis 1-forms are antisymmetric), but what about the rule from high school math, $\frac{\partial}{\partial x}\frac{\partial}{\partial y} f= \frac{\partial}{\partial y}\frac{\partial}{\partial x} f$?}
\fi

And similarly for $F^{\mu\nu}$, and thus their product is as well.

\pagebreak
\subsection{Noether current}

%Addressing Ceyda's comment, we first define a Noether current from David Tong's QFT Lecture Notes.

Following the notation of \cite{tong}, let $X_a(\phi_a)$ be some function of one of the fields, $\phi_a$.

The transformations 

\begin{equation}\delta \phi_a(x) = X_a(\phi)\end{equation}

are symmetries if 

\begin{equation}\delta \mathcal{L} = \partial_\mu F^\mu,\end{equation}

i.e. the Lagrangian changes my a total-directive/four-divergence, where $F^\mu(\phi)$ are some arbitrary functions of $\phi$.

Noether's theorem then reads,

\begin{equation}\partial_\mu j^\mu = 0\end{equation}

where $j^\mu$ is the Noether current,

\begin{equation}j^\mu = \frac{\partial \mathcal{L}}{\partial(\partial_\mu \phi_a)} X_a(\phi) - F^\mu(\phi)\end{equation}

First, lets identity the Noether current for the global U(1) gauge theory.

We have 

\begin{equation}\mathcal{L} = \partial_\mu\bar{\phi}\partial^\mu \phi - m^2\bar{\phi}\phi\end{equation}

and the transformations,

\begin{eqnarray} \notag
\phi &\rightarrow& e^{i\alpha}\phi
\\
\bar{\phi} &\rightarrow& e^{-i\alpha}\bar{\phi}
\end{eqnarray}

And so in the context of $\delta \phi_a(x) = X_a(\phi)$, we have (with $\alpha$ infinitesimal)

\begin{eqnarray} \notag
\delta\phi &=& X_\phi = i\alpha\phi 
\\
\delta\bar{\phi} &=& X_{\bar{\phi}} = -i\alpha\bar{\phi} 
\end{eqnarray}

Since $\delta \mathcal{L} = 0$, there will be no total derivative term, $\partial_\mu F^\mu$ in this case. (This is not the case if our symmetry of concern was spacetime translations.

So let's begin:

\begin{eqnarray} \notag
j^\mu &=& \frac{\partial \mathcal{L}}{\partial(\partial_\mu \phi_a)} X_a(\phi) - F^\mu(\phi) \\ \notag
&=& \frac{\partial \mathcal{L}}{\partial(\partial_\mu \phi)} X_\phi + \frac{\partial \mathcal{L}}{\partial(\partial_\mu \bar{\phi})} X\bar{\phi} \\ \notag
&=& \partial^\mu(\bar{\phi})i\alpha\phi - \partial^\mu(\phi) i\alpha\bar{\phi} \\
&=& i\alpha(\partial^\mu(\bar{\phi})\phi - \partial^\mu(\phi) \bar{\phi})
\end{eqnarray}

Consider the following, 

Let $J^\mu$ be the current we just found, and let $j^\mu$ be the same expression without the $\alpha$ so that $J^\mu = \alpha j^\mu.$ By Noether's theorem, the current we found is conserved, $\partial_\mu J^\mu=0$. Notice that $\alpha$ does not depend on spacetime, notably, we aren't working with $\alpha(x)$. Thus the above equation becomes,

\begin{equation}\partial_\mu J^\mu= \partial_\mu (\alpha j^\mu) = \alpha \partial_\mu j^\mu = 0\end{equation}

and by dividing both sides by $\alpha$ we have that $\partial_\mu j^\mu=0$. And so following this common convention \cite{peskinschroeder} \cite{tong}, we write our current as $j^\mu$ instead of $J^\mu$. Notably, 

\begin{equation} j^\mu = i(\partial^\mu(\bar{\phi})\phi - \partial^\mu(\phi) \bar{\phi})\end{equation}

For the local U(1) case, we have 

\begin{eqnarray} \notag
\mathcal{L} &=& D_\mu \bar{\phi}D^\mu \phi - m^2 \bar{\phi}\phi \\ \notag
D_\mu &=& \partial_\mu-iA_\mu \\ \notag
X_\phi &=& i\alpha(x)\phi \\
X_{\bar{\phi}} &=& -i\alpha(x)\bar{\phi}
\end{eqnarray}

So we expand $\mathcal{L}$ to get, 
\begin{eqnarray} \notag
\mathcal{L} &=& D_\mu \bar{\phi}D^\mu \phi - m^2 \bar{\phi}\phi \\ \notag
&=& (\partial_\mu\bar{\phi}+iA_\mu\bar{\phi})(\partial^\mu\phi-iA^\mu\phi) - m^2\bar{\phi}\phi\\ 
&=& \partial_\mu\bar{\phi}\partial^\mu\phi -\partial_\mu\bar{\phi}iA^\mu\phi + iA_\mu\bar{\phi}\partial^\mu\phi + A_\mu\bar{\phi}A^\mu\phi - m^2\bar{\phi}\phi
\end{eqnarray}

\begin{eqnarray} \notag
j^\mu &=& \frac{\partial \mathcal{L}}{\partial(\partial_\mu \phi_a)} X_a(\phi) - F^\mu(\phi) \\ \notag
&=& \bigg(\frac{\partial \mathcal{L}}{\partial(\partial_\mu \phi)}\bigg) X_\phi + \bigg(\frac{\partial \mathcal{L}}{\partial(\partial_\mu \bar{\phi})}\bigg) X\bar{\phi} \\ \notag
&=& (\partial^\mu\bar{\phi} + iA^\mu\bar{\phi})i\alpha(x)\phi - (\partial^\mu\phi - iA^\mu\phi)i\alpha(x)\bar{\phi} \\ \notag
&=& \alpha(x)(i\phi\partial^\mu\bar{\phi} - A^\mu\bar{\phi}\phi -i\alpha(x)\bar{\phi}\partial^\mu\phi - A^\mu\bar{\phi}\phi) \\ \notag
&=& i(\phi\partial^\mu\bar{\phi} + iA^\mu\bar{\phi}\phi) - i(\bar{\phi}\partial^\mu\phi + A^\mu\bar{\phi}\phi) \\ 
&=& i(\phi D^\mu\bar{\phi} - \bar{\phi} D^\mu \phi)
\end{eqnarray}

Because the U(1) algebra is 1-dimensional, i.e. our gauge theory has 1 symmetry transformation parameter, there exists 1 conserved Noether charge.

\pagebreak

\subsection{Global, local, rigid, and spacetime}
\begin{figure}[H]
\centering
\includegraphics[scale=0.4]{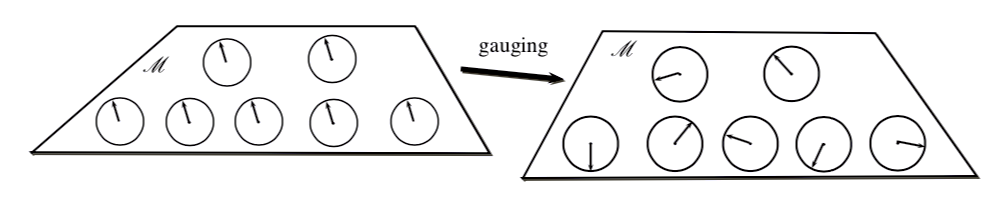}
\captionsetup{format=hang}
\caption{A global symmetry (on the left) in a theory implies the field can be rotated in the same way in every spacetime point and leave the Lagrangian invariant, whereas for a theory to have a gauge/local symmetry (on the right) i.e., the field must be able to be rotated differently at every spacetime point while leaving the Lagrangian invariant \cite{Roelthesis}.}
\label{fig:gauging}
\end{figure}

The gauging procedure we just outlined involved taking a ``global'' $(\phi(x) \rightarrow e^{i\alpha}\phi(x))$ to a ``local'' $(\phi(x) \rightarrow e^{i\alpha(x)}\phi(x))$ symmetry. A neat way to visualize this is with the following picture, see Figure \ref{fig:gauging}

There exists another distinction in categorizing symmetries — ``internal (rigid)'' versus ``spacetime.''
\newline
An example of a global internal symmetry is $\phi(x) \rightarrow e^{i\alpha}\phi(x)$.
\newline
An example of a local internal symmetry is $A_\mu(x) \rightarrow A_\mu(x) + \partial_\mu (\alpha(x))$.
\newline
An example of a global spacetime symmetry is $\phi(x) \rightarrow \phi(\Lambda x + a)$.
\newline
An example of a local spacetime symmetry comes from General Relativity.

\pagebreak

\section{Digging deeper into gauge theory}
\label{chap:diggingDeeper}
\subsection{Gauge covariant derivative}

Where in the world did the motivation for $D_\mu = \partial_\mu-iA_\mu$ come from?

\subsubsection{Electricity and magnetism}

For some explanation, we examine the source material of the old legends \cite{YM}, and two pedagogical attempts at motivating the form of the covariant derivative from E \& M \cite{fadeev} \cite{griffiths}.

As Yang and Mills put it \cite{YM}: 
\begin{displayquote}
``In accordance with the discussion in the previous section, we require, in analogy with the electromagnetic case, that all derivatives of $\psi$ appear in the following combination:
\begin{equation}
(\partial_\mu -ie B_\mu) \psi."
\end{equation}
\end{displayquote}

However, in the previous section, they only state something more conservative, notably that:

\begin{displayquote}
``To preserve invariance one notices that in electrodynamics it is necessary to counteract the variation of $\alpha$ with x, y, z, and t by introducing the electromagnetic field $A_\mu$, which changes under a gauge transformation as 

\begin{equation}
A_\mu{}^{'} = A_\mu + \frac{1}{e} \frac{\partial \alpha}{\partial x_\mu}.
\end{equation}

In an entirely similar manner we introduce a B field in the case of the isotopic gauge transformation to counter-act the dependence of S on x, y, s, and t. ... The field equations satisfied by the twelve independent components of the B field, which we shall call the \textbf{b} field, and their interaction with any field having an isotopic spin are essentially fixed, in much the same way that the free electromagnetic field and its interaction with charged fields are essentially determined by the requirement of gauge invariance."
\end{displayquote}

This is all well and good, we know from E\& M that , ``one is free to add any function to $\mathbf{A}$ whose curl is zero, i.e. is the gradient of a scalar, and the physical quantity $\mathbf{B}$ is left unchanged since the curl of a gradient is zero", but how this informs our ansatz for the covariant derivative remains unclear.

Fadeev and Slavnov are similarly cryptic on page 4 \cite{fadeev}:

\begin{displayquote}
``The electromagnetic field interacts with charged fields, which are described by complex functions $\psi(x)$. In the equations of motion the field $A_\mu(x)$ always appears in the following combination:
\begin{equation}
\nabla_\mu \psi = (\partial_\mu - A_\mu) \psi = (\partial_\mu - iA_\mu) \psi."
\end{equation}
\end{displayquote}

Griffiths make an explicit reference to WHERE in E \& M this ansatz come from on page 360 \cite{griffiths}:

\begin{displayquote}
``The subsitution of $D_\mu$ for $\partial_\mu$, then, is a beautifully simple device for converting a \textit{globally} invariant Lagrangian into a \textit{locally} invariant one; we call it the \textit{`minimal coupling rule'.} $^\dagger$''
\end{displayquote}

Where the $\dagger$ points to an accompanying footnote that expound on this:

\begin{displayquote}
``The minimal coupling rule is much older than the principle of local gauge invariance. In terms of momentum ($p_\mu \leftrightarrow i\hbar \partial_\mu$) reads $p_\mu \rightarrow p_\mu - i(q/c)A_\mu$, and is a well-known trick in classical electrodynamics for obtaining the equation of motion for a charged particle in the presence of electromagnetic fields. It amounts, in this sense, to a sophisticated formulation of the Lorentz force law. In modern particle theory we prefer to regard local gauge invariance as fundamental and minimal coupling as the vehicle for achieving it."
\end{displayquote}

This occurs in the ``minimal coupling Hamiltonian,'' 

\begin{equation}H= \frac{1}{2m} (\mathbf{p}-q\mathbf{A})^2 +q\phi\end{equation}

which is indeed used in quantum mechanics, see Sakurai's eq 2.7.28 where the dynamics of the Schrodinger equation for such a Hamiltonian are discussed \cite{sakurai}. 

However this \textbf{use} of the covariant derivative is not satisfying. The more helpful discussion in Sakurai is laid out a few pages further into the chapter on page 141. It goes as follows:

\begin{displayquote}
``Consider some function of position at $\mathbf{x}: F(\mathbf{x})$. At a neighboring point we obviously have

\begin{equation}F(\mathbf{x} + d\mathbf{x}) \cong F(\mathbf{x}) + (\nabla F) \cdot d\mathbf{x}. \end{equation}

But suppose we apply a scale change as we go from $\mathbf{x}$ to $\mathbf{x} + d\mathbf{x}$ as follows:

\begin{equation}1 |_{\text{at } \mathbf{x}} \rightarrow [1 + \mathbf{\Sigma}(\mathbf{x})\cdot d\mathbf{x}] | _{\text{at } \mathbf{x}+d\mathbf{x}}.\end{equation}

We must then rescale $F(\mathbf{x})$ as follows:

\begin{equation}F(\mathbf{x}+d\mathbf{x}) |_\text{rescaled} \cong F(\mathbf{x})+[(\nabla + \mathbf{\Sigma})F]\cdot d\mathbf{x}\end{equation}

... The combination $\nabla + \mathbf{\Sigma}$ is similar to the gauge-invariant combination

\begin{equation} \nabla - (\frac{i e}{\hbar c})\mathbf{A}."\end{equation}
\end{displayquote}

This discussion by Sakurai (and the discussion directly below this on page 141 regarding Weyl's \textit{geometrization} of electromagnetism) has led me to a much better way of thinking about the underpinning of the gauge \textit{covariant derivative}.

\pagebreak
\subsubsection{General relativity}
\label{subsecGR}

The Gradient of a Tensor is Not a Tensor \cite{moore}.

For some four-vector $A^\mu$, the components of the gradient of the four-vector transform when changing coordinate systems:

\begin{eqnarray} \notag
\partial^{'}{}_\nu A^{' \mu} &=& \frac{\partial A^{' \mu}}{\partial x^{' \nu}} \\ \notag
&=& \frac{\partial }{\partial x^{' \nu}}\bigg(\frac{\partial x^{' \mu}}{\partial x^ \alpha}A^ \alpha \bigg) \\ \notag
&=& \frac{\partial x^\beta}{\partial x^{' \nu}}\frac{\partial }{\partial x^ \beta} \bigg(\frac{\partial x^{' \mu}}{\partial x^{' \alpha}}A^ \alpha \bigg) \\
&=& \frac{\partial x^\beta}{\partial x^{' \nu}}\frac{\partial^2 x^{' \mu}}{\partial x^ \beta \partial x^\alpha}A^ \alpha + \frac{\partial x^\beta}{\partial x^{' \nu}} \frac{\partial x^{' \mu}}{\partial x^\alpha} (\partial_\beta A^ \alpha ).
\end{eqnarray}

The second term here is precisely what a ``one upper one lower" tensor transformed like. But this first term will not be zero in any physical circumstance in which the coordinate transformation factor ($\partial x^{' \mu}/\partial x^\nu$) is not a constant.

As Carroll puts it, not only do we need a manifold and a metric, but we need a connection to do (i.e. take derivatives in) GR \cite{carrollbook}. This \textit{connection} accounts for the first term in our expression above, reconciling the fact that we are must take into account the different bases of different tangent spaces that vectors live in as we take derivatives. 

\begin{displayquote}
``Now that we know how to take covariant derivatives, let's step back and put this in the context of differentiation more generally. We think of a derivative as a way of quantifying how fast something is changing. In the case of tensors, the crucial issue is `changing with respect to what?' An ordinary function defines a number at each point in spacetime, and it is straightforward to compare two different numbers, so we shouldn't be surprised that the partial derivative of functions remained valid on arbitrary manifolds. But a tensor is a map from vectors and dual vectors to the real numbers, and it's not clear how to compare such maps at different points in spacetime. Since we have successfully constructed a covariant derivative, can we think of it as somehow measuring the rate of change of tensors? The answer turns out to be yes: the covariant derivative quantifies the instantaneous rate of change of a tensor field in comparison to what the tensor would be if it were `parallel transported.' In other words, a connection defines a specific way of keeping a tensor constant (along some path), on the basis of which we can compare nearby tensors."
\end{displayquote}

This definition of the covariant derivative in GR follows \cite{moore}.

Firstly, we define the connection:

\begin{equation} \frac{\partial \mathbf{e}_\alpha}{\partial x^\mu} = \Gamma^\nu_{\mu\alpha} \mathbf{e}_\nu\end{equation}

\begin{displayquote}
``The partial derivative here is the differential change in the basis vector $\mathbf{e}_\alpha$ as we move from P to a point a differential displacement $\partial x^\mu$ along along a curve where the other coordinates are constant, divided by that differential displacement $\partial x^\mu$. ... The four coefficients $\Gamma^\nu_{\mu\alpha}$ appearing in that sum also depend on the component direction of the displacement (specified by $\mu$) and which particular basis vector $\mathbf{e}_\alpha$ is being examined..."
\end{displayquote}

And as PhysicsPages adds \cite{physicspages}:

\begin{displayquote}
``Since the derivative of a vector is another vector,and the basis vectors span the space,  we can express this derivative as a linear combination of the basis vectors at the point at which the derivative is taken."
\end{displayquote}

Considering the change in a vector $\mathbf{A}=A^\mu \mathbf{e}_\mu$ as we move an arbitrary infinitesimal displacement $d$ (whose components are $dx^\alpha$), we have

\begin{eqnarray} \notag
d \mathbf{A} &=& d(A^\mu \mathbf{e}_\mu) \\ \notag
&=& d(A^\mu) \mathbf{e}_\mu + A^\mu d(\mathbf{e}_\mu) \\ \notag
&=& \bigg( \frac{\partial A^\mu}{\partial x^\sigma} d x^\sigma \bigg) \mathbf{e}_\mu + A^\mu  \frac{\partial \mathbf{e}_\mu}{\partial x^\alpha} d x^\alpha\\ \notag
&=& \bigg(\frac{\partial A^\mu}{\partial x^\sigma} d x^\sigma\bigg) \mathbf{e}_\mu + A^\mu  \Gamma^{ \nu}_{\mu\alpha} \mathbf{e}_\nu d x^\alpha\\ \notag
&=& \bigg(\frac{\partial A^\nu}{\partial x^\alpha} d x^\alpha\bigg) \mathbf{e}_\nu + A^\mu  \Gamma^{ \nu}_{\mu\alpha} \mathbf{e}_\nu d x^\alpha\\ \notag
&=& \big(\frac{\partial A^\nu}{\partial x^\alpha} + A^\mu  \Gamma^{ \nu}_{\mu\alpha}\big) \mathbf{e}_\nu d x^\alpha\\
&=& \nabla_\alpha A^\nu  \mathbf{e}_\nu d x^\alpha
\end{eqnarray}

where we have the definition of the covariant derivative:

\begin{equation} \nabla_\alpha A^\nu = \frac{\partial A^\nu}{\partial x^\alpha} + \Gamma^{ \nu}_{\mu\alpha} A^\mu. \end{equation}

Compare this to the U(1) \textbf{gauge} covariant derivative:

\begin{equation} D_\mu \psi(x) = \partial_\mu \psi (x) + i e A_\mu \psi(x).\end{equation}

$A_\mu$ \textbf{and} $\Gamma^{ \nu}_{\mu\alpha}$ \textbf{are BOTH CONNECTIONS.}

Schwartz notes this eloquently on page 489 of \cite{schwartz}

\begin{displayquote}
``In this way, the gauge field is introduced as a connection, allowing us to compare field values at different points, despite their arbitrary phases. Another example of a connection that you might be familiar with from general relativity is the Christoffel connection, which allows us to compare field values at different points, despite their different local coordinate systems."
\end{displayquote}

\begin{figure}[H]
\centering
\includegraphics[scale=0.4]{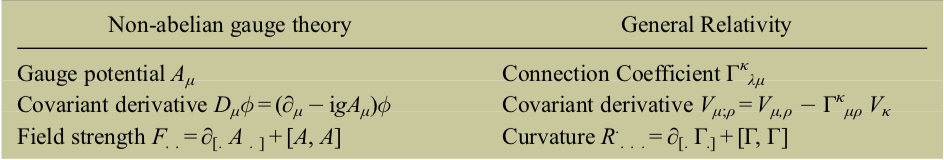}
\captionsetup{format=hang}
\caption{Ryder's comparison between non-abelian gauge theory and GR.}
\label{table:ryder}
\end{figure}

Also, see Figure \ref{table:ryder} for a table from Lewis Ryder's GR textbook \cite{ryder} comparing gauge theory to GR.

For a detailed study of this, see the following work relating these connections via the theory of \textit{fiber/vector bundles}:
\begin{itemize}
    \item section 7.10 on page 190 of Renteln \cite{MTF},
    \item section 1.8 on page 56 of Nakahara \cite{nakahara}, and 
    \item chapter 20 on page 523 of Frankel \cite{frankel}.
\end{itemize}

Given the extent of this deviation into the nether regions of understanding everything thing I do in physics based on the math underneath, I will conclude (having been finally satisfied as to the origin of this magic) with Carroll's short breadcrumb trial down this road on page \cite{carrollbook}:

\begin{displayquote}
``In the language of noncoordinate bases, it is possible to compare the formalism of connections and curvature in Riemannian geometry to that of gauge theories in particle physics. In both situations, the fields of interest live in vector spaces that are assigned to each point in spacetime. In Riemannian geometry the vector spaces include the tangent space, the cotangent space, and the higher tensor spaces constructed from these. In gauge theories, on the other hand, we are concerned with `internal' vector spaces. The distinction is that the tangent space and its relatives are intimately associated with the manifold itself, and are naturally defined once the manifold is set up; the tangent space, for example, can be thought of as the space of directional derivatives at a point. In contrast, an internal vector space can be of any dimension we like, and has to be defined as an independent addition to the manifold. In math jargon, the union of the base manifold with the internal vector spaces (defined at each point) is a fiber bundle, and each copy of the vector space is called the `fiber' (in accord with our definition of the tangent bundle).

Besides the base manifold (for us, spacetime) and the fibers, the other important ingredient in the definition of a fiber bundle is the `structure group,' a Lie group that acts on the fibers to describe how they are sewn together on overlap- ping coordinate patches. Without going into details, the structure group for the tangent bundle in a four-dimensional spacetime is generally GL$(4, R)$, the group of real invertible 4 x 4 matrices; if we have a Lorentzian metric, this may be reduced to the Lorentz group S0(3, 1). Now imagine that we introduce an internal three-dimensional vector space, and sew the fibers together with ordinary rotations; the structure group of this new bundle is then S0(3). A field that lives in this bundle might be denoted $\phi^A(x^\mu)$, where A runs from one to three; it is a three-vector (an internal one, unrelated to spacetime) for each point on the manifold. We have freedom to choose the basis in the fibers in any way we wish; this means that `physical quantities' should be left invariant under local S0(3) transformations such as
\begin{equation}\phi^A(x^\mu) \rightarrow \phi^{A^{'}}(x^\mu) = O^{A^{'}}{}_A(x^\mu)\phi^A(x^\mu)\end{equation}
where $O^{A^{'}}{}_A(x^\mu)$ is a matrix in S0(3) that depends on spacetime. Such transformations are known as gauge transformations, and theories invariant under them are called `gauge theories.'
For the most part it is not hard to arrange things such that physical quantities are invariant under gauge transformations. The one difficulty arises when we consider partial derivatives, $\partial_\mu \phi^A$. Because the matrix $O^{A^{'}}{}_A(x^\mu)$ depends on spacetime, it will contribute an unwanted term to the transformation of the partial derivative. By now you should be able to guess the solution: introduce a connection to correct for the inhomogeneous tern in the transformation law. We therefore define a connection on the fiber bundle to be an object $A_\mu{}^A{}_B$, with two `group indices' and one spacetime index."
\end{displayquote}

\pagebreak
\subsection{SU(2) Yang-Mills theory}

We consider here a classical SU(2) Yang-Mills theory constructed with two massless complex scalars $\phi_1$ and $\phi_2$ in the fundamental representation \cite{EPP}.

In components, 

\begin{equation} \mathcal{L} = \partial_\mu \phi_1{}^*(x) \partial^\mu \phi_1(x) + \partial_\mu \phi_2{}^*(x) \partial^\mu \phi_2 (x)\end{equation}

or in the ``vector" representation as objects in the fundamental representation, i.e. with $\vec{\phi} = \begin{pmatrix}
\phi_1 \\
\phi_2 
\end{pmatrix}$ and $\vec{\phi}^\dagger=\begin{pmatrix}
\phi_1{}^* & \phi_2{}^* 
\end{pmatrix}$, we have 

\begin{equation} \mathcal{L} = (\partial_\mu \vec{\phi})^\dagger \cdot \partial^\mu \vec{\phi}\end{equation}

where we've dropped the notation that makes explicit the spacetime dependence of the fields. Just recall that $\vec{\phi}=\phi^a$ is always understood to mean $\vec{\phi}(x)=\phi^a(x)$.

A group element of SU(2) (and its matrix representation) can be written as 

\begin{equation} U=e^{i\boldsymbol{\theta}\cdot \mathbf{T}} \equiv U=e^{i\theta^a T^a} \sim U=(e^{i\theta^a T^a})_{ij}\end{equation}

where $a=\{1,2,3\}$ (since SU(2) is $(2)^2-1=3$-dimensional), $i,j=\{1,2\}$ are matrix indices, $\theta^a$ are the parameters/phases of SU(2), and $T^a$ are the generators defined by

\begin{equation}\mathbf{T}=T^a=\frac{\boldsymbol{\sigma}}{2}=\frac{\sigma^a}{2}\end{equation}

where $\sigma^a$ are the Pauli matrices.

For $U \in$ SU(2) we have $U U^\dagger = \mathbf{1}$ and det(U)=1.

The su(2) Lie algebra's structure constants are the Levi-Civita symbols $\epsilon^{abc}$,

\begin{equation}[T^a, T^b]=i f^{abc} T^c = i \epsilon^{abc} T^c\end{equation}

It can be checked very quickly that under a global SU(2) transformation,

\begin{eqnarray} \notag
\vec{\phi} &\rightarrow& U \vec{\phi} \\
\vec{\phi}^\dagger &\rightarrow& \vec{\phi}^\dagger U^\dagger 
\end{eqnarray}

the 2 complex scalar Lagrangian above is left invariant since U doesn't depend on spacetime.

When we gauge the symmetry, i.e. $\theta^a \rightarrow \theta^a(x)$ and consequently $U \rightarrow U(x)$, our first step is to look for a covariant derivative (or equivalently, look for a connection) that allows us to take derivatives along a path whose end points may have different bases (or equivalently along different sections of the fiber bundle). 

``Covariant" means, transforms the same as the field itself. So we are looking for a covariant derivative that satisfies,

\begin{equation}D_\mu \vec{\phi} \rightarrow U(x) (D_\mu \vec{\phi}).\end{equation}

As we motivated in the previous section, we make the ansatz that the gauge field $A_\mu$ will be our connection, and so our covariant derivative will be of the form,

\begin{equation}D_\mu = \partial_\mu - i g A_\mu\end{equation}

where $g$ is the coupling constant of this theory. This quantity represents the strength of the force that the field of the particular gauge theory exerts. For instance $e$ was the coupling for the U(1) theory since that theory describes the electromagnetic field and the strength of the electromagnetic force is characterized by the electromagnetic charge.  Since the Standard Model is SU(3) x SU(2) x U(1), representing the strong, weak, and E\&M sectors respectively, the $g$ above is the weak force's coupling constant.

$A_\mu$ is the gauge field/connection for the theory in vector form. In component form, it is 
\begin{equation}A_\mu = A_\mu{}^a T^a\end{equation}

where $T^a$ are the SU(2) generators as before.

Pause to compare this to the general construction via Freedman-Van Proeyen that appears in Section \ref{subsec:covderiv}.

Now, if this ansatz ($D_\mu = \partial_\mu - i g A_\mu$) is correct, we will have $D_\mu \vec{\phi} \rightarrow U(x) (D_\mu \vec{\phi}).$ As will become clear, we need the gauge field/connection to transform to make this work. We will derive how it needs to transform based on the desire to satisfy $D_\mu \vec{\phi} \rightarrow U(x) (D_\mu \vec{\phi})$ given the ansatz. 

\begin{eqnarray} \notag
D_\mu \vec{\phi} &\rightarrow& (D_\mu \vec{\phi})^{'} \\ \notag
&=& \partial_\mu(\vec{\phi})^{'} - i g A_\mu{}^{ '}(\vec{\phi})^{ '} \\ \notag
&=& \partial_\mu(U(x)\vec{\phi})  - i g A_\mu{}^{ '}(U(x)\vec{\phi}) \\
&=& \partial_\mu(U(x))\vec{\phi} + U(x)\partial_\mu(\vec{\phi}) - i g A_\mu{}^{ '}U(x)\vec{\phi} 
\end{eqnarray}

To motivate the next step, recall what we're after,

\begin{eqnarray} \notag
D_\mu \vec{\phi} &\rightarrow& U(x) (D_\mu \vec{\phi}) \\
&=& U(x) (\partial_\mu\vec{\phi} - i g A_\mu\vec{\phi})
\end{eqnarray}

We will shoot for this term starting from the second term $U(x)\partial_\mu(\vec{\phi})$ in the last line of the computation above. We proceed by adding zero in the form of adding and subtracting $i g U(x) A_\mu\vec{\phi}$.

\begin{eqnarray} \notag
D_\mu \vec{\phi} &\rightarrow& (D_\mu \vec{\phi})^{'} \\ \notag
&=& \partial_\mu(U(x))\vec{\phi} + U(x)\partial_\mu(\vec{\phi}) - i g A_\mu{}^{ '}U(x)\vec{\phi} \\ \notag
&=& \partial_\mu(U(x))\vec{\phi} + U(x)\partial_\mu(\vec{\phi}) + [-i g U(x) A_\mu\vec{\phi}+i g U(x) A_\mu\vec{\phi}] - i g A_\mu{}^{ '}U(x)\vec{\phi} \\ \notag
&=&  [U(x)\partial_\mu(\vec{\phi}) - i g U(x) A_\mu\vec{\phi}]+ \partial_\mu(U(x))\vec{\phi} + i g U(x) A_\mu\vec{\phi}- i g A_\mu{}^{ '}U(x)\vec{\phi} \\ \notag
&=&  U(x) (D_\mu \vec{\phi})+ \partial_\mu(U(x))\vec{\phi} + i g U(x) A_\mu\vec{\phi}- i g A_\mu{}^{ '}U(x)\vec{\phi} \\
&=&  U(x) (D_\mu \vec{\phi})+ [\partial_\mu U(x) + i g U(x) A_\mu- i g A_\mu{}^{ '}U(x)]\vec{\phi} 
\end{eqnarray}

It is clear that our goal now reduces to getting rid of the second term. What is left at our disposal? How the gauge field/connection transforms. Notice we still have $A_\mu{}^{ '}$ in the last line. So we work on satisfying the vanishing of the second term.

\begin{eqnarray} \notag
[\partial_\mu U(x) + i g U(x) A_\mu- i g A_\mu{}^{ '}U(x)]\vec{\phi} &=& 0 \\
\partial_\mu U(x) + i g U(x) A_\mu- i g A_\mu{}^{ '}U(x) &=& 0 
\end{eqnarray}

Now we right multiply by $U^\dagger=U^{-1}$ (recall $U\in$ SU(2)) and multiply by $\frac{i}{g}$ (on both sides).

\begin{eqnarray} \notag
\partial_\mu U(x) + i g U(x) A_\mu- i g A_\mu{}^{ '}U(x) &=& 0 \\ \notag
\frac{i}{g}[\partial_\mu U(x)]U(x)^{-1} + \bigg(\frac{i}{g}\bigg)i g U(x) A_\mu [U(x)^{-1}] - \bigg(\frac{i}{g}\bigg) i g A_\mu{}^{ '}U(x)[U(x)^{-1}] &=& 0 \\
\frac{i}{g}[\partial_\mu U(x)]U(x)^{-1} - U(x) A_\mu U(x)^{-1} +  A_\mu{}^{ '} &=& 0 
\end{eqnarray}

Well now we have it!

\begin{equation}A_\mu{}^{ '} = U(x) A_\mu U(x)^{-1} -\frac{i}{g}[\partial_\mu U(x)]U(x)^{-1} \end{equation}

In other words, in order for our covariant derivative ansatz to be correct, the gauge field/connection must transform as is written above. 

We can write the covariant derivative in two different ways, in the compact vector notation, or expanded with the field indices and matrix indices made explicit where the $T^a$ are embedded in the fundamental representation,

\begin{eqnarray} \notag
D_\mu \vec{\phi} &=& \partial_\mu\vec{\phi} - i g A_\mu\vec{\phi} \\ \notag
&=& \partial_\mu\vec{\phi} - i g A_\mu{}^a T^a \vec{\phi} \\
&=& \partial_\mu \phi_i - i g A_\mu{}^a (T^a)_{ij} \phi_j 
\end{eqnarray}

For completeness, we can also write the transformation of the fields $\vec{\phi}$ and $A_\mu$ given an infinitesimal transformation. (This is possible because SU(2) is a connected Lie group. It is topologically the 3-sphere as can be seen from its Lie algebra's isomorphism between su(2) and so(3).) So (suppressing the spacetime dependence of the group elements, parameters, matter fields, and gauge fields/connections) we have that an element $U\in$SU(2) can be expressed infinitesimally (in the fundamental representation as a matrix too) as,

\begin{eqnarray} \notag
U = e^{i \theta^a T^a} &\approx& 1 + i \theta^a T^a + \mathcal{O}((\theta^a)^2) \\
U = (e^{i \theta^a T^a})_{ij} &\approx& \delta_{ij} + i \theta^a (T^a)_{ij} + \mathcal{O}((\theta^a)^2) 
\end{eqnarray}

So then the transformations of $\vec{\phi}$ and $A_\mu$ read

\begin{eqnarray} \notag
\vec{\phi} = \phi_i &\rightarrow& (e^{i \theta^a T^a})_{ij}\phi_j \\
&\approx& \delta_{ij}\phi_j + i \theta^a (T^a)_{ij}\phi_j + ...
\end{eqnarray}

for the \hypertarget{inf}{infinitesimal transformation} of $A_\mu$, $\mathcal{O}((\theta^a)^2)$ terms will be  ignored, 

\begin{eqnarray} \notag
A_\mu = A_\mu{}^a T^a &\rightarrow& U A_\mu{}^a T^a U^{-1} -\frac{i}{g}(\partial_\mu U)U^{-1}\\ \notag
&\approx& (1 + i \theta^b T^b + ...) A_\mu{}^a T^a (1 - i \theta^c T^c + ...) -\frac{i}{g}(\partial_\mu (1 + i \theta^b T^b + ...))(1 - i \theta^c T^c + ...)\\
&=& (A_\mu{}^a T^a + i \theta^b A_\mu{}^a T^b T^a  - i \theta^c A_\mu{}^a T^a T^c + ...) + \frac{1}{g}(\partial_\mu (\theta^b T^b + ...))
\end{eqnarray}

in the third term change the c index to b, and in the fourth term change the b index to a 

\begin{eqnarray} \notag
A_\mu{}^a T^a &\rightarrow& U A_\mu{}^a T^a U^{-1} -\frac{i}{g}(\partial_\mu U)U^{-1}\\ \notag
&\approx& (A_\mu{}^a T^a + i \theta^b A_\mu{}^a T^bT^a  - i \theta^c A_\mu{}^a T^a T^c + ...) + \frac{1}{g}(\partial_\mu (\theta^b T^b + ...))\\ \notag
&=& A_\mu{}^a T^a + i \theta^b A_\mu{}^a T^bT^a  - i \theta^b A_\mu{}^a T^a T^b + \frac{1}{g}\partial_\mu (\theta^a T^a)\\
&=& A_\mu{}^a T^a + i \theta^b A_\mu{}^a [T^b,T^a]  + \frac{1}{g}\partial_\mu (\theta^a T^a)
\end{eqnarray}

keep in mind something important about the third term here. The PARAMETER depends on spacetime, but the generator just not. The Lie algebra is an INTERNAL space and is not a VECTOR internal space as the Carroll quote states in Section \ref{subsecGR}. That internal vector space he refers to is a Lie group structure, NOT a Lie algebra structure.

Thus the term $\partial_\mu (\theta^a T^a)$ is actually $\partial_\mu (\theta^a(x)) T^a$. Continuing, we have

\begin{eqnarray} \notag
A_\mu{}^a T^a &\rightarrow& U A_\mu{}^a T^a U^{-1} -\frac{i}{g}(\partial_\mu U)U^{-1}\\ \notag
&\approx& A_\mu{}^a T^a + i \theta^b A_\mu{}^a [T^b,T^a]  + \frac{1}{g}\partial_\mu (\theta^a) T^a\\ 
&=& A_\mu{}^a T^a - (i)(i) \theta^b A_\mu{}^a f^{abc} T^c  + \frac{1}{g}\partial_\mu (\theta^a) T^a
\end{eqnarray}

reindexing a with c and vice versa in the second term, and then using the antisymmetry of structure constant to make 3 pairwise index switches,

\begin{eqnarray} \notag
A_\mu{}^a T^a &\rightarrow& U A_\mu{}^a T^a U^{-1} -\frac{i}{g}(\partial_\mu U)U^{-1}\\ \notag
&\approx& A_\mu{}^a T^a - (i)(i) \theta^b A_\mu{}^a f^{abc} T^c  + \frac{1}{g}\partial_\mu (\theta^a) T^a\\ \notag
&=& A_\mu{}^a T^a + \theta^b A_\mu{}^c f^{cba} T^a  + \frac{1}{g}\partial_\mu (\theta^a) T^a\\ \notag
&=& A_\mu{}^a T^a - \theta^b A_\mu{}^c f^{abc} T^a  + \frac{1}{g}\partial_\mu (\theta^a) T^a\\
A_\mu^a &\rightarrow& A_\mu{}^a - \theta^b A_\mu{}^c f^{abc}   + \frac{1}{g}\partial_\mu (\theta^a) 
\end{eqnarray}

Having proven everything is constructed nicely, we can write the Lagrangian for our theory with covariant derivatives to account for the local gauging,

\begin{equation} \mathcal{L} = (D_\mu \vec{\phi})^\dagger D^\mu \vec{\phi}
\label{anchor} \end{equation}

However, apparently this is incomplete. To quote several authors on it,

\begin{displayquote}
the above Lagrangian ``...is invariant under local gauge transformations, but we have been obliged to introduce three [for SU(2)] new vectors fields $A_\mu{}^a$, and they will require their own \textit{free} Lagrangian..." — Griffiths page 364 \cite{griffithsParticles}
\end{displayquote}

\begin{displayquote}

``To complete the construction of a locally invariant Lagrangian, we must find a kinetic energy term for the field $A_\mu$: a locally invariant term that depends on $A_\mu$ and its derivatives, but not on $\psi$." — Peskin and Schroeder page 483 \cite{peskinschroeder}

... ``Using the covariant derivative, we can build the most general gauge invariant Lagrangians involving $\psi_i$. But to write a complete Lagrangian, we must also find gauge-invariant terms that depend only on $A_\mu{}^a$. To do this, we construct the analogue of the electromagnetic field tensor." — Peskin and Schroeder page 488
\end{displayquote}

\begin{displayquote}
``We can now immediately write a gauge invariant Lagrangian, namely [the above Lagrangian]
but the gauge potential $A_\mu$ does not yet have dynamics of its own. In the familiar example
of U(1) gauge invariance, we have written the coupling of the electromagnetic potential
$A_\mu$ to the matter field $\phi$, but we have yet to write the Maxwell term $-\frac{1}{4}F_{\mu\nu}F^{\mu\nu}$ in the Lagrangian. Our first task is to construct a field strength $F_{\mu\nu}$ out of $A_\mu$." — Zee page 255 \cite{zeeQM}
\end{displayquote}

\begin{displayquote}
``The difference between this Lagrangian [the one above with covariant derivatives] and the original globally gauge-invariant Lagrangian [our original 2 complex scalar fields Lagrangian with normal partial derivatives] is seen to be the \textit{interaction Lagrangian}... This term introduces interactions between the n scalar fields just as a consequence of the demand for local gauge invariance. \textbf{However, to make this interaction physical and not completely arbitrary, the mediator A(x) needs to propagate in space. ...} [i.e. we must describe the dynamics of $A_\mu$, as the other authors put it]
\newline
The picture of a classical gauge theory developed in the previous section is almost complete, except for the fact that to define the covariant derivatives D, one needs to know the value of the gauge field A (x) at all space-time points. Instead of manually specifying the values of this field, it can be given as the solution to a field equation. Further requiring that the Lagrangian that generates this field equation is locally gauge invariant as well, one possible form for the gauge field Lagrangian is ..." [the Lagrangian given here is the one we will develop next.] — Wikipedia \cite{wiki}
\end{displayquote}

An additional aspect of the Wikipedia article that is helpful is the breaking up of the full Lagrangian for pure Yang-Mills as follows

\begin{equation}\mathcal{L} = \mathcal{L}_{\text{local}} + \mathcal{L}_{\text{kinetic}} = \mathcal{L}_{\text{global}} + \mathcal{L}_{\text{interacting}} + \mathcal{L}_{\text{kinetic}}\end{equation}

where the local Lagrangian is the local invariant one with covariant derivatives, the global Lagrangian is the global invariant one with normal derivatives, the interacting Lagrangian is the difference between the local and global, and the kinetic Lagrangian is the one with the field strength/curvature that we will develop momentarily.
\pagebreak
\subsubsection{Maxwell Lagrangian via varying actions to reproduce \textit{known} equations of motion}

Before looking into a more generic construction of the field strength/curvature for a given gauge theory, we will look at the electromagnetic field strength and it's features.

Page 244 of Zee gives a succinct derivation (from variation of an action) of the Lorentz force law for a charged particle moving in the presence of an electromagnetic field \cite{zee}.

Varying the action 

\begin{eqnarray} \notag
S &=& - \int m \sqrt{-\eta_{\mu\nu}dx^\mu dx^\nu} + V(x)dt \\ \notag
&=&- \int m \sqrt{-\eta_{\mu\nu}dx^\mu dx^\nu} + A_\mu(x)dx^\mu \\
&=&- m \int d\tau \sqrt{-\eta_{\mu\nu}\frac{dx^\mu}{d\tau} \frac{dx^\nu}{d\tau}} + \int d\tau A_\mu(x(\tau))\frac{dx^\mu}{d\tau} 
\end{eqnarray}

results in 

\begin{equation} m \frac{d^2 x^\mu}{d\tau^2} = F^\mu{}_\nu \frac{dx^\nu}{d\tau}\end{equation}

As Zee says on page 248,

\begin{displayquote}
``Electrodynamics should be a mutual dance between particles and field. The field causes the charged particles to move, and the charged particles should in turn generate the field. ... the first half of this dynamics [were shown above]. Now we have to describe the second half; in other words, we are going to look for the action governing the dynamics of $A_\mu(x)$."
\end{displayquote}

Moreover, Zee points out that on the next page

\begin{displayquote}
``Whatever emerges from varying a gauge invariant action has to be gauge invariant. The gauge potential A is not gauge invariant, but the field strength $F$ [$=\partial_\mu A_\nu - \partial_\nu A_\mu$] is."
\end{displayquote}

Indeed

\begin{eqnarray} \notag
F_{\mu\nu} &=& \partial_\mu (A_\nu(x)) - \partial_\nu (A_\mu(x))
\\ \notag
&\rightarrow& \partial_\mu (A_\nu(x)+ \partial_\nu(\alpha(x))) - \partial_\nu (A_\mu(x)+ \partial_\mu(\alpha(x)))
\\ \notag
&=& \partial_\mu A_\nu(x)+ \partial_\mu\partial_\nu(\alpha(x)) - \partial_\nu A_\mu(x) - \partial_\nu\partial_\mu(\alpha(x))
\\ \notag
&=& \partial_\mu (A_\nu(x)) - \partial_\nu (A_\mu(x)) 
\\
&=& F_{\mu\nu}
\end{eqnarray}

Since we want a Lagrangian that is gauge invariant, this object $F_{\mu\nu}$ is obviously something we should take seriously to construct an action. To make this object Lorentz invariant, we need to saturate the indices. Squaring the field strength like $F_{\mu\nu}F^{\mu\nu}$ and varying it gives

\begin{equation}\delta (F^{\mu\nu}F_{\mu\nu}) = 2 F^{\mu\nu} \delta F_{\mu\nu} = 4 F^{\mu\nu} \partial_\mu \delta A_\nu\end{equation}

Finally, in order to get nice resemblance to the free Maxwell's equations when varying the action, the $F^{\mu\nu}F_{\mu\nu}$ gets a $-\frac{1}{4}$ in the Lagrangian.

Indeed, Maxwell's equations (of motions) can be derived from the Maxwell Lagrangian  $\mathcal{L}=-\frac{1}{4}F^{\mu\nu}F_{\mu\nu}$.

But what about generalizing this? We worked with actions/variations/equations of motion in this above derivation of the Maxwell (kinetic) Lagrangian. How can we do it for a different gauge group? We do not a priori know the equations of motion for the gauge fields we introduced when gauging the symmetry.

\pagebreak
\subsubsection{Kinetic Lagrangian via differential forms, fiber bundles, and geometry}
\label{introdifforms}

Note, we call this part of the full Lagrangian the ``kinetic Lagrangian" in agreement with the discussion following equation \discolorlinks{\ref{anchor}}. The term accounts for the dynamics of the gauge field itself.

As we are faced with another impass, we turn to Carroll's bread crumb trail, i.e. a more general structure underlying what's going on here. As Zee puts it,
\begin{displayquote}
``At the same time, the fact that (11) emerges so smoothly clearly indicates a profound underlying mathematical structure. Indeed, there is a one-to-one translation between the physicist’s language of gauge theory and the mathematician’s language of fiber bundles." \cite{zeeQM}
\end{displayquote}

See Sections \ref{shittydiff}, \ref{zeevielbein}, and \ref{subsec:returnofmanylegs} for more on differential forms. 

From the connection one-form, $A=A_\mu dx^\mu$, we can define a curvature two-form 

\begin{equation}F = dA + \frac{1}{2} [A \wedge A]\end{equation}

Where $[A \wedge A]$ is a Lie algebra-valued form, but since the Lie algebras we work with are all matrix algebras, we can write this as simply $[A,A]$. In E{\&}M, the group is U(1), which is abelian, so $[A,A]=0.$ So for E{\&}M, we have $F = dA.$

Keep in mind that F is still a two-form, and so can be written in components as 

\begin{equation}F=\frac{1}{2} F_{\mu\nu}dx^\mu dx^\nu\end{equation}

Let's see what becomes of $F = dA$ written out in components (not that differential forms are anti-symmetric so that the $dx^\mu dx^\nu = - dx^\nu dx^\mu$)

\begin{eqnarray} \notag
F &=& dA \\ \notag
&=& \partial_\mu A_\nu dx^\mu dx^\nu \\ \notag
&=& \frac{1}{2}\partial_\mu A_\nu dx^\mu dx^\nu + \frac{1}{2}\partial_\mu A_\nu dx^\mu dx^\nu \\ \notag
&=&\frac{1}{2}\partial_\mu A_\nu dx^\mu dx^\nu + \frac{1}{2}\partial_\nu A_\mu dx^\nu dx^\mu \\ \notag
&=& \frac{1}{2}\partial_\mu A_\nu dx^\mu dx^\nu - \frac{1}{2}\partial_\nu A_\mu dx^\mu dx^\nu \\ \notag
&=& \frac{1}{2}(\partial_\mu A_\nu - \partial_\nu A_\mu ) dx^\mu dx^\nu
\\
&=& \frac{1}{2} F_{\mu\nu} dx^\mu dx^\nu
\end{eqnarray}

and so we identify $F_{\mu\nu}$ with $\partial_\mu A_\nu - \partial_\nu A_\mu$.

But this was for an abelian algebra, if it isn't abelian, F remains 
\begin{equation}F = dA + \frac{1}{2} [A,A] = dA + A_\mu A_\nu dx^\mu dx^\nu = dA + A^2\end{equation}

With this in mind, let's see if this F two-form even does the job for us, notably, is it gauge invariant (and can we make a gauge invariant term out of it for our Lagrangian)

We will neglect the complex i and the coupling constant to make things quicker, and take the transformation of the gauge field to be

\begin{equation}A  \rightarrow U A U^{-1} - (d U)U^{-1}\end{equation}

where U here is a 0-form so that $dU=\partial_\mu U dx^\mu$.

Hitting both sides of the transformation with an exterior derivative, noting that you pick up a minus sign whenever you pull the d through another 1-form ($dA=-Ad$ and $d(dU)=-(dU)d$, and that the boundary of a boundary is zero ($dd=0$), we have 

\begin{equation}dA  \rightarrow (dU) A U^{-1} +U(dA)U^{-1} - UA(d U^{-1}) + (d U)(dU^{-1})\end{equation}

We can also square both sides of the transformation, noting the following 

\begin{eqnarray} 
\label{eq:id}
\notag
U U^{-1} &=& 1 \\ \notag
d(U U^{-1}) &=& d(1) \\ \notag
d(U)U^{-1} + U (d U^{-1})&=& 0 \\
d(U)U^{-1} &=& - U (d U^{-1})
\end{eqnarray}

so we have,

\begin{eqnarray} \notag
A^2  &\rightarrow& (U A U^{-1} - (d U)U^{-1})(U A U^{-1} - (d U)U^{-1}) \\ \notag
&=& U A^2 U^{-1} - U A U^{-1}(d U)U^{-1} - (d U)AU^{-1} + [(d U)U^{-1}][(d U)U^{-1}] \\ \notag
&=& U A^2 U^{-1} + U A U^{-1}U(dU^{-1}) - (d U)AU^{-1} + [U(dU^{-1})][U(dU^{-1})] \\ \notag
&=& U A^2 U^{-1} + U A (dU^{-1}) - (d U)AU^{-1} + Ud(dU^{-1}) \\
&=& U A^2 U^{-1} + U A (dU^{-1}) - (d U)AU^{-1} - (dU)(dU^{-1}) 
\end{eqnarray}

Then we can add the transformations for $dA$ and $A^2$ to get 

\begin{eqnarray} \notag
dA + A^2  &\rightarrow& U(dA +A^2)U^{-1} \\
F  &\rightarrow& U (F) U^{-1}
\end{eqnarray}

So in all,

\begin{eqnarray} \notag
F  &=& dA +A^2 \\ \notag
&=& dA + \frac{1}{2} [A,A] \\ \notag
&=& \partial_\mu A_\nu dx^\mu dx^\nu + \frac{1}{2} [A_\mu,A_\nu]dx^\mu dx^\nu \\ \notag
&=& \frac{1}{2}(\partial_\mu A_\nu - \partial_\nu A_\mu)dx^\mu dx^\nu + \frac{1}{2} [A_\mu,A_\nu]dx^\mu dx^\nu \\ \notag
&=& \frac{1}{2}(\partial_\mu A_\nu - \partial_\nu A_\mu + [A_\mu,A_\nu])dx^\mu dx^\nu \\
&=& \frac{1}{2}(F_{\mu\nu})dx^\mu dx^\nu 
\end{eqnarray}

and we can restore the complex i and coupling constant g to write this as 

\begin{equation}F_{\mu\nu}=\partial_\mu A_\nu - \partial_\nu A_\mu -ig [A_\mu,A_\nu]\end{equation}

and if we write $F_{\mu\nu}$ as $F_{\mu\nu}{}^a T^a$, $A_\mu$ as $A_\mu{}^a T^a$, and using the commutation relation $[T^a,T^b]=i f^{abc}T^c$, we have 

\begin{equation}F_{\mu\nu}{}^a=\partial_\mu A_\nu{}^a - \partial_\nu A_\mu{}^a +g f^{abc} A_\mu{}^b A_\nu{}^c \end{equation}

%One point I am confused about is how this is gauge covariant. I was under the impression that gauge covariant meant that it transforms as the field (and at this point, the covariant derivative too) i.e. object= U object. The extra inverse U on the other side of $F_{\mu\nu}$ confuses me.

Recall that we can make a Lorentz invariant object out of $F_{\mu\nu}$ for the Lagrangian by saturating the indices, 

\begin{equation}F_{\mu\nu} F^{\mu\nu} \rightarrow U F_{\mu\nu} F^{\mu\nu} U^{-1} \end{equation}

and then we can take the trace of this. Recall $F_{\mu\nu}= F_{\mu\nu}{}^a T^a$, the cyclic property of the trace (tr$(ABCD)$=tr$(DABC)$) and also note that we can normalize generators of a Lie algebra in anyway we want, so we choose $trace(T^a T^b)=\frac{1}{2} \delta^{ab}$. Then we have 

\begin{eqnarray} \notag
tr(U F_{\mu\nu} F^{\mu\nu} U^{-1}) &=& tr(U^{-1} U F_{\mu\nu} F^{\mu\nu} ) \\ \notag
&=& tr(F_{\mu\nu}{}^a T^a F^{\mu\nu}{}^b T^b) \\ \notag
&=& F_{\mu\nu}{}^a F^{\mu\nu}{}^b tr(T^a T^b) \\
&=& \frac{1}{2} F_{\mu\nu}{}^a F^{\mu\nu}{}^a 
\end{eqnarray}

Since we don't know the equations of motions for the gauge field (i.e. there is no reason to account for factors that arise in varying the action), there really is no motivation for this additional factor, but to respect convention and mirror Maxwell,

\begin{equation}\mathcal{L}= -\frac{1}{2} tr(F_{\mu\nu} F^{\mu\nu}) = -\frac{1}{4} F_{\mu\nu}{}^a F^{\mu\nu}{}^a \end{equation}

As one additional point to add on to Zee's neat use of differential forms here, Wikipedia has a nice paragraph about constructing the E{\&}M field strength through fiber bundles:

\begin{displayquote}
``An elegant and intuitive way to formulate Maxwell's equations is to use complex line bundles or a principal U(1)-bundle, on the fibers of which U(1) acts regularly. The principal U(1)-connection $\nabla$ on the line bundle has a curvature $F=\nabla^2$ which is a two-form that automatically satisfies $dF = 0$ and can be interpreted as a field-strength. If the line bundle is trivial with flat reference connection d we can write $\nabla = d + A$ and $F = dA$ with A the 1-form composed of the electric potential and the magnetic vector potential. "
\cite{wikiEMmath}
\end{displayquote}
\pagebreak
\subsubsection{Kinetic Lagrangian via commutator of covariant derivatives}
\label{pesky}

If the previous section taught us anything it should be that treating the field strength as curvature was fruitful. Apart from the curvature two-form we used in the previous section, there exists one more object that defines curvature that it can be at least pedagogically useful to relate the field-strength to. The Riemann tensor.

Recall the first revelation between geometry and gauge theory:

the definition of the covariant derivative from GR reads

\begin{equation} \nabla_\alpha A^\nu = \frac{\partial A^\nu}{\partial x^\alpha} + \Gamma^{ \nu}_{\mu\alpha} A^\mu. \end{equation}

and the (U(1)) covariant derivative from gauge theory reads

\begin{equation} D_\mu \psi(x) = \partial_\mu \psi (x) + i e A_\mu \psi(x)\end{equation}

where both the Christoffel symbols and the gauge field play the role of a connection stemming from an underlying fiber bundle narrative. 

Now onto curvature in a GR setting. The notion of parallel transport (which is allows once the manifold has a connection defined on it so that there is some idea of what ``parallel" means) is the most intuitive thing to keep in mind when thinking about how to define curvature.

\begin{figure}[H]
\centering
\includegraphics[scale=0.1]{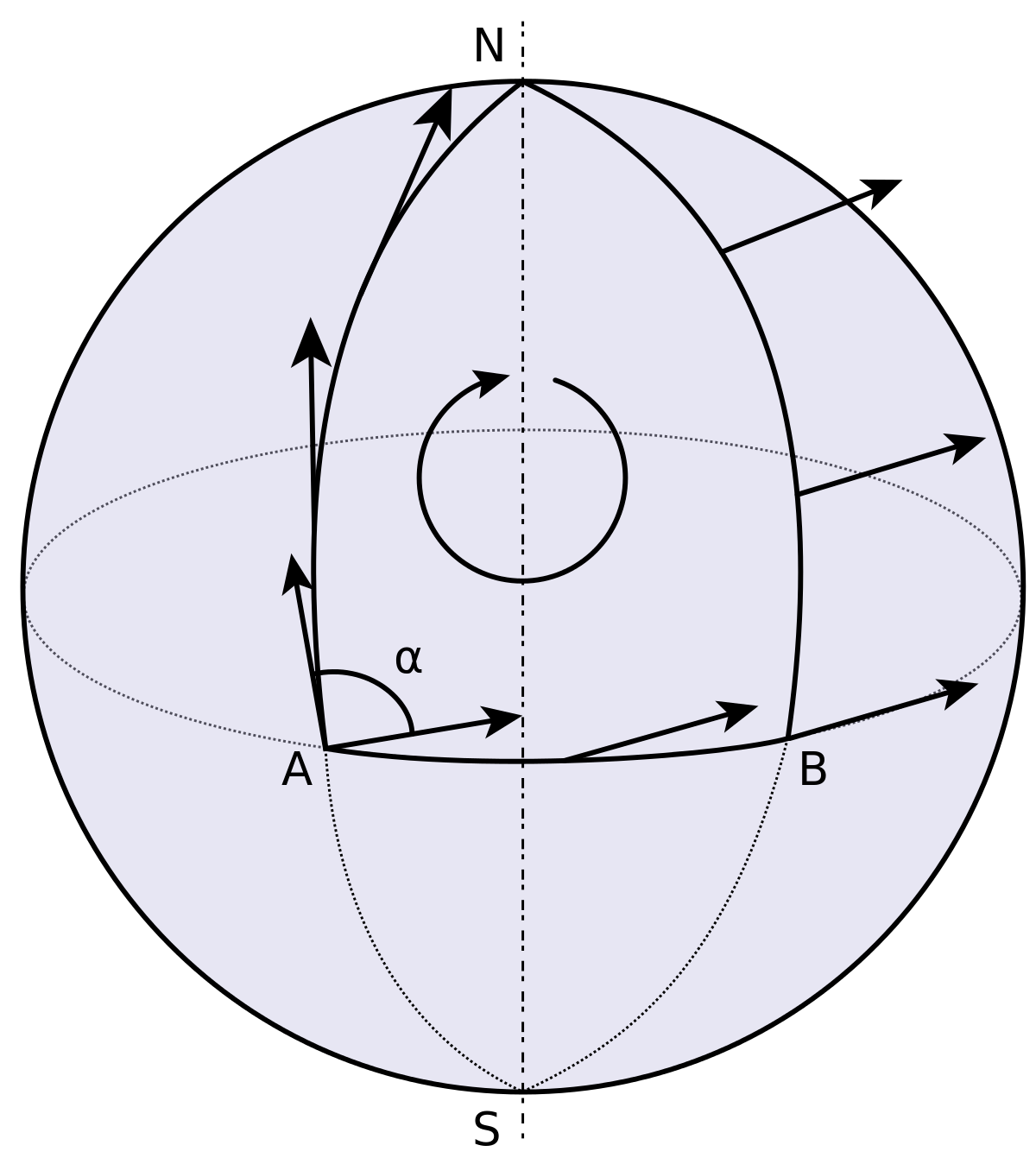}
\captionsetup{format=hang}
\caption{Parallel transport around a curved manifold \cite{wikiparallel}.}
\end{figure}

This notion of parallel transporting a vector about a closed loop can be viewed from another perspective as well that Carroll puts very well:

\begin{displayquote}
``Knowing what we do about parallel transport, we could very carefully perform the necessary manipulations to see what happens to the vector under this operation, and the result would be a formula for the curvature tensor in terms of the connection coefficients. It is much quicker, however, to consider a related operation, the commutator of two covariant derivatives. The relationship between this and parallel transport around a loop should be evident; the covariant derivative of a tensor in a certain direction measures how much the tensor changes relative to what it would have been if it had been parallel transported (since the covariant derivative of a tensor in a direction along which it is parallel transported is zero). The commutator of two covariant derivatives, then, measures the difference between parallel transporting the tensor first one way and then the other, versus the opposite ordering." page 75 of \cite{carrollnotes}
\end{displayquote}

\begin{figure}[H]
\centering
\includegraphics[scale=0.5]{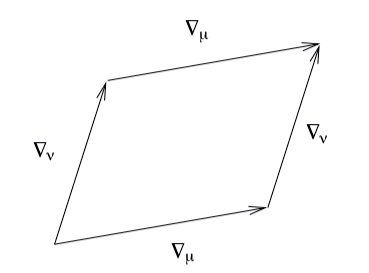}
\captionsetup{format=hang}
\caption{Parallel transport around a curved manifold via the commutator of covariant derivatives \cite{carrollnotes}.}
\end{figure}

The Riemann curvature tensor is subsequently defined as precisely this object. For a torsionless connection, 

\begin{equation}[\nabla_\mu, \nabla_\nu] V^\rho = R^\rho{}_{\sigma \mu \nu} V^\sigma\end{equation}

or worked out in terms of the connection and its derivatives, 

\begin{equation} R^\rho{}_{\sigma \mu \nu} = \partial_\mu \Gamma^{\rho}_{\nu \sigma} - \partial_\nu \Gamma^{\rho}_{\mu \sigma} + \Gamma^{\rho}_{\mu \lambda}\Gamma^{\lambda}_{\nu \sigma} -\Gamma^{\rho}_{\nu \lambda}\Gamma^{\lambda}_{\mu \sigma}\end{equation}

The similarities between the gauge covariant derivative and the GR covariant derivative should be plenty sufficient to motivate trying the same construction of curvature/field strengths in terms of commutators of covariant derivatives.

We will not cover path integrals here, but according to Schwartz and Pallante, this very same commutator of covariant derivatives construction can be motivated via Wilson lines/loops as well \cite{schwartz} \cite{EPP}.

Mimicking the GR derivative, but respecting hermitian operators by including an i and also  including the gauge coupling constant, the gauge field strength/curvature is defined as 

\begin{equation}\frac{i}{g}[D_\mu, D_\nu] =F_{\mu\nu}\end{equation}

Schwartz connects his Wilson loop motivation to geometry as well,

\begin{displayquote}
``This has a nice geometric interpretation: it is the difference between what you get from $D_\mu D_\nu$, which compares values for fields separated in the $\nu$ direction followed by a separation in the $\mu$ direction, to what you get from doing the comparison in the other order. Equivalently, it is the result of comparing field values around an infinitesimal closed loop in the $\mu$-$\nu$ plane, as shown in Figure 25.1. This is, not coincidentally, also the limit of the Wilson loop around a small rectangular path as in Eq. (25.51), as we discuss further in Section 25.5." page 490 of \cite{schwartz}
\end{displayquote}

\begin{figure}[H]
\centering
\includegraphics[scale=0.3]{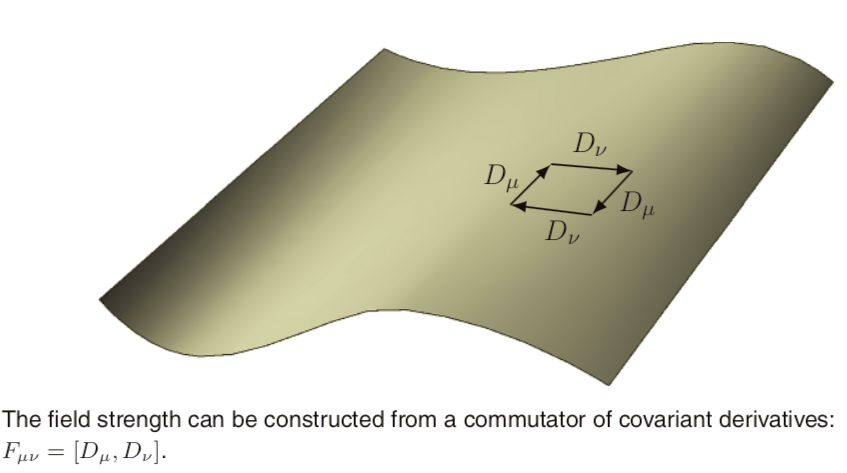}
\captionsetup{format=hang}
\caption{Construction of the gauge field strength/curvature in much the same way as with the Riemann curvature, via the commutator of covariant derivatives \cite{schwartz}.}
\end{figure}

Let's put it to work!

Then's start off with the U(1) local theory. We will let the object act on a scalar field in the theory,

\begin{eqnarray} \notag
\frac{i}{g}[D_\mu, D_\nu] \phi &=& [\partial_\mu -ig A_\mu,\partial_\nu -ig A_\nu] \phi \\
&=& [\partial_\mu,\partial_\nu]\phi - ig [\partial_\mu, A_\nu]\phi - ig [A_\mu, \partial_\nu]\phi -g^2[A_\mu,A_\nu] \phi
\end{eqnarray}

Note that the first term and last term vanish. The first vanishes for any theory because of the symmetry of second derivatives, $\partial_\mu \partial_\nu \phi= \partial_\nu \partial_\mu \phi$. The last term vanishes for this specific theory because U(1) is abelian and so the gauge fields commute.

\begin{eqnarray} \notag
[D_\mu, D_\nu] \phi &=& [\partial_\mu -ig A_\mu,\partial_\nu -ig A_\nu] \phi \\ \notag
&=& - ig [\partial_\mu, A_\nu]\phi - ig [A_\mu, \partial_\nu]\phi \\ \notag
&=& - ig (\partial_\mu (A_\nu \phi)- A_\nu\partial_\mu \phi) - ig (A_\mu \partial_\nu \phi - \partial_\nu (A_\mu \phi)) \\ \notag
&=& - ig (\partial_\mu (A_\nu) \phi + A_\nu \partial_\mu \phi - A_\nu\partial_\mu \phi) - ig (A_\mu \partial_\nu \phi - \partial_\nu (A_\mu) \phi - A_\mu \partial_\nu \phi) \\ \notag
&=& - ig \partial_\mu (A_\nu) \phi +ig \partial_\nu (A_\mu) \phi \\
&=& -ig (\partial_\mu A_\nu  - \partial_\nu A_\mu) \phi 
\end{eqnarray}

now if we multiply both sides by $\frac{i}{g}$ we get the field strength

\begin{eqnarray} \notag
[D_\mu, D_\nu] \phi &=& [\partial_\mu -ig A_\mu,\partial_\nu -ig A_\nu] \phi \\ \notag
&=& -ig (\partial_\mu A_\nu  - \partial_\nu A_\mu) \phi \\ \notag
\frac{i}{g} [D_\mu, D_\nu] \phi &=& \frac{i}{g} \bigg(-ig (\partial_\mu A_\nu  - \partial_\nu A_\mu) \phi\bigg) \\ \notag
&=& (\partial_\mu A_\nu  - \partial_\nu A_\mu) \phi \\ \notag
&=& F_{\mu\nu}\phi \\
\frac{i}{g} [D_\mu, D_\nu]&=& F_{\mu\nu}
\end{eqnarray}

The only difference in the SU(2) non-abelian case is that the commutator of the gauge fields is non-zero. So we get,

\begin{eqnarray} \notag
[D_\mu, D_\nu] \vec{\phi} &=& [\partial_\mu -ig A_\mu,\partial_\nu -ig A_\nu] \vec{\phi} \\ \notag
&=& - ig [\partial_\mu, A_\nu]\vec{\phi} - ig [A_\mu, \partial_\nu]\vec{\phi} -g^2[A_\mu,A_\nu] \vec{\phi} \\ \notag
&=& - ig \partial_\mu (A_\nu) \vec{\phi} +ig \partial_\nu (A_\mu) \vec{\phi} -g^2[A_\mu,A_\nu] \vec{\phi}\\ \notag
&=& (-ig (\partial_\mu A_\nu  - \partial_\nu A_\mu) -g^2[A_\mu,A_\nu])) \vec{\phi} \\ \notag
\frac{i}{g}[D_\mu, D_\nu] \vec{\phi} &=& \frac{i}{g}(-ig (\partial_\mu A_\nu  - \partial_\nu A_\mu) -g^2[A_\mu,A_\nu])) \vec{\phi} \\ \notag
&=& (\partial_\mu A_\nu  - \partial_\nu A_\mu) -ig[A_\mu,A_\nu])) \vec{\phi} \\ \notag
\frac{i}{g}[D_\mu, D_\nu] &=& \partial_\mu A_\nu  - \partial_\nu A_\mu) -ig[A_\mu,A_\nu] \\
&=& F_{\mu\nu}  
\end{eqnarray}

And then of course since the gauge fields are not singular, i.e. they are in vector notation above and they could be written out as $A_\mu= A_\mu{}^a T^a$, we similarly can write out the above vector form of $F_{\mu\nu}$ in components $F_{\mu\nu}= F_{\mu\nu}{}^a T^a$ using the commutation relations for the Lie algebra, $[T^a,T^b]=if^{abc}T^c$ as follows

\begin{equation}F_{\mu\nu}{}^a = \partial_\mu A_\nu{}^a  - \partial_\nu A_\mu{}^a + g f^{abc}A_\mu{}^b A_\nu{}^c\end{equation}

To conclude, as we \hyperlink{inf}{did with $A_\mu$}, we can write the transformation of the field strength/curvature in an infinitesimal form (where $\mathcal{O}((\theta^a)^2)$ terms will be ignored as we did before).

\begin{eqnarray} \notag
F_{\mu\nu} &\rightarrow& U F_{\mu\nu} U^{-1}\\ \notag
F_{\mu\nu}{}^a T^a &\rightarrow& e^{i\theta^bT^b} F_{\mu\nu}{}^aT^a e^{-i\theta^cT^c}\\ \notag
&\approx& (1 + i \theta^b T^b + ...) F_{\mu\nu}{}^aT^a (1 - i \theta^c T^c + ...)\\ \notag
&=& F_{\mu\nu}{}^a T^a + i \theta^b F_{\mu\nu}{}^a T^b T^a  - i \theta^c F_{\mu\nu}{}^a T^a T^c + ... \\ \notag
&=& F_{\mu\nu}{}^a T^a + i \theta^b F_{\mu\nu}{}^a [T^b,T^a]  \\ \notag
&=& F_{\mu\nu}{}^a T^a - (i)(i) \theta^b F_{\mu\nu}{}^a f^{abc} T^c\\ \notag
&=& F_{\mu\nu}{}^a T^a + \theta^b F_{\mu\nu}{}^c f^{cba} T^a\\ \notag
&=& F_{\mu\nu}{}^a T^a - \theta^b F_{\mu\nu}{}^c f^{abc} T^a  \\
F_{\mu\nu}{}^a &\rightarrow& F_{\mu\nu}{}^a - \theta^b F_{\mu\nu}{}^c f^{abc}
\end{eqnarray}

\pagebreak

\subsection{Freedman-Van Proeyen translation}
\label{fvplanguage}

Because it will be important as we shift to gauging the Poincaré, where Freedman-Van Proeyen's nomenclature reigns supreme, lets set up a dictionary between what we've done above and F-VP's conventions. Most notable is the matrix exponential convention for relating Lie algebra/group elements, and the explicit detailing of the adjoint representation of matrix Lie algebra elements, $(t_a)^i{}_j=f_{aj}{}^i$. The left column is our work, and the right column is F-VP.

\begin{align*}
U(x)  &= e^{i\theta^a(x)T^a}             &  U(x) &= e^{-\theta^a(x)(t_a)^i{}_j}  \\ 
[T^a,T^b] &= if^{abc}T^c  &  [t^a,t^b] &= f_{ab}{}^c t^c \\
\delta \phi_i &= i\theta^a(T^a)_{ij} \phi_j  &  \delta \phi^i & 
    \begin{aligned}[t]
         &= -\theta^a(t_a)^i{}_j \phi^j \\
        &= -\theta^af_{aj}{}^i \phi^j \\
        &= \theta^af_{ja}{}^i \phi^j
    \end{aligned} \\
A_\mu  &= A_\mu{}^a T^a            &  A_\mu &=  A_\mu{}^a t_a \\ 
A_\mu  &\rightarrow   UA_\mu U^{-1}  - \frac{i}{g} [\partial_\mu U] U^{-1} 
&  
A_\mu  &\rightarrow   UA_\mu U^{-1}- \frac{1}{g} [\partial_\mu U] U^{-1} \\
D_\mu \phi_i &= \partial_\mu \phi_i -igA_\mu{}^a(T^a)_{ij}\phi^j & D_\mu \phi^i &
\begin{aligned}[t]
         &= \partial_\mu \phi^i +g(t_a)^i{}_jA_\mu{}^a\phi_j \\
        &= \partial_\mu \phi^i +gf_{aj}{}^iA_\mu{}^a\phi^j 
    \end{aligned} \\
    \delta A_\mu{}^a &= \frac{1}{g}\partial_\mu (\theta^a)-\theta^bA_\mu{}^cf^{abc} &
    \delta A_\mu{}^a & 
    \begin{aligned}[t]
    &= \frac{1}{g}\partial_\mu (\theta^a)+\theta^cA_\mu{}^b f_{bc}{}^a \\
    &= \frac{1}{g}\partial_\mu (\theta^a)+\theta^bA_\mu{}^c f_{cb}{}^a \\
    &= \frac{1}{g}\partial_\mu (\theta^a)-\theta^bA_\mu{}^c f_{bc}{}^a
    \end{aligned} \\
    F_{\mu\nu}{}^a &= \partial_\mu A_\nu{}^a - \partial_\nu A_\mu{}^a +gf^{abc}A_\mu{}^bA_\nu{}^c & F_{\mu\nu}{}^a &= \partial_\mu A_\nu{}^a - \partial_\nu A_\mu{}^a +gf_{bc}{}^aA_\mu{}^bA_\nu{}^c \\
\delta F_{\mu\nu}^a &= -\theta^bF_{\mu\nu}^cf^{abc} &
    \delta F_{\mu\nu}^a & 
    \begin{aligned}[t]
    &= \theta^cF_{\mu\nu}^b f_{bc}{}^a \\
    &= \theta^bF_{\mu\nu}^c f_{cb}{}^a \\
    &= -\theta^bF_{\mu\nu}^c f_{bc}{}^a
    \end{aligned} 
\end{align*}

\pagebreak

\section{Vectors, differential forms, and vielbein}
\label{sec:manylegs}

Here we take a detour into geometry to learn about the incredibly elegant mathematics of differential forms, manifolds, and frame fields/tetrad/vielbein. We use Sean Carroll's textbook \cite{carrollbook} (and the lecture notes that precipitated that textbook \cite{carrollnotes}) as well as Anthony Zee's textbook \cite{zee}.

\subsection{(dual)Vectors review}
Vectors $V$ live in the tangent space.

They have components $V^\mu$, basis elements $\hat{e}_{(\mu)}$, and can be given a coordinate basis with $\partial_\mu$.

Dual vectors/one-forms $\omega$ live in the cotangent space.

They have components $\omega_\mu$, basis elements $\hat{\theta}^{(\mu)}$, and can be given a coordinate basis with $dx^\mu$.

\pagebreak
\subsection{Differential forms synopsis}
\label{shittydiff}

In hindsight, the information in this section is too formal. It is much more useful pedagogically to get one's hands dirty. See Sections \ref{introdifforms}, \ref{zeevielbein}, and \ref{subsec:returnofmanylegs}

A differential p-form is a (0,p) tensor that is completely antisymmetric.

A p-form ``A'' plus (``wedge'') a q-form ``B'' is a (p + q)-form,

\begin{equation}(A \wedge B)_{\mu_1 ... \mu_{p+q}} = \frac {(p+q)!}{p!q!} A_{[\mu_1 ... \mu_p}B_{\mu_{p+1}... \mu_{p+q}]}\end{equation}

For example, given 1-forms (dual-vectors) A and B,

\begin{eqnarray} \notag
(A \wedge B)_{\mu\nu} &=& \frac{(1+1)!}{1!1!}( A_{[\mu}B_{\nu]})
\\ \notag
&=& 2 (\frac{1}{2!} A_\mu B_\nu - A_\nu B_\mu)
\\
&=& A_\mu B_\nu - A_\nu B_\mu
\end{eqnarray}

Notice that for the above case, $A \wedge B= -(B \wedge A)$, and in general for p-form A and q-form B, $A \wedge B= (-1)^{pq}(B \wedge A)$.

We define the exterior derivative as,

\begin{equation}(dA)_{\mu_1 ... \mu_{p+1}} = (p+1)\partial_{[\mu_1}A_{\mu_2 ... \mu_{p+1}]}\end{equation}

For example,

\begin{equation}(d\phi)_\mu = \partial_\mu \phi\end{equation}

For a p-form $\omega$ and a q-form $\eta$,

\begin{equation}d(\omega \wedge \eta) = d\omega \wedge \eta + (-1)^p\omega \wedge d\eta\end{equation}

We define the Hodge star operator on an n-dimensional manifold as,

\begin{equation}(\star A)_{\mu_1 ... \mu_{n-p}} = \frac{1}{p!} \epsilon ^{\nu_1 ... \nu_p}{}_\nu\end{equation}
\pagebreak
\subsection{Learning many legs via Carroll}
\label{ewcarroll}

In hindsight, the information in this section is too formal. It is much more useful pedagogically to get one's hands dirty. See Sections \ref{zeevielbein}, \ref{subsec:returnofmanylegs}, and \ref{final steps}

$\{\hat{e}_{(a)}\}$ = vielbein/orthonormal basis vectors

$\{\hat{\theta}^{(a)}\}$ = vielbein/orthonormal basis dual-vectors

They are orthonormal, i.e. 
\begin{eqnarray} \notag
g(\hat{e}_{(a)},\hat{e}_{(b)}) &=& \eta_{ab} \text{(Minkowski space)} \\
 &=& \delta^a_b \text{(Euclidean space)} 
\end{eqnarray}

This notation is a bit confusing, as it is not clear how this matches up with the notion of the metric as an inner product that one learns in GR,

\begin{equation}g(v,u) = \sum_{i,j} g_{ij}v^iu^j\end{equation}

since $g(u,v)$ is a map $V(\text{vector }) \times V(\text{vector}) \rightarrow F(\text{scalar})$, and not $V(\text{vector }) \times V(\text{vector}) \rightarrow \text{(tensor)}$.

They also satisfy $\hat{\theta}^{(a)}(\hat{e}_{(b)}) = \delta^a_b$ as all vector/dual vector combos do.

The vielbein (``components" — just vielbein), $e_\mu{}^a$, can express the standard basis vectors in terms of the orthonormal basis vectors

\begin{equation} \hat{e}_{(\mu)}= e_\mu{}^a\hat{e}_{(a)} \end{equation}

and can express the orthonormal basis dual-vectors in terms of the standard basis dual-vectors

\begin{equation}\hat{\theta}^{(a)} = e_\mu{}^a \hat{\theta}^{(\mu)}\end{equation}

The inverse vielbeins, defined as, $e^\mu{}_a$, and satisfying

\begin{eqnarray} \notag
e_\mu{}^a e^\mu{}_b &=& \delta^a_b \\
e^\mu{}_a e_\nu{}^a &=& \delta^\mu_\nu
\end{eqnarray}

can express the orthonormal basis vectors in terms of the standard basis vectors

\begin{equation} \hat{e}_{(a)}= e^\mu{}_a\hat{e}_{(\mu)} \end{equation}

and can express the standard basis dual-vectors in terms of the orthonormal basis dual-vectors

\begin{equation}\hat{\theta}^{(\mu)} = e^\mu{}_a \hat{\theta}^{(a)}\end{equation}

They also provide a way to take a vector or general tensor between coordinate and orthonormal bases,

\begin{eqnarray} \notag
V &=& V^\mu \hat{e}_{(\mu)} \\
V &=& V^a \hat{e}_{(a)},
\end{eqnarray}
where $V^a=e_\mu{}^a V^\mu$.

\begin{eqnarray} \notag
T^a{}_b &=& e_\mu{}^a T^\mu{}_b = e_\mu{}^a e^\nu{}_b T^\mu{}_\nu \\
&=& e^\nu{}_b T^a{}_\nu.
\end{eqnarray}

Greek indices like $\mu$ and $\nu$ are called curved indices, and Latin indices like $a$ and $b$ are called flat indices.

Recall the standard coordinate transformation corresponding to a Lorentz transformation,

\begin{equation}x^{\mu'}=\Lambda^{\mu'}{}_\nu x^\nu\end{equation}

Coordinates changing implies that the (co)tangent vectors (i.e. the (dual)vector components and the basis vectors) change as well,

\begin{eqnarray} \notag
V^{\mu'}&=&\Lambda^{\mu'}{}_\nu V^\nu  \\ \notag
\hat{e}_{(\nu')}&=&\Lambda^\mu{}_{\nu'} \hat{e}_{(\mu)} \\\notag
\omega_{\nu'}&=&\Lambda^{\mu}{}_{\nu'} \omega_\mu \\
\hat{\theta}^{(\mu')}&=&\Lambda^{\mu'}{}_{\nu} \hat{\theta}^{(\nu)},
\end{eqnarray}

where the following are satisfied,

\begin{eqnarray} \notag
\Lambda^{\mu'}{}_\nu\Lambda^{\nu}{}_{\gamma'} &=& \delta^{\mu'}_{\gamma'} \\
\Lambda^{\mu}{}_{\nu'}\Lambda^{\nu'}{}_{\gamma} &=& \delta^{\mu}_{\gamma}.
\end{eqnarray}

Components of the metric tensor for flat spacetime have the same numerical value for all Cartesian-like coordinate systems that are connected by Lorentz transformations,

\begin{eqnarray} \notag
\eta_{\mu\nu}= \Lambda^{\gamma'}{}_\mu\Lambda^{\sigma'}{}_{\nu} \eta_{\gamma'\sigma'} \\
\eta_{\phi'\delta'}= \Lambda^{\mu}{}_{\phi'}\Lambda^{\nu}{}_{\delta'} \eta_{\mu\nu}
\end{eqnarray}

Orthonormal basis vectors transform, not with general coordinate transformations (GCTs), but with local Lorentz transformations (LLTs),

\begin{equation}\hat{e}_{(a)} \rightarrow \hat{e}_{(a')} = \Lambda^a{}_{a'} \hat{e}_{(a)},\end{equation}

where the LLTs satisfy 

\begin{equation}\Lambda^a{}_{a'}\Lambda^b{}_{b'}\eta_{ab}=\eta_{a'b'}\end{equation}

GCTs and LLTs can be performed together,

\begin{equation}T^{a'\mu'}{}_{b'\nu'}=\Lambda^{a'}{}_{a}\Lambda^{\mu'}{}_\mu\Lambda^b{}_{b'}\Lambda^b{}_{v'} T^{a\mu}{}_{b\nu}\end{equation}

The covariant derivative of a tensor with curved indices uses the Christoffel symbols,

\begin{equation}\nabla_\sigma T^\mu_\nu = \partial_\sigma T^\mu_\nu + \Gamma^\mu_{\sigma\lambda}T^\lambda{}_\nu -\Gamma^\lambda_{\sigma\nu}T^\mu{}_\lambda\end{equation}

while the covariant derivative of a tensor with flat indices uses the spin connection symbols,

\begin{equation}\nabla_\sigma T^a_b = \partial_\sigma T^a_b + \omega_\sigma{}^a{}_c T^c{}_b -\omega_\sigma{}^c{}_b T^a{}_c.\end{equation}

The Christoffel symbols can be written in terms of the normal and inverse vielbeins + the spin connections,

\begin{equation}\Gamma^\nu_{\mu\lambda} = e^\nu{}_a\partial_\mu e_\lambda{}^a + e^\nu{}_a e_\lambda{}^b\omega_\mu{}^a{}_b.\end{equation}

Likewise, the spin connections can be written in terms of the normal and inverse vielbeins + the Christoffel symbols,

\begin{equation}\omega_\mu{}^a{}_b = e_\nu{}^a e^\lambda{}_b \Gamma^\nu_{\mu\lambda}- e^\lambda{}_b \partial_\mu e_\lambda{}^a.\end{equation}

Similarly to the covariant derivative of the metric tensor, we have,

\begin{eqnarray} \notag
\nabla_\mu e_\nu{}^a &=& \partial_\mu e_\nu{}^a - \Gamma^\lambda_{\mu\nu}e_\lambda{}^a + \omega_\mu{}^a{}_b e_\nu{}^b\\
&=& 0.
\end{eqnarray}

\iffalse
\textbf{Questions after taking these notes before heading off to study differential forms/exterior calculus/etc.:}

\begin{itemize}
    \item Q1 — Is there any particular reason Carroll is using that annoying notation for LLTs? Why $\Lambda^a{}_a'$ and not $\Lambda^a{}_b'$ like with our old-fashioned Lorentz transformations?
    \item Q2 — Notice in my equation for ``GCTs and LLTs can be performed together'' I have Lorentz transformations $\Lambda^{\mu'}{}_\mu$. Carroll writes the GCTs here as $\frac{\partial x^{\mu'}}{\partial x^\mu}$. My undergrad books has these being synonymous, but I wonder if that it truly the case. It makes sense to have then written as Carroll does to signify transformations more general than Lorentz.
    \item Q3 — Are there are analogous ways to determine the spin connections like we do for the Christoffel symbols? For the Christoffel symbols we have the 2 forms of the geodesic equation we can compare, and the formula from the derivatives of the metric.
\end{itemize}
\fi

\pagebreak
\subsection{Some notes after our vielbein/differiental forms question session}

Notation:

$\{\hat{e}_{(a)}\} \equiv \{e_a\}$: vielbein/orthonormal basis vectors

$\{\hat{\theta}^{(a)}\} \equiv \{e^a\}$: vielbein/orthonormal basis (one-forms) dual-vectors

Vectors $V$ live in the tangent space, and can be expressed in the following ways:

\begin{eqnarray} \notag
V &=& V^\mu \hat{e}_{(\mu)} \text{ (in a generic/canonical basis)} \\\notag
&=& V^\mu \partial_\mu \text{ (in a coordinate basis)} \\\notag
&=& V^a \hat{e}_{(a)} \text{ (in a non-coordinate basis)} \\
&\equiv& V^a e_a 
\end{eqnarray}
where $V^a= e_\mu{}^a V^\mu$, and $e_a=e^\mu{}_a \partial_\mu$.

Dual vectors/one-forms $\omega$ live in the cotangent space, and can be expressed in the following ways:

\begin{eqnarray} \notag
\omega &=& \omega_\nu \hat{\theta}^{(\nu)} \text{ (in a generic/canonical basis)} \\\notag
&=& \omega_\nu dx^\nu \text{ (in a coordinate basis)} \\\notag
&=& \omega_b \hat{\theta}^{(b)} \text{ (in a non-coordinate basis)} \\\notag
&\equiv& \omega_b \hat{e}^{(b)} \\
&\equiv& \omega_b e^b 
\end{eqnarray}
where $\omega_b= e^\nu{}_b \omega_\nu$, and $e^b=e_\nu{}^b dx^\nu$.

\pagebreak
\subsection{Learning many legs via Zee and getting our hands dirty.}
\label{zeevielbein}

\paragraph{Exercise 1} Use the Vielbein formalism to calculate the Riemann tensor, the Ricci tensor, and the Ricci scalar for the unit (or, with radius $\ell$) round metric on $\mathbb{S}^2$. Can you generalize the result to higher-order spheres $\mathbb{S}^N$?

The metric of a unit-radius 2-sphere is

$ds^2=d\theta^2+\sin^2(\theta)d\phi^2$,

i.e. $g_{\theta\theta}=1$ and $g_{\phi\phi}=\sin^2(\theta)$.

By definition of the vielbein, we have

$g_{\mu\nu}=\eta_{ab}e^a{}_\mu e^b{}_\nu$

In our case, we take $\eta_{ab}=\delta^a_b$, since the 2-sphere is locally-flat.

In our case, for $\mu=\nu=\theta$ we have 

\begin{eqnarray} \notag
g_{\theta\theta} &=& \eta_{ab}e^a{}_\theta e^b{}_\theta \\\notag
1 &=& \delta^a_b e^a{}_\theta e^b{}_\theta \\\notag
&=& (e^a{}_\theta)^2 \\\notag
 1 &=& e^a{}_\theta \\
1 &=& e^1{}_\theta
\end{eqnarray}

where $a$ runs from 1 to 2 so we label this vielbein without loss of generality as $a=1$.

And for $\mu=\nu=\phi$ we have 

\begin{eqnarray} \notag
g_{\phi\phi} &=& \eta_{ab}e^a{}_\phi e^b{}_\phi \\\notag
\sin^2(\theta)&=& \delta^a_b e^a{}_\phi e^b{}_\phi \\\notag
&=& (e^a{}_\phi)^2 \\\notag
\sin\theta &=& e^a{}_\phi \\
\sin\theta &=& e^2{}_\phi
\end{eqnarray}

Recall that the $e^a{}_\mu$ are components of a one-form $e^a=e^a{}_\mu dx^\mu$.

Cartan's first structure equation, $de + \omega e=0$, but with indices brought out of suppression and the spin connection $\omega$ brought to the other side, reads

\begin{equation}de^a=-\omega^a{}_be^b\end{equation}

where $\omega^a{}_b$ are the spin connection one-forms,

\begin{equation}\omega^a{}_b=\omega_\mu{}^a{}_{b}dx^\mu\end{equation}.

We have $e^1{}_\theta=1$ and $e^2{}_\phi=\sin\theta$, so then

\begin{eqnarray} \notag
e^1&=&e^1{}_\theta d\theta=1 \\
e^2&=&e^2{}_\phi d\phi= \sin\theta d\phi
\end{eqnarray}

Then differentiated we have 

\begin{eqnarray} \notag
d(e^1)&=&\partial_\nu (1) dx^\nu d\theta =0 \\\notag
d(e^2)&=& \partial_\nu(\sin\theta) dx^\nu d\phi \\\notag
&=& \partial_\theta(\sin\theta) d\theta d\phi \\
&=& \cos\theta d\theta d\phi
\end{eqnarray}

Before we use Cartan's structure equation to determine the spin connections, note that 

\begin{equation}\omega^b{}_c=\omega^{bc}=-\omega^{cb}\end{equation}

Thus 

1) we raise the indices on the spin connections indiscriminately, and

2) the second equality (showing antisymmetry) tells us that $\omega^{aa}$=0, since 

\begin{eqnarray} \notag
\omega^{aa}&=&-\omega^{aa} \\\notag
\omega^{aa}+\omega^{aa} &=& 0\\\notag
2\omega^{aa} &=& 0 \\
\omega^{aa} &=& 0 
\end{eqnarray}

This implies the basis vectors of the spin connections are antisymmetric as well,

$dx^\mu dx^\nu= -dx^\nu dx^\mu$, and moreover, as above,

$dx^\mu dx^\mu =0$.

We will put off the $a=1$ case for just a moment.

For the $a=2$ case, we use $w^{aa}=0$ to obtain,

\begin{eqnarray} \notag
de^a &=& -\omega^a{}_be^b \\\notag
de^a &=& -\omega^{ab}e^b \\\notag
de^2 &=& -\omega^{2b}e^b \\\notag
\cos\theta d\theta d\phi &=& -\omega^{21}e^1 -\omega^{22}e^2 \\\notag
&=& -\omega^{21} d\theta - 0 \\
\omega^{21} &=& - \cos\theta d\phi
\end{eqnarray}

Something is amiss here. In this calculation, Zee got $\omega^{21} = \cos\theta d\phi$. This is not an error according to Zee's errata, and so moving forward we will assume $\omega^{21} = \cos\theta d\phi$ for future calculations but we urge the reader to take note of this error.

Now for the $a=1$ case, we use the following:

1) that $w^{aa}=0$, 
\newline
2) that $dx^\mu dx^\mu=0$, as well as 
\newline 
3) the result of the $a=2$ case to check for self-consistency, notably, that $\omega^{12}=-\cos\theta d\phi$

\begin{eqnarray} \notag
de^a &=& -\omega^a{}_be^b \\\notag
de^a &=& -\omega^{ab}e^b \\\notag
de^1 &=& -\omega^{1b}e^b \\\notag
0 &=& -\omega^{11}e^1 -\omega^{12}e^2 \\\notag
0 &=& 0 + \cos\theta d\phi \sin \theta d\phi \\\notag
0 &=& \cos\theta  \sin \theta (d\phi d\phi) \\\notag
0 &=&  \cos\theta  \sin \theta (0) \\
0 &=& 0
\end{eqnarray}

Now that we have the vielbein and the spin connections, we are prepared to computer the Riemann tensor and other curvature quantities.

First, we will write Cartan's second structure equation, $R= d\omega +\omega^2,$ with all indices restores,

\begin{equation}R^a{}_b= d\omega^a{}_b + \omega^a{}_c\omega^c{}_b\end{equation}

Recall that we can raise indices indiscriminately here, so long as we remember what to sum over. So then

\begin{equation}R^{ab}= d\omega^{ab} + \omega^{ac}\omega^{cb}\end{equation}

$R^{ab} \equiv R^{ab}{}_{\mu\nu} \propto R_{ab\mu\nu}$ is antisymmetric in a and b. This may enough to say that $R^{aa}$ is always zero as we did with the $\omega^{aa}$ and $dx^\mu dx^\mu$ before, but we can also argue that $R^{11}=R^{22}=0$ because the first term would be $d(0)=0$ and the second would be zero because of $d\phi d\phi=0$.

Since $R^{11}=R^{22}=0$, and $R^{12}=-R^{21}$, there is only one quantity to compute. 

\begin{equation}R^{12}= d(\omega^{12}) + \omega^{1c}\omega^{c2}\end{equation}

Note that the second term is $0$ for both $c=1,2$ since $\omega^{11}=\omega^{22}=0$. So then

\begin{eqnarray} \notag
R^{12}&=& d(\omega^{12}) \\\notag
&=& d(-\cos\theta d\phi) \\\notag
&=& \partial_\nu(-\cos\theta) dx^\nu d\phi \\\notag
&=& \partial_\theta(-\cos\theta) d\theta d\phi \\\notag
&=& \sin\theta d\theta d\phi
\end{eqnarray}

Notice that, since $e^1=d\theta$ and $e^2=\sin\theta d\phi$, we have $R^{12}=e^1e^2$.

Also, writing the $e^a$ out in components, $R^{12}=e^1{}_\theta e^2{}_\phi d\theta d\phi$.

Here, Zee ``expands the 2-form $R^{12}$'' to obtain $R^{12}=e^1e^2=\frac{1}{2}(R^{12}{}_{12}e^1e^2 + R^{12}{}_{21}e^2e^1)$. It is not immediately clear this was done. 

However, the curvature 2-form $R^{ab}$ can also be written out in components as

\begin{equation}R^{ab}=\frac{1}{2}R^{ab}{}_{\mu\nu}dx^\mu dx^\nu\end{equation}

In our case, we  have,

\begin{equation}R^{12}=\frac{1}{2}R^{12}{}_{\mu\nu}dx^\mu dx^\nu\end{equation}

Because of $dx^\mu dx^\nu = - dx^\nu dx^\mu$ ($dx^\mu dx^\mu =0$), as well as the Riemann tensor being antisymmetric in the 3rd and 4th indices, we have 

\begin{eqnarray} \notag
R^{12}&=& \frac{1}{2}R^{12}{}_{\mu\nu}dx^\mu dx^\nu \\\notag
&=& \frac{1}{2}(R^{12}{}_{\theta\theta}dx^\theta dx^\theta + R^{12}{}_{\theta\phi}dx^\theta dx^\phi+ R^{12}{}_{\phi\theta}dx^\phi dx^\theta+R^{12}{}_{\phi\phi}dx^\phi dx^\phi) \\\notag
&=& \frac{1}{2}(0 + R^{12}{}_{\theta\phi}d\theta d\phi+ R^{12}{}_{\phi\theta}d\phi d\theta+0) \\\notag
&=& \frac{1}{2}(R^{12}{}_{\theta\phi}d\theta d\phi- R^{12}{}_{\phi\theta}d\theta d\phi) \\\notag
&=& \frac{1}{2}(R^{12}{}_{\theta\phi}d\theta d\phi+ R^{12}{}_{\theta\phi}d\theta d\phi) \\
&=& R^{12}{}_{\theta\phi}d\theta d\phi
\end{eqnarray}

and if we compare this with $R^{12}=e^1{}_\theta e^2{}_\phi d\theta d\phi$, we have

\begin{eqnarray} \notag
R^{12}{}_{\theta\phi}&=&e^1{}_\theta e^2{}_\phi\\\notag
&=& (1)(\sin\theta) \\
&=& \sin\theta
\end{eqnarray}

If we want to write this without a mix of flat and curved indices, we just hit it with inverse vielbein components, 

\begin{eqnarray} \notag
e^\theta{}_1e^\phi_2(R^{12}{}_{\theta\phi})&=&(\sin\theta)e^\theta{}_1e^\phi_2\\
R^{12}{}_{12}&=& (\sin\theta) \frac{1}{1}\frac{1}{\sin\theta} \\ 
&=& 1
\end{eqnarray}

Thus, we have all in all

Vielbein $e^1{}_\theta=1$ and $e^2{}_\phi=\sin\theta$

Spin connection $\omega^{12}=-\cos\theta d\phi$, recall this error from earlier. Zee gets $\omega^{21} = \cos\theta d\phi$ here

Riemann tensor $R^{12}=e^1e^2 = \sin\theta d\theta d\phi$

Ricci tensor $R^{12}{}_{12}=1$

Ricci Scalar $R^{ab}{}_{ab}=R^{12}{}_{12} + R^{21}{}_{21} = 2$

\pagebreak
\paragraph{Exercise 3*} Use the Vielbein formalism to calculate the Riemann tensor, the Ricci tensor, and the Ricci scalar for a generic, conformally flat metric
\begin{align*}
    ds^2 = \Omega^2(x)( dx^2 + dy^2)
\end{align*}
in terms of $\Omega(x)$. Show that \textbf{Exercises 1} and \textbf{2} are special cases. Can you generalize to arbitrary dimensions?

The above metric tells us that $g_{xx}=\Omega^2(x)$ and $g_{yy}=\Omega^2(x)$.

By definition of the vielbein, we have

$g_{\mu\nu}=\eta_{ab}e^a{}_\mu e^b{}_\nu$

Take $\eta_{ab}=\delta^a_b$, since the metric is locally-flat.

Let $a=\{1,2\}$

In our case we have, for $\mu=\nu=x$ and $\mu=\nu=y$,

\begin{eqnarray} \notag
g_{xx} &=& \eta_{ab}e^a{}_x e^b{}_x \\\notag
\Omega^2(x) &=& \delta^a_b e^a{}_x e^b{}_x \\\notag
&=& (e^a{}_x)^2 \\\notag
\Omega(x) &=& e^a{}_x \\
&=& e^1{}_x
\end{eqnarray}

and 
\begin{eqnarray} \notag
g_{yy} &=& \eta_{ab}e^a{}_y e^b{}_y \\\notag
\Omega^2(x) &=& \delta^a_b e^a{}_y e^b{}_y \\\notag
&=& (e^a{}_y)^2 \\\notag
\Omega(x) &=& e^a{}_y \\
&=& e^2{}_y
\end{eqnarray}

We will neglect the x dependence of $\Omega$ until later on.

Recall that the $e^a{}_\mu$ are components of a one-form $e^a=e^a{}_\mu dx^\mu$.

Cartan's first structure equation, $de + \omega e=0$, but with indices brought out of suppression and the spin connection $\omega$ brought to the other side, reads

\begin{equation}de^a=-\omega^a{}_be^b\end{equation}

where $\omega^a{}_b$ are the spin connection one-forms,

\begin{equation}\omega^a{}_b=\omega^a{}_{b\mu}dx^\mu\end{equation}.

We have $e^1{}_x=\Omega$ and $e^2{}_y=\Omega$, so then

\begin{eqnarray} \notag
e^1&=&e^1{}_x dx=\Omega dx \\
e^2&=&e^2{}_y dy= \Omega dy
\end{eqnarray}

We write $\partial_x(\Omega)$ as $\dot{\Omega}$

Recall that $dx^\mu dx^\mu =0$ since $dx^\mu dx^\nu = - dx^\nu dx^\mu$.

Then differentiating we have 

\begin{eqnarray} \notag
d(e^1)&=&\partial_\nu (\Omega) dx^\nu dx \\\notag
&=&\partial_x (\Omega) dx dx \\\notag
&=&\partial_x (\Omega) 0 \\
&=& 0
\end{eqnarray}

and 
\begin{eqnarray} \notag
d(e^2)&=& \partial_\nu(\Omega) dx^\nu dy \\\notag
&=& \partial_x(\Omega) dx dy \\
&=& \dot{\Omega} dx dy
\end{eqnarray}

Before we use Cartan's structure equation to determine the spin connections, note that 

\begin{equation}\omega^b{}_c=\omega^{bc}=-\omega^{cb}\end{equation}

Thus 

1) we raise the indices on the spin connections indiscriminately, and

2) the second equality (showing antisymmetry) tells us that $\omega^{aa}$=0, since 

\begin{eqnarray} \notag
\omega^{aa}&=&-\omega^{aa} \\\notag
\omega^{aa}+\omega^{aa} &=& 0\\\notag
2\omega^{aa} &=& 0 \\
\omega^{aa} &=& 0 
\end{eqnarray}

This implies the basis vectors of the spin connections are antisymmetric as well,

$dx^\mu dx^\nu= -dx^\nu dx^\mu$, and moreover, as above,

$dx^\mu dx^\mu =0$.

We will put off the $a=1$ case for just a moment.

For the $a=2$ case, we use $w^{aa}=0$ to obtain,

\begin{eqnarray} \notag
de^a &=& -\omega^a{}_be^b \\\notag
de^a &=& -\omega^{ab}e^b \\\notag
de^2 &=& -\omega^{2b}e^b \\\notag
\dot{\Omega} dx dy &=& -\omega^{21}e^1 -\omega^{22}e^2 \\\notag
&=& -\omega^{21} \Omega dx - 0 \\
\omega^{21} &=& - \frac{\dot{\Omega}}{\Omega} dy
\end{eqnarray}

Now for the $a=1$ case, we use the following:

1) that $w^{aa}=0$, 
\newline
2) that $dx^\mu dx^\mu=0$, as well as 
\newline 
3) the result of the $a=2$ case to check for self-consistency, notably, that $\omega^{12}= \frac{\dot{\Omega}}{\Omega} dy$

\begin{eqnarray} \notag
de^a &=& -\omega^a{}_be^b \\\notag
de^a &=& -\omega^{ab}e^b \\\notag
de^1 &=& -\omega^{1b}e^b \\\notag
0 &=& -\omega^{11}e^1 -\omega^{12}e^2 \\\notag
0 &=& 0 - \frac{\dot{\Omega}}{\Omega} dy \Omega dy \\\notag
0 &=& -\dot{\Omega}(dy dy) \\\notag
0 &=&  -\dot{\Omega} (0) \\
0 &=& 0
\end{eqnarray}

Now that we have the vielbein and the spin connections, we are prepared to computer the Riemann tensor and other curvature quantities.

First, we will write Cartan's second structure equation, $R= d\omega +\omega^2,$ with all indices restores,

\begin{equation}R^a{}_b= d\omega^a{}_b + \omega^a{}_c\omega^c{}_b\end{equation}

Recall that we can raise indices indiscriminately here, so long as we remember what to sum over. So then

\begin{equation}R^{ab}= d\omega^{ab} + \omega^{ac}\omega{cb}\end{equation}

Since $R^{11}=R^{22}=0$, and $R^{21}=-R^{12}$, there is only one quantity to compute. 

\begin{equation}R^{21}= d(\omega^{21}) + \omega^{2c}\omega^{c1}\end{equation}

Note that the second term is $0$ for both $c=1,2$ since $\omega^{11}=\omega^{22}=0$. True for any 2-dimensional theory, no?

So then (restoring the x dependence of $\Omega(x)$ to make things clear) by the quotient rule we have,

\begin{eqnarray} \notag
R^{21}&=& d(\omega^{21}) + \omega(x)^{2c}\omega^{c1}\\\notag
&=& d\bigg(- \frac{\dot{\Omega(x)}}{\Omega9x)} dy\bigg) + 0 \\\notag
&=& -\partial_\nu\bigg(\frac{\dot{\Omega(x)}}{\Omega(x)}\bigg) dx^\nu dy \\\notag
&=& -\partial_x\bigg(\frac{\dot{\Omega(x)}}{\Omega(x)}\bigg) dx dy \\\notag
&=& -\bigg(\frac{\ddot{\Omega}\Omega-(\dot{\Omega})^2}{\Omega^2}\bigg) dx dy \\
&=& \frac{-\ddot{\Omega}\Omega+(\dot{\Omega})^2}{\Omega^2} dx dy
\end{eqnarray}

The curvature 2-form $R^{ab}$ can also be written out in components as

\begin{equation}R^{ab}=\frac{1}{2}R^{ab}{}_{\mu\nu}dx^\mu dx^\nu\end{equation}

In our case, we  have (recalling the antisymmetry of the Riemann tensor in the 3rd and 4th indices, and the antisymmetry of the basis oneforms)

\begin{eqnarray} \notag
R^{ab} &=& \frac{1}{2}R^{ab}{}_{\mu\nu}dx^\mu dx^\nu \\\notag
&=&\frac{1}{2}(R^{ab}{}_{xx}dx dx +R^{ab}{}_{xy}dx dy + R^{ab}{}_{yx}dy dx+R^{ab}{}_{yy}dy dy) \\\notag
&=&\frac{1}{2}(0 +R^{ab}{}_{xy}dx dy - R^{ab}{}_{xy}dy dx+0) \\\notag
&=&\frac{1}{2}(2 R^{ab}{}_{xy}dx dy) \\
&=& R^{ab}{}_{xy}dx dy
\end{eqnarray}

Specifically,

\begin{equation}R^{21}= R^{21}{}_{xy}dx dy \end{equation}

and if we compare this with $R^{21}=\frac{-\ddot{\Omega}\Omega+(\dot{\Omega})^2}{\Omega^2} dx dy $, we have

\begin{equation}R^{21}{}_{xy}= \frac{-\ddot{\Omega}\Omega+(\dot{\Omega})^2}{\Omega^2} \end{equation}

If we want to write this without a mix of flat and curved indices, we just hit it with inverse vielbein components, 

\begin{eqnarray} \notag
e^x{}_2e^y_1(R^{21}{}_{xy})
&=&
\bigg(
\frac{-\ddot{\Omega}\Omega+(\dot{\Omega})^2}{\Omega^2} 
\bigg)e^x{}_2e^y_1\\\notag
R^{21}{}_{21}&=& \bigg(\frac{-\ddot{\Omega}\Omega+(\dot{\Omega})^2}{\Omega^2} \bigg) \frac{1}{\Omega}\frac{1}{\Omega} \\ 
&=& \frac{-\ddot{\Omega}\Omega+(\dot{\Omega})^2}{\Omega^4}
\end{eqnarray}

Thus, we have all in all

Vielbein $e^1{}_x=\Omega$ and $e^2{}_y=\Omega$

Spin connection $\omega^{21}=\omega^{12}=-\frac{\dot{\Omega}}{\Omega} dy$

Riemann tensor $\frac{-\ddot{\Omega}\Omega+(\dot{\Omega})^2}{\Omega^2} dx dy$

Ricci tensor $R^{21}{}_{21}=\frac{-\ddot{\Omega}\Omega+(\dot{\Omega})^2}{\Omega^4}$

Ricci Scalar $R = R^{ab}{}_{ab}=R^{21}{}_{21} + R^{12}{}_{12} = 2\bigg(\frac{-\ddot{\Omega}\Omega+(\dot{\Omega})^2}{\Omega^4}\bigg)$

From here we can compute the generic (vacuum) Einstein equation in 2 dimensions.

Note that the metric is diagonal, and the Ricci tensor is symmetric. For $G=c=1$, we have 

\begin{eqnarray} \notag
R_{\mu\nu} -\frac{1}{2}R g_{\mu\nu} &=& 0\\\notag
R_{xx} -\frac{1}{2}R g_{xx} + R_{xy} -\frac{1}{2}R g_{xy} + R_{yx} -\frac{1}{2}R g_{yx} + R_{yy} -\frac{1}{2}R g_{yy} &=& 0 \\\notag
0 -\frac{1}{2}R g_{xx} + R_{xy} - 0 + R_{yx} - 0 + 0 -\frac{1}{2}R g_{yy} &=& 0 \\\notag
-\frac{1}{2}R (g_{xx}+g_{yy}) + 2(R_{xy}) &=& 0 \\\notag
-\frac{1}{2}\bigg( 2\bigg(\frac{-\ddot{\Omega}\Omega+(\dot{\Omega})^2}{\Omega^4}\bigg)\bigg) (2 \Omega^2) + 2\bigg(\frac{-\ddot{\Omega}\Omega+(\dot{\Omega})^2}{\Omega^4}\bigg) &=& 0 \\\notag
-\Omega^2 +1 &=&0 \\
\Omega^2 &=& 1
\end{eqnarray}

\pagebreak

\section{Pure Lorentz algebra gauge theory}
\label{chap:lgauge}

As an additional step before returning to full-blown Poincaré we will apply what we learned through tackling SU(2) Yang-Mills in Chapter \ref{chap:diggingDeeper} to solely the Lorentz algebra.

\subsection{Lorentz symmetries and the Lorentz group/algebra}
\label{subsec:ls}

First, note that we started off with a \textit{globally} SU(2) invariant Lagrangian before we gauged the symmetry in SU(2) Yang-Mills. This is even easier if the global symmetry we start with is Lorentz — nearly everything Lagrangian in QFT is Lorentz invariant (barring 1st-order dynamics), one just needed to make sure all Lorentz indices are contracted with Lorentz-invariant objects like the metric.

A quick recap of how global Lorentz symmetries manifest themselves. Given the Lorentz transformation of spacetime, 

\begin{equation}x^\mu \rightarrow \Lambda^\mu{}_\nu x^\nu \end{equation}

the following transformations follow (noting the following identity)

\begin{equation}\Lambda^\mu{}_\rho \Lambda^\nu{}_\sigma \eta^{\rho\sigma} = \eta^{\mu\nu} \rightarrow (\Lambda^{-1})^\rho{}_\mu (\Lambda^{-1})^\sigma{}^\nu \eta^{\mu\nu} = \eta^{\rho\sigma}\end{equation}

\begin{eqnarray} \notag
\phi(x) &\rightarrow& \phi(\Lambda^{-1}x) \\\notag
\partial_\mu \phi(x) &\rightarrow& (\Lambda^{-1})^\nu{}_\mu \partial_\nu \phi(\Lambda^{-1}x) \\\notag
\eta^{\mu\nu} \partial_\mu \phi \partial_\nu \phi &\rightarrow& \eta^{\rho\sigma} \partial_\rho \phi(\Lambda^{-1}x) \partial_\sigma \phi(\Lambda^{-1}x) \\\notag
\mathcal{L}(x) &\rightarrow& \mathcal{L}(\Lambda^{-1}x) \\
V^\mu(x) &\rightarrow& \Lambda^\mu{}_\nu V^\nu(x)(\Lambda^{-1}x)
\end{eqnarray}

Focusing solely on the ``pure Lorentz group", i.e. the connected components that contains the identity element, sometimes denoted $L_+{}^\uparrow$ or $SO^+(3,1),$ we can express the group elements as exponential of the algebra elements \cite{boer}.

\begin{equation} \Lambda = e^{\frac{i}{2}\omega_{\mu\nu}M^{\mu\nu}}\end{equation}

Along the way we will play close attention to the similarities to the SU(2) Yang-Mills case. Noted that the $\omega_{\mu\nu}$ parameters are synonymous with the $\theta^a$ parameters, and the $M^{\mu\nu}$ generators are synonymous with the $T^a$ generators.

The antisymmetric $M^{\mu\nu}$ Lorentz generators obey the following commutation relation

\begin{equation}[M^{\mu\nu},M^{\rho\sigma}]=i(\eta^{\nu\rho}M^{\mu\sigma}+\eta^{\mu\sigma}M^{\nu\rho}-\eta^{\mu\rho}M^{\nu\sigma}-\eta^{\nu\sigma}M^{\mu\rho})\end{equation}

In analogy with the $[T^a,T^b]=if^{abc}T^c$ of Yang-Mills, the structure constants can be written out with some antisymmetrization deftness \cite{SUGRA}

\begin{equation}[M_a,M_b]=i f_{ab}{}^c M_c \rightarrow [M_{\mu\nu},M_{\rho\sigma}]=\frac{i}{2} f_{[\mu\nu][\rho\sigma]}{}^{[\kappa\tau]} M_{\kappa\tau} \label{struc1}\end{equation}

where the structure constants are

\begin{equation} f_{[\mu\nu][\rho\sigma]}{}^{[\kappa\tau]} = 8 \eta_{[\rho[\nu}\delta^{[\kappa}_{\mu]}\delta^{\tau]}_{\sigma]} \label{struc2}\end{equation}

Let's proceed and see if using those are even necessary.

\pagebreak
\subsection{Gauging the Lorentz algebra}

In analogy with going from global to local SU(2)

\begin{equation} U=e^{i\theta^aT^a} \rightarrow U(x)=e^{i\theta^a(x)T^a}\end{equation}

we gauge the global Lorentz in the same fashion

\begin{equation} \Lambda=e^{\frac{i}{2}\omega_{\mu\nu}M^{\mu\nu}} \rightarrow \Lambda(x)=e^{\frac{i}{2}\omega_{\mu\nu}(x)M^{\mu\nu}}\end{equation}

What was our next step in the Yang-Mills case? We saw that the normal partial derivatives introduced an extra term that spoiled the covariant nature of the derivative of the field, so we introduced a gauge field/connection that enabled us to construct a new derivative that was indeed covariant.

Where did we get this gauge field from? Recall that it was a 1-form connection. More specifically it was a \textit{Lie-algebra valued} 1-form

\begin{equation}A_\mu = A_\mu{}^a T^a\end{equation}

What will our new connection be for the Lorentz Lie algebra? In hindsight after inspiration from the work with Freedman-Van Proeyen, we know the spin connection ought to be our first candidate. Let's dig a little deeper to work out the details and motivate it clearly.

\pagebreak
\subsection{The return of the vielbein, differential forms, and Cartan's structure equations}
\label{subsec:returnofmanylegs}

The vielbein 1-form $e^a=e_\mu{}^a(x)dx^\mu$ obeys

\begin{equation} g_{\mu\nu}(x)=\eta_{ab}e_\mu{}^a(x) e_\nu{}^b(x)\end{equation}

Given a local Lorentz transformations (LLT), i.e. satisfies $\Lambda^a{}_b(x) \Lambda^c{}_d (x) \eta_{ac} = \eta_{bd}$, we can construct new solutions to the above ``square root of the metric" equation

\begin{equation}
\label{eq:llt}e_\mu{}^{a'}(x) = (\Lambda^{-1})^a{}_b(x) e_\mu{}^b(x)\end{equation}

In other words, all choices of vielbein that are related by LLTs are totally equivalent. As a consequence of this, the vielbein \textit{and all geometric quantities derived from it} must be used in such a way that is covariant according to the above transformation of the vielbein. We will come back to this when we establish that the spin connection is constructed out of the vielbein and is not independent.

Notice that theses LLTs are exactly what we are interested in during this gauging global symmetries process.

Let's see if derivatives of the vielbein transform nicely. Let d be the exterior derivative

\begin{eqnarray} \notag
de^{a'} &=& d((\Lambda^{-1})^a{}_b  e^b) \\
&=& d((\Lambda^{-1})^a{}_b)  e^b + (\Lambda^{-1})^a{}_b  de^b
\end{eqnarray}

It does not transform nicely (as a vector according to equation \discolorlinks{\ref{eq:llt}}, $X^a=(\Lambda^{-1})^a{}_b X^b$) the first term spoils it.

Take a moment to compare this to YM. The extra term arises in both when we naively try to use an old form of a derivative after gauging a symmetry.

While $de^a$ doesn't transform covariantly, we can define a new object that does

\begin{equation}T^a=de^a+\omega^a{}_be^b \label{torsion1}
\end{equation}

provided this new 1-form $\omega^{ab}=\omega_\mu{}^{ab}$ transforms like

\begin{equation} \omega^a{}_b{}^{'} \rightarrow (\Lambda^{-1})^a{}_c d(\Lambda^c{}_b) + (\Lambda^{-1})^a{}_c \omega^c{}_d \Lambda^d{}_b\end{equation}

Take a moment to compare this to how the gauge field/connection $A_\mu$ needed to transform so that the covariant derivative worked,

\begin{equation}A_\mu{}^{ '} = -\frac{i}{g}[\partial_\mu U(x)]U(x)^{-1} + U(x) A_\mu U(x)^{-1} \end{equation}

Remarkable, no? U(x) there were elements of the SU(2) group, and $\Lambda(x)$ here are elements of the Lorentz group.

Proving this $T^a$ transforms nicely (recall equation \discolorlinks{\ref{eq:id}})

\begin{eqnarray} \notag
T^{a'}  &=& (de^{a} + \omega^a{}_b{}e^{b})^{'} \\\notag
&=& de^{a'} + \omega^a{}_b{}^{'}e^{b'} \\\notag
&=& d((\Lambda^{-1})^a{}_b)  e^b + (\Lambda^{-1})^a{}_b  de^b + [(\Lambda^{-1})^a{}_c d(\Lambda^c{}_b) + (\Lambda^{-1})^a{}_c \omega^c{}_d \Lambda^d{}_b][(\Lambda^{-1})^b{}_c e^c] \\\notag
&=& (\Lambda^{-1})^a{}_b  de^b + d((\Lambda^{-1})^a{}_b)  e^b + (\Lambda^{-1})^a{}_c d(\Lambda^c{}_b)(\Lambda^{-1})^b{}_c e^c + (\Lambda^{-1})^a{}_c \omega^c{}_d \Lambda^d{}_b(\Lambda^{-1})^b{}_c e^c \\\notag
&=& (\Lambda^{-1})^a{}_b  de^b + d((\Lambda^{-1})^a{}_b)  e^b + (\Lambda^{-1})^a{}_c d(\Lambda^c{}_b)(\Lambda^{-1})^b{}_c e^c + (\Lambda^{-1})^a{}_c \omega^c{}_d \delta^d_c e^c \\\notag
&=& (\Lambda^{-1})^a{}_b  de^b + d((\Lambda^{-1})^a{}_b)  e^b - (\Lambda^{-1})^a{}_c \Lambda^c{}_b d((\Lambda^{-1})^b{}_c) e^c + (\Lambda^{-1})^a{}_c \omega^c{}_d e^d \\\notag
&=& (\Lambda^{-1})^a{}_b  de^b + d((\Lambda^{-1})^a{}_b)  e^b - \delta^a_b d((\Lambda^{-1})^b{}_c) e^c + (\Lambda^{-1})^a{}_c \omega^c{}_d e^d \\\notag
&=& (\Lambda^{-1})^a{}_b  de^b + d((\Lambda^{-1})^a{}_b)  e^b - d((\Lambda^{-1})^a{}_b) e^b + (\Lambda^{-1})^a{}_b \omega^b{}_c e^c \\\notag
&=& (\Lambda^{-1})^a{}_b de^b + (\Lambda^{-1})^a{}_b  \omega^b{}_c e^c \\
&=& (\Lambda^{-1})^a{}_b T^b
\end{eqnarray}

With inspiration from Cartan's first equation, $T^a= T_{\mu\nu}{}^a$ is the torsion two form. The torsionless 1st Cartan's equation reads $de^a+\omega^a{}_b e^b=0$.

The form of this nicely transforming $T^a$ and its resemblance to the gauge covariant derivative $dX \rightarrow dX +AX$ motivates labeling the spin connection as the gauge field/connection for the Lorentz algebra. In analogy with $D_\mu = \partial_\mu -i A_\mu$, we define (dropping the i for personal reasons) $D_\mu = \partial_\mu - \omega_\mu$. Recall that the connection $A_\mu$ is a \textit{Lie algebra-valued} connection, $A_\mu = A_\mu{}^aT^a$. Taking into account anti-symmetry of the generators of the Lorentz algebra, we can do the same for $\omega_\mu$ so that 

\begin{equation}D_\mu = \partial_\mu - \omega_\mu= \partial_\mu - \frac{1}{2}\omega_\mu{}^{ab}M_{ab}\end{equation}

Before moving onto the curvature/field strength, let's recap what steps we took in SU(2) Yang-Mills

\begin{enumerate}
    \item Gauged the symmetry, $U \rightarrow U(x)$
    \item Introduced a gauge field/connection/ Lie algebra-valued 1-form $A_\mu$
    \item Defined a new derivative with the gauge field
    \item Determined the transformation of the gauge field by requiring that the new derivative is covariant, i.e. it transforms like the field itself
    \item Accounted for the gauge field having dynamics of its own by defining a field strength/curvature that we could use to construct a term in the Lagrangian describing the gauge fields dynamics
\end{enumerate}

In that last step we took two approaches, the differential forms and curvature 2-form approach, and the commutator of two covariant derivatives approach. Let's do the same for Lorentz now.

\pagebreak
\subsection{Lorentz curvature/field strength}
\label{sec:lcurv}

Recall in SU(2) that the field strength/curvature was also Lie-algebra valued, $F_{\mu\nu}=F_{\mu\nu}{}^a T^a$. Similarly for the Lorentz curvature (again, taking into account antisymmetry of the Lorentz generators)

\begin{equation}R_{\mu\nu}=\frac{1}{2} R_{\mu\nu}{}^{ab}M_{ab}\end{equation}

In a similar fashion to defining the curvature 2-form, $F=dA+\frac{1}{2}[A,A]$, we define the same for the new connection $\omega$

\begin{equation}R=d\omega +\frac{1}{2}[\omega,\omega] = \frac{1}{2}(\partial_\mu \omega_\nu -\partial_\nu \omega_\mu + [\omega_\mu,\omega_\nu])dx^\mu dx^\nu\end{equation}

such that $R_{\mu\nu}=\partial_\mu \omega_\nu -\partial_\nu \omega_\mu + [\omega_\mu,\omega_\nu]$. We can expand the $\omega_\mu$ in this expression such that

\begin{eqnarray} \notag
R_{\mu\nu}&=&\partial_\mu \Big(\frac{1}{2}\omega_\nu{}^{ab}M_{ab}\Big) -\partial_\nu \Big(\frac{1}{2}\omega_\mu{}^{ab}M_{ab}\Big) + \bigg[\Big(\frac{1}{2}\omega_\mu{}^{ab}M_{ab}\Big),\Big(\frac{1}{2}\omega_\nu{}^{ab}M_{ab}\Big)\bigg] \\\notag
&=& \frac{1}{2}M_{ab}(\partial_\mu \omega_\nu{}^{ab} -\partial_\nu\omega_\mu{}^{ab}) + \Big[\big(\frac{1}{2}\omega_\mu{}^{ab}M_{ab}\big),\big(\frac{1}{2}\omega_\nu{}^{cd}M_{cd}\big)\Big] \\
&=& \frac{1}{2}M_{ab}(\partial_\mu \omega_\nu{}^{ab} -\partial_\nu\omega_\mu{}^{ab}) + \frac{1}{4} \omega_\mu{}^{ab}\omega_\nu{}^{cd} [M_{ab},M_{cd}]   
\end{eqnarray}

where in the last line the one-half's and $\omega$'s were pulled out because they are constants.

Some care is required in evaluating the commutator $[M_{ab},M_{cd}]$. Since we saturated the indices with another antisymmetric object, the $\omega$'s, we have taken into account the antisymmetry already. And so, the commutation relation (dropping the i)

\begin{equation}[M_{\mu\nu},M_{\rho\sigma}]=\eta_{\nu\rho}M_{\mu\sigma}+\eta_{\mu\sigma}M_{\nu\rho}-\eta_{\mu\rho}M_{\nu\sigma}-\eta_{\nu\sigma}M_{\mu\rho}\end{equation}

is simplified. Notably, the 1st/3rd terms and the 2nd/4th terms are the same if the antisymmetry in $\mu$ and $\nu$ have been accounted for. In our case it has, since we saturated the a and b with $\omega_\mu{}^{ab}$. So we instead have the commutation relation

\begin{equation}[M_{\mu\nu},M_{\rho\sigma}]= 2\eta_{\nu\rho}M_{\mu\sigma}+2\eta_{\mu\sigma}M_{\nu\rho}\end{equation}

Implementing this above we continue and get

\begin{eqnarray} \notag
R_{\mu\nu}
&=& \frac{1}{2}M_{ab}(\partial_\mu \omega_\nu{}^{ab} -\partial_\nu\omega_\mu{}^{ab}) + \frac{1}{4} \omega_\mu{}^{ab}\omega_\nu{}^{cd} [M_{ab},M_{cd}]   \\\notag
&=& \frac{1}{2}M_{ab}(\partial_\mu \omega_\nu{}^{ab} -\partial_\nu\omega_\mu{}^{ab})+\frac{1}{4} \omega_\mu{}^{ab}\omega_\nu{}^{cd} (2 \eta_{bc}M_{ad} + 2 \eta_{ad}M_{bc}) \\\notag
&=& \frac{1}{2}M_{ab}(\partial_\mu \omega_\nu{}^{ab} -\partial_\nu\omega_\mu{}^{ab})
+\frac{1}{2} (\omega_\mu{}^{ab}\omega_\nu{}^{cd} \eta_{bc}M_{ad} + \omega_\mu{}^{ab}\omega_\nu{}^{cd}\eta_{ad}M_{bc}) \\\notag
&=& \frac{1}{2}M_{ab}(\partial_\mu \omega_\nu{}^{ab} -\partial_\nu\omega_\mu{}^{ab})
+\frac{1}{2} (\omega_\mu{}^a{}_c\omega_\nu{}^{cd} M_{ad} + \omega_\mu{}_d{}^b\omega_\nu{}^{cd} M_{bc}) \\\notag
&=& \frac{1}{2}M_{ab}(\partial_\mu \omega_\nu{}^{ab} -\partial_\nu\omega_\mu{}^{ab})
+\frac{1}{2} (\omega_\mu{}^{ac}\omega_\nu{}_c{}^b M_{ab} + \omega_\nu{}^{cd} \omega_\mu{}_d{}^b M_{bc}) \\\notag
&=& \frac{1}{2}M_{ab}(\partial_\mu \omega_\nu{}^{ab} -\partial_\nu\omega_\mu{}^{ab})
+\frac{1}{2} (\omega_\mu{}^{ac}\omega_\nu{}_c{}^b M_{ab} + \omega_\nu{}^{ac} \omega_\mu{}_c{}^b M_{ba}) \\\notag
&=& \frac{1}{2}M_{ab}(\partial_\mu \omega_\nu{}^{ab} -\partial_\nu\omega_\mu{}^{ab})
+\frac{1}{2} (\omega_\mu{}^{ac}\omega_\nu{}_c{}^b -\omega_\nu{}^{ac} \omega_\mu{}_c{}^b) M_{ab} \\
&=& \frac{1}{2}(\partial_\mu \omega_\nu{}^{ab} -\partial_\nu\omega_\mu{}^{ab}+\omega_\mu{}^{ac}\omega_\nu{}_c{}^b -\omega_\nu{}^{ac} \omega_\mu{}_c{}^b) M_{ab} 
\end{eqnarray}

where in the 5th to 6th lines we reindexed $d\rightarrow c$ and $c\rightarrow a$.

Using $R_{\mu\nu}=\frac{1}{2} R_{\mu\nu}{}^{ab}M_{ab}$ we identify the curvature as 

\begin{align} \notag
R_{\mu\nu}{}^{ab} &= \partial_\mu \omega_\nu{}^{ab} -\partial_\nu\omega_\mu{}^{ab}+\omega_\mu{}^{ac}\omega_\nu{}_c{}^b -\omega_\nu{}^{ac} \omega_\mu{}_c{}^b  \\ \notag
&= 2 \partial_{[\mu} \omega_{\nu]}{}^{ab} -\omega_\mu{}^{ca}\omega_\nu{}_c{}^b +\omega_\nu{}^{ca} \omega_\mu{}_c{}^b  \\ 
&= 2 ( \partial_{[\mu} \omega_{\nu]}{}^{ab} - \omega_{[\mu}{}^{ca}\omega_{\nu]}{}_c{}^b)  \label{LLTcurve}
\end{align}

\pagebreak

\section{Abstract symmetry transformations and gauge theory}
\label{gauge}

In preparation for the language that we will encounter when working with the Poincaré symmetry transformation, we use appendix B of Andringa's thesis \cite{Roelthesis}, and chapter 11 of Freedman and Van Proeyen's textbook \cite{SUGRA} to generalize the work of Chapters \ref{chap:gauge1}, \ref{chap:diggingDeeper}, and \ref{chap:lgauge} to abstract symmety transformations and gauge theory.

\subsection{Global symmetry transformations}
An infinitesimal symmetry transformation is determined by 

1) a parameter, call it $\epsilon^A$, and 

2) an operation, call it $\delta(\epsilon)$.

The operation $\delta(\epsilon)$

1) depends linearly on the parameter $\epsilon^A$, and 

2) acts on fields, i.e. $\delta(\epsilon)\phi^i$.

For some global symmetry, $\epsilon^A$ does not depend on the spacetime $x^\mu$.

Another way to say ``$\delta(\epsilon)$ depends linearly on the parameter $\epsilon^A$,'' is to write 
\begin{equation}\delta(\epsilon) =\epsilon^A T_A\end{equation}

where the $T_A$ are some operations on fields. (The $T_A$ operate on fields just like $\delta(\epsilon)$ so they are kind of like basis elements for the symmetry transformations $\delta(\epsilon)$.) $T_A$ are also called the field-space generators of the symmetry transformation.

Let $\{(t_A)^i{}_j\}$ be the matrix generators of a representation of some Lie algebra.

This Lie algebra (LA) is defined by $[t_A,t_B]=f_{AB}{}^C t_C$.

The action of $T_A$ on the fields is defined with the LA basis elements,

\begin{equation}T_A(\phi^i)=-(t_A)^i{}_j \phi^i\end{equation}.

So then we have 

\begin{eqnarray} \notag
\delta(\epsilon)\phi^i &=& \epsilon^A T_A(\phi^i) \\
&=& -\epsilon^A (t_A)^i{}_j (\phi^j)
\end{eqnarray}

Then the product of two symmetry transformations reads,

\begin{eqnarray} 
\notag
\delta(\epsilon_1)\delta(\epsilon_2)\phi^i &=& \epsilon_1{}^A T_A(\epsilon_2{}^B T_B\phi^i) \\\notag
&=& \epsilon_1{}^A T_A(-\epsilon_2{}^B (t_B)^i{}_j \phi^j) \\\notag
&=& -\epsilon_1{}^A \epsilon_2{}^B (t_B)^i{}_j T_A \phi^j \\\notag
&=& -\epsilon_1{}^A \epsilon_2{}^B (t_B)^i{}_j (-(t_A)^j{}_k \phi^k) \\
&=& \epsilon_1{}^A \epsilon_2{}^B (t_B)^i{}_j (t_A)^j{}_k \phi^k 
\end{eqnarray}

Note that the $T_A$ act on fields, and $(t_B)^i{}_j$ is just a matrix, so $T_A$ doesn't act on it. Also note that matrix multiplication is associative.

Using $[t_A,t_B]=f_{AB}{}^C t_C$, we have 

\begin{equation}[\delta(\epsilon_1),\delta(\epsilon_2)]\phi^i = \epsilon_1{}^A \epsilon_2{}^B f_{AB}{}^C T_C \phi^i\end{equation}

And then we have

\begin{equation}[T_A,T_B]=f_{AB}{}^C T_C,\text{ and}\end{equation}

\begin{equation}[\delta(\epsilon_1),\delta(\epsilon_2)]=\delta(\epsilon_3{}^C),\end{equation} 

where $\epsilon_3{}^C  =\epsilon_1{}^A \epsilon_2{}^B f_{AB}{}^C$.

\pagebreak
\subsection{Local symmetry transformations}

If we want to work with local transformations, then in the same way that we went from $\phi \rightarrow \phi ' = e^{i\alpha}\phi$ to $\phi \rightarrow \phi ' = e^{i\alpha(x)}\phi$, now we let $\epsilon^A$ depend on the spacetime $x^\mu.$ Thus, everywhere $\epsilon^A$ is written, $\epsilon^A(x)$ is implied.

Recall the discussion in Section \ref{subsecGR} where we needed to introduce the gauge field/connection $A^\mu$ to compare fields are different points in spacetime. We generalize this and introduce (for each symmetry transformation, labeled by A,B,C,etc.) the gauge field/connection $B_\mu{}^A$.

Recall that $\delta(\epsilon)$ acts on fields. Well $B_\mu{}^A$ is a field too, so we have 

\begin{equation}\delta(\epsilon)B_\mu{}^A = \partial_\mu \epsilon^A + \epsilon^C B_\mu{}^B f_{BC}{}^A\end{equation}

Lets compare this general formula to our example before,

\begin{equation}\delta(\alpha)A_\mu = \partial_\mu (\alpha(x)) \end{equation}

where we see that our $\alpha(x)$ is synonymous with $\epsilon^A(x)$. (Note that this notation, $\delta(A_\mu)$ means the variation, $A_\mu ' - A_\mu$, of the gauge field from its transformation $A_\mu \rightarrow A_\mu + \partial_\mu \alpha$. This notation will change below.)

Moreover, note that in our example, the symmetry was that of the U(1) Lie group. The Lie algebra of U(1) is 1-dimensional, notably the only generator is the phase (think of U(1) as the circle group, it can be parametrized by $e^i\alpha$, where $\alpha$ is the Lie algebra element). The notion of a structure constant does not make sense in a 1-dimensional Lie algebra, the Lie algebra needs to be at least 2-dimensional for the structure constant formula ($[a,b]=xa$ for some constant x for example) to make sense. Thus is makes sense that in our previous U(1) example, the formula for the result of the symmetry transformation acting on the gauge field didn't have the second term containing the structure constant. 
\iffalse
Using the ``graded" Jacobi identity, $\epsilon_2{}^B\epsilon_1{}^A\epsilon_3{}^C f_{AB}{}^D f_{CD}{}^E$ + (cyclic permutations of 1 to 2 and 2 to 3) , as well as the anti-commutativity of the $\epsilon^A$ parameters (??? I thought the parameters didn't commute only for fermionic systems???) we can compute

\begin{equation}[\delta(\epsilon_1),\delta(\epsilon_2)] B_\mu{}^A= \partial_\mu \epsilon_3{}^A + \epsilon_3{}^C B_\mu{}^B f_{BC}^A\end{equation} 

where $\epsilon_3{}^C$ (also $\epsilon_3{}^A$?) is given by $\epsilon_2{}^B \epsilon_1{}^A f_{AB}{}^C.$
\fi
\pagebreak
\subsection{Covariant derivatives and curvatures}
\label{subsec:covderiv}

A normal partial is not covariant because it picks up a term that doesn't transform like the field $\phi$ does. Another way to say it isn't covariant, is to say that it involved derivatives of the gauge parameter. For instance, in the following

\begin{eqnarray} \notag
\delta(\epsilon) \partial_\mu \phi &=& \partial_\mu \delta(\epsilon) \phi \\\notag
&=& \partial_\mu (\epsilon^A T_A \phi) \\
&=& \partial_\mu (\epsilon^A) T_A \phi + \epsilon^A \partial_\mu (T_A \phi)
\end{eqnarray}

the second term is totally kosher and transforms just like the field, but the first term (containing a derivative of the gauge parameters) ruins the covariance.

Consider the same equation from the U(1) example,

\begin{eqnarray} \notag
\partial_\mu\phi(x) &\rightarrow& \partial_\mu(e^{i\alpha(x)}\phi(x))
\\
&=& \partial_\mu(e^{-i\alpha(x)})\phi(x) +e^{i\alpha(x)}\partial_\mu (\phi(x))
\end{eqnarray}

We define the generic covariant gauge derivative,

\begin{eqnarray} \notag
D_\mu &=& \partial_\mu - \delta(B_\mu) \\
&=& \partial_\mu - B_\mu^A T_A
\end{eqnarray}

As we were warned about above, the term $\delta(B_\mu)$ is not the variation of the gauge field's transformation. Rather, it is read as using the gauge field as a \textit{parameter} for the symmetry transformation previously defined, $\delta(\epsilon)\phi^i=\epsilon^A T_A \phi^i$. Thus, one simply replaces $\epsilon$ with $B_\mu$, so that $\delta(B_\mu)\phi^i=B_\mu{}^A T_A \phi^i$.

Before testing the covariance of this new derivative, let's use $[\delta(\epsilon_1),\delta_(\epsilon_2)]=\delta(\epsilon_3{}^C)$ (where $\epsilon_3{}^C  =\epsilon_1{}^A \epsilon_2{}^B f_{AB}{}^C$) in a fancy way.

Replace $\epsilon_1$ with $B_\mu$, replace $\epsilon_2$ with $\epsilon$ and let both sides act on $\phi$.

\begin{eqnarray} 
\label{eq:above}
\notag
[\delta(\epsilon_1),\delta(\epsilon_2)]\phi &=& \delta(\epsilon_1{}^A \epsilon_2{}^B f_{AB}{}^C)\phi \\\notag
\delta(B_\mu)\delta(\epsilon)\phi - \delta(\epsilon)\delta(B_\mu)\phi&=& \delta(B_\mu{}^A \epsilon^B f_{AB}{}^C)\phi \\\notag
\delta(B_\mu)(\epsilon^A T_A \phi) - \delta(\epsilon)(B_\mu{}^A T_A \phi)&=& B_\mu{}^A \epsilon^B f_{AB}{}^C T_C \phi \\
\epsilon^A\delta(B_\mu)( T_A \phi) - B_\mu{}^A\delta(\epsilon)(T_A \phi)&=& B_\mu{}^A \epsilon^B f_{AB}{}^C T_C \phi 
\end{eqnarray}

To test the covariance of $D_\mu$ we use the following:

\begin{itemize}
    \item $\delta(\epsilon) \partial_\mu \phi = \partial_\mu \delta(\epsilon) \phi$ (3rd equality below)
    \item $\delta(\epsilon)B_\mu{}^A = \partial_\mu \epsilon^A + \epsilon^C B_\mu{}^B f_{BC}{}^A$ (4th equality below)
    \item the negative, reindiced $(B\rightarrow C, A\rightarrow B, C\rightarrow A)$, and rearranged version of equation \discolorlinks{\ref{eq:above}}, 
\end{itemize}

\begin{eqnarray} \notag
\epsilon^A\delta(B_\mu)( T_A \phi) - B_\mu{}^A\delta(\epsilon)(T_A \phi)&=& B_\mu{}^A \epsilon^B f_{AB}{}^C T_C \phi \\
\epsilon^A\delta(B_\mu)( T_A \phi) - B_\mu{}^A\delta(\epsilon)(T_A \phi)&=& \epsilon^B B_\mu{}^A f_{AB}{}^C T_C \phi \\
-\epsilon^B B_\mu{}^A f_{AB}{}^C T_C \phi &=& -\epsilon^A\delta(B_\mu)( T_A \phi) + B_\mu{}^A\delta(\epsilon)(T_A \phi) \\
-\epsilon^C B_\mu{}^B f_{BC}{}^A T_A \phi &=& -\epsilon^A\delta(B_\mu)( T_A \phi) + B_\mu{}^A\delta(\epsilon)(T_A \phi) \\
\end{eqnarray}

\hspace{1cm} which we use in the 7th equality.

And so, testing the covariance of $D_\mu$ we have

\begin{eqnarray} \notag
\delta(\epsilon) D_\mu \phi &=& \delta(\epsilon) (\partial_\mu - \delta(B_\mu))\phi\\\notag
&=&  \delta(\epsilon)\partial_\mu \phi - \delta(\epsilon)B_\mu{}^A T_A\phi\\\notag
&=& \partial_\mu(\delta(\epsilon) \phi) - \delta(\epsilon)(B_\mu{}^A) T_A\phi -  B_\mu{}^A \delta(\epsilon)(T_A\phi)\\\notag
&=& \partial_\mu(\epsilon^A T_A \phi) - \partial_\mu \epsilon^A T_A\phi - \epsilon^C B_\mu{}^B f_{BC}{}^A T_A\phi -  B_\mu{}^A \delta(\epsilon)(T_A\phi)\\\notag
&=& \partial_\mu(\epsilon^A) T_A \phi + \epsilon^A \partial_\mu( T_A \phi) - \partial_\mu \epsilon^A T_A\phi - \epsilon^C B_\mu{}^B f_{BC}{}^A T_A\phi -  B_\mu{}^A \delta(\epsilon)(T_A\phi)\\\notag
&=& \epsilon^A \partial_\mu( T_A \phi) - \epsilon^C B_\mu{}^B f_{BC}{}^A T_A\phi -  B_\mu{}^A \delta(\epsilon)(T_A\phi)\\\notag
&=& \epsilon^A \partial_\mu( T_A \phi) -\epsilon^A\delta(B_\mu)( T_A \phi) + B_\mu{}^A\delta(\epsilon)(T_A \phi) -  B_\mu{}^A \delta(\epsilon)(T_A\phi)\\\notag
&=& \epsilon^A \partial_\mu( T_A \phi) -\epsilon^A\delta(B_\mu)( T_A \phi) \\\notag
&=& \epsilon^A [\partial_\mu( T_A \phi) -\delta(B_\mu)] T_A \phi \\
&=& \epsilon^A D_\mu T_A \phi
\end{eqnarray}

which is precisely the form we would like for a covariant derivative. (Think back to the U(1) case if you'd like, $D_\mu(\phi(x))=e^{i\alpha(x)}D_\mu(\phi(x))$.)

While the construction of curvatures is much more nature from the perspective of Section \ref{introdifforms} and \ref{sec:lcurv}, the covariant derivative can be indeed be used to construct curvatures as we saw in Section \ref{pesky}.

The commutator of the covariant derivatives reads 

\begin{equation}[D_\mu,D_\nu]=-\delta(R_{\mu\nu})\end{equation}

where

\begin{equation}R_{\mu\nu}{}^A= \partial_\mu B_\nu{}^A -\partial_\nu B_\mu{}^A+ B_\nu{}^C B_\mu{}^B f_{BC}{}^A\end{equation}

\pagebreak

\section{Poincaré algebra gauge theory}
\label{sec:pgauge}

\subsection{Poincaré group/algebra review and general coordinate transformations}

Here, we update the group/algebra notation from Section \ref{subsec:ls}, and update the  general coordinate transformation (GCT) notation from Section \ref{ewcarroll} using the conventions of Freedman-Van Proeyen \cite{SUGRA}. Our metric is $(-+++)$ and we neglect imaginary $i$'s in algebra commutation relations.

A Lorentz transformation doesn't change position in spacetime, but acts as a kind of ``rotation" (at each point of spacetime for LLTs),

\begin{equation}x^\mu \longrightarrow x^{\mu'}=(\Lambda^{-1})^\mu{}_\nu x^\nu.\end{equation}

A Poincaré transformation also transforms the spacetime coordinates themselves

\begin{equation}x^\mu \longrightarrow x^{\mu'}=(\Lambda^{-1})^\mu{}_\nu (x^\nu -a^\nu)\end{equation}

\iffalse
In order to hash out some of the notation in Freedman-Van Proeyen that confused me at first, I will include some notation from Professor Daniël Boer's Lectures Notes in Lie Groups in Physics \cite{boer}.

An infinitesimal Lorentz transformation can be written as 

\begin{equation}\Lambda^\mu{}_\nu =\delta^\mu{}_\nu + \omega^\mu{}_\nu\end{equation}

where $\omega_{\mu\nu}$ is antisymmetric, $\omega_{\mu\nu}=-\omega_{\nu\mu}$.

All Lorentz algebra elements, infinitesimal rotations $J^i$ and Lorentz boosts $K^i$ can be succinctly combined in the object $M^{\mu\nu}$ which we can represent as a matrix as 

\begin{equation}(M^{lm})^j{}_k=i (g^
{lj}g^m{}_k - g^{mj}g^l{}_k)\end{equation}

where on the left hand side, it is helpful to think of l and m as indices/labels, and the j and k as rows/columns for the whole matrix.

\fi

Thus, in addition to the Lorentz rotations and boosts of the Lorentz group, the full-blown Poincaré group also includes spacetime translations. And so the Poincaré algebra has 4 more symmetry transformation generators $P_\mu$ in addition to the (anti-symmetric) $M_{\mu\nu}$ of the pure Lorentz algebra. The algebra is given by the following commutation relations 

\begin{align}
    [P_\mu,P_\nu] &=0 
    \label{poin1} \\ 
    \notag
    [M_{\mu \nu },P_\rho] &= \eta_{\nu \rho}P_\mu-\eta_{\mu \rho}P_\nu \\  \notag
    &= \eta_{\rho\nu}P_\mu -\eta_{\rho\mu }P_\nu \\  
    &= 2 \eta_{\rho[\nu}P_{\mu]} \label{poin2}\\
    \notag
    [M_{\mu \nu},M_{\rho\sigma}] &= \eta_{\nu \rho}M_{\mu \sigma}+\eta_{\mu \sigma}M_{\nu \rho}-\eta_{\mu \rho}M_{\nu \sigma}-\eta_{\nu \sigma}M_{\mu \rho} \\ \notag
    &= -\eta_{\mu \rho}M_{\nu \sigma}+\eta_{\mu \sigma}M_{\nu \rho}+\eta_{\nu \rho}M_{\mu \sigma}-\eta_{\nu \sigma}M_{\mu \rho} \\ \notag
    &= \eta_{\mu \rho}M_{\sigma\nu }-\eta_{\mu \sigma}M_{\rho\sigma}-\eta_{\nu \rho}M_{\sigma\mu }+\eta_{\nu \sigma}M_{\rho\mu } \\ \notag
    &= 2(\eta_{\mu[\rho}M_{\sigma]\nu } -\eta_{\nu [\rho}M_{\sigma]\mu}) \\ 
    &= 4\eta_{[\mu[\rho}M_{\sigma]\nu ]} \label{poin3}
\end{align}

An element of the (connected component of the) Lorentz group can be written as

\begin{equation}
U(\Lambda) = e^{-\frac{1}{2}\lambda^{\mu\nu}M_{\mu\nu}}
\end{equation}

where $M_{\mu\nu}$ are the Lorentz generators, and an element of the translation subgroup of the Poincaré group can be written as

\begin{equation}
U(a) = e^{a^\mu P_\mu} \label{oppoin}
\end{equation}

where $P_\mu$ are the translation generators and can be identified with partial derivatives $\partial_\mu$. See page 81 of \cite{QFTGA} for a nice explanation of this.

We can also represent $M_{\mu\nu}$ as follows for an infinitesimal transformation

\begin{equation}M_{\mu\nu}=x_\mu \partial_\nu - x_\nu \partial_\mu\end{equation}

If we match the anti-symmetric $M_{\mu\nu}$ with the anti-symmetric parameter $\lambda^{\mu\nu}$, we can write

\begin{equation}
\label{eq:4}
-\frac{1}{2}\lambda^{\mu\nu}M_{\mu\nu} = \lambda^\mu{}_\nu x^\nu \partial_\mu
\end{equation}

The transformations of scalars fields under Lorentz transformations follow from the expression of the group elements as a exponentiation of the generators, and equation \discolorlinks{\ref{eq:4}} can be used to express it in an alternative form

\begin{eqnarray} 
\label{eq1}
\notag
\delta(\lambda) \phi(x) &=& -\frac{1}{2}\lambda^{\mu\nu}M_{\mu\nu} \phi(x) \\
&=& \lambda^{\mu\nu} x_\nu \partial_\mu \phi(x)
\end{eqnarray}

Similarly for translations,

\begin{eqnarray} 
\label{eq2}
\notag
\delta(a) \phi(x) &=& a^\mu P_{\mu} \phi(x) \\
&=& a^\mu \partial_\mu \phi(x)
\end{eqnarray}

A general coordinate transformation (GCT) takes the form \cite{SUGRA}

\begin{equation}
x^\mu \rightarrow x^{' \mu} (x^\nu)
\end{equation}

where the coordinates are related by the Jacobian matrix $\frac{\partial x^{' \mu}}{\partial x^\nu}$.

A spacetime-dependent scalar field transforms so that the change in coordinates is negated 

\begin{equation}
\phi^{'} (x^{'}) = \phi(x)
\end{equation}

A vector transforms as

\begin{equation}
V_{\mu}{}^{'}(x^{'}) = \frac{\partial x^{\nu}}{\partial x^{'\mu}} V_\nu (x)
\end{equation}

Now consider an infinitesimal GCT, where

\begin{equation}
x^\mu \rightarrow x^{' \mu} = x^\mu - \xi^\mu(x)
\end{equation}

Then the changes in the above scalar field/vector become

\begin{eqnarray} \notag
\delta_{gct}\phi &=& \phi '(x) - \phi(x) \\ \notag
&=& L_\xi \phi \\ 
&=& \xi^\mu(x) \partial_\mu \phi
\end{eqnarray}

\begin{eqnarray} \notag
\delta_{gct} V_\mu (x) &=& V_{\mu}{}^{'}(x) - V_{\mu}(x) \\ \notag
&=& L_\xi V_{\mu} \\ 
&=& \xi^\rho(x) (\partial_\rho V_{\mu}(x)) + V_{\rho}(\partial_\mu \xi^\rho(x)) \label{gctvector}
\end{eqnarray}

where $L_\xi$ is shorthand for Lie derivative.

Combining equations \discolorlinks{\ref{eq1}} and \discolorlinks{\ref{eq2}} allows us to express a local ($a^\mu,\lambda^{\mu\nu} \rightarrow a^\mu(x),\lambda^{\mu\nu}(x))$ Poincaré transformation as a GCT in this framework,

\begin{eqnarray} 
\notag
\delta(a^\mu,\lambda^{\mu\nu}) \phi(x) &=& (a^\mu(x) P_\mu -\frac{1}{2}\lambda^{\mu\nu}(x)M_{\mu\nu}) \phi(x) \\\notag
&=& (a^\mu(x) \partial_\mu +\lambda^{\mu\nu}(x) x_\nu \partial_\mu) \phi(x)\\\notag
&=& (a^\mu(x) +\lambda^{\mu\nu}(x) x_\nu) \partial_\mu \phi(x) \\\notag
&=& (\xi^\mu(x)) \partial_\mu \phi(x) \\\notag
&=& L_\xi \phi(x) \\
\delta(\xi^\mu) \phi(x) &=& \delta_{gct} \phi(x) \label{gctphi}
\end{eqnarray}

where we generalized the spacetime translation vector $a^\mu(x)$ to curved spacetime with $\xi^\mu(x)= a^\mu(x)+ \lambda^{\mu\nu}(x) x_\nu$. So we will have GCTs parametrized by $\xi^\mu(x)$ and LLTs parametrized by $\lambda^{ab}(x)$.

\pagebreak
\subsection{Naive Poincaré gauge theory}
\label{naive}

Before we move onto the Poincaré algebra, let us type up Freedman-Van Proeyen's gauge theory nomenclature. We did this for SU(2) in Section \ref{fvplanguage} (recall that Freedman-Van Proeyen's notation is the right column there), but now we will concern ourselves with completely arbitrary symmetry transformations as we studied in Chapter \ref{gauge}

\begin{eqnarray}
    \delta(\epsilon) &=& \epsilon^A T_A \\ 
T_a \phi^b &=& - f_{ac}{}^b \phi^c \\ 
\delta(\epsilon)B_\mu{}^a &=& \partial_\mu \epsilon^a + \epsilon^c B_\mu{}^b f_{bc}{}^a \label{transgauge}\\ \notag
D_\mu \phi^i &=& (\partial_\mu - \delta(B_\mu))\phi^i \\ 
&=& \partial_\mu \phi^i - B_\mu^a T_a \phi^i \label{gaugcovder}\\ 
R_{\mu\nu}{}^a &=& 2 \partial_{[\mu} B_{\nu]}{}^a + B_\nu{}^c B_\mu{}^b f_{bc}{}^a \label{gaugecurvature}\\
\delta (\epsilon)R_{\mu\nu}^a &=& \epsilon^c R_{\mu\nu}^b f_{bc}{}^a
\end{eqnarray}

While before for SU(2) and Lorentz we only had one gauge field to create a Lie algebra-valued connection/1-form with, nothing stops us from  making one for Poincaré where we have two gauge fields. The space of 1-forms is a vector space, and addition is defined there. So a sum of 1-forms can indeed be a 1-form.

Summarizing the SU(2), Lorentz, and Poincaré connection 1-forms side-by-side we have

\begin{align}
&\text{SU(2)} & A_\mu &= A_\mu{}^a T_a \\
&\text{Lorentz} & \omega_\mu &= \frac{1}{2} \omega_\mu{}^{ab} M_{ab} \\
&\text{Poincaré} & A_\mu &
\begin{aligned}[t] 
         &= e_\mu + \omega_\mu  \\
        &=  e_\mu{}^a P_a + \frac{1}{2} \omega_\mu{}^{ab} M_{ab} 
    \end{aligned} \label{poin1form}
\end{align}

where we have assigned a gauge field $e_\mu{}^a$ to the P-translations $P_a$ with parameters $\xi^a(x)$, and assigned a gauge field $\omega_\mu{}^{ab}$ to the LLTs $M_{ab}$ with parameters $\lambda^{ab}(x)$.

Moreover, the curvature two-form prescription we following in Sections \ref{introdifforms} and \ref{sec:lcurv} can in principle apply here as well. We can define the curvature 2-form, $R=dA+\frac{1}{2}[A,A]$, just with a new connection given by equation \discolorlinks{\ref{poin1form}}

\begin{align} \notag
R&=d A +\frac{1}{2}[A,A] \\\notag
&= \frac{1}{2}(\partial_\mu A_\nu -\partial_\nu A_\mu + [A_\mu,A_\nu])dx^\mu dx^\nu \\
&= \frac{1}{2}(\partial_\mu (e_\nu+\omega_\nu) -\partial_\nu (e_\nu+\omega_\nu) + [(e_\nu+\omega_\nu),(e_\nu+\omega_\nu)])dx^\mu dx^\nu
\end{align}

Recall that the curvature 2-form is still Lie-algebra valued. So instead of the $F_{\mu\nu}=F_{\mu\nu}{}^a T^a$ or $R_{\mu\nu}=\frac{1}{2} R_{\mu\nu}{}^{ab}M_{ab}$ that we had for SU(2) and Lorentz pure gauge theory respectively, now we have 

\begin{align} \notag
    R_{\mu\nu} &= R_{\mu\nu}{}^a P_a + \frac{1}{2} R_{\mu\nu}{}^{ab}M_{ab} \\
    R_{\mu\nu}(A) &= R_{\mu\nu}{}^a (e) P_a + \frac{1}{2} R_{\mu\nu}{}^{ab} (\omega) M_{ab}
\end{align}

where $R_{\mu\nu}{}^a (e)$ and $R_{\mu\nu}{}^{ab} (\omega)$ are the curvatures associated to each gauge field of the algebra. 

Notice that we said above that this curvature two-form prescription can \textit{in principle} apply here. By this we mean that, while everything above is indeed correct, it is no longer the path of least resistance for calculating curvatures as it was in the SU(2)/Lorentz case. When the number of gauge fields exceeds one, it is more convenient (and indeed totally sufficient) to compute the curvatures \textit{for each} gauge field, like $R_{\mu\nu}{}^a (e)$ and $R_{\mu\nu}{}^{ab} (\omega)$. In this case, the generic formula for gauge theory curvatures, equation \discolorlinks{\ref{gaugecurvature}} is a much smoother process.

\iffalse
Thomas's thesis gives a neat looking alternative to equation \discolorlinks{\ref{transgauge}} \cite{Thomasthesis}

\begin{equation}\delta A_\mu = \partial\xi + [A_\mu,\xi]\end{equation}

where 

\begin{equation}\xi = \xi^a P_a + \frac{1}{2}\lambda^{ab}M_{ab}\end{equation}

Try this out at some point in the future.
\fi

The important results for this section are the transformations of the gauge fields and the curvatures 

\begin{equation}\delta (\xi) e_\mu{}^a = \partial_\mu \xi^a -  \xi^b \omega_\mu{}^{ab} + \lambda^{ab} e_\mu{}^b \label{gtvielbein}
\end{equation}

\begin{equation}\delta (\lambda) \omega_\mu{}^{ab} = \partial_\mu \lambda^{ab} + 2\lambda^{c[a}\omega_\mu{}^{b]c} \label{transomega} \end{equation}

\begin{equation}R_{\mu\nu}{}^{a} (e)=2\partial_{[\mu} e_{\nu]}{}^{a} - 2 \omega_{[\mu}{}^a{}_b e_{\nu]}{}^b \label{Re} \end{equation}

\begin{equation}R_{\mu\nu}{}^{ab} (\omega)=2\partial_{[\mu} \omega_{\nu]}{}^{ab} - 2 \omega_{[\mu}{}^a{}_c \omega_{\nu]}{}^{cb} \label{Romega} \end{equation}

Note that equation \discolorlinks{\ref{Romega}} is the same as equation \discolorlinks{\ref{LLTcurve}}.

Equations \discolorlinks{\ref{gtvielbein}} and \discolorlinks{\ref{transomega}} follows from equation \discolorlinks{\ref{transgauge}}, and equations \discolorlinks{\ref{Re}} and \discolorlinks{\ref{Romega}} follows from equation \discolorlinks{\ref{gaugecurvature}}. We'll show this for the P-translations since they will be used later on in Section \ref{sec:soldeq}.

Firstly, the structure constant expressions from Chapter 5, equations \discolorlinks{\ref{struc1}} and \discolorlinks{\ref{struc2}} will actually be useful now that we are using equations \discolorlinks{\ref{transgauge}} and \discolorlinks{\ref{gaugecurvature}}. They are written in table 11.2 on page 218 of \cite{SUGRA}.

From \discolorlinks{\ref{poin2}} we have
\begin{equation}
    f_{[ab],[cd]}{}^{[ef]} = 8 \eta_{[c[b}\delta_{a]}^{[e}\delta_{d]}^{f]}
\end{equation}

and from \discolorlinks{\ref{poin3}} we have

\begin{equation}
    f_{a,[bc]}{}^d = 2 \eta_{a[b}\delta_{c]}^d \label{PMstruc}
\end{equation}

Note that, is a similar matter to as it was mentioned in Section \ref{sec:lcurv} regarding simplifying the commutation relation of the Lorentz algebra, if an antisymmetric object $A^{bc}$ appears alongside  \discolorlinks{\ref{PMstruc}}, then we would have 

\begin{equation}
    A^{bc}f_{a,[bc]}{}^d = \eta_{ab}\delta_{c}^d \label{antisymedit}
\end{equation}

Equation \discolorlinks{\ref{gtvielbein}} follows from \discolorlinks{\ref{transgauge}} as follows, where we utilize \discolorlinks{\ref{antisymedit}}

\begin{align} 
\delta(\epsilon)B_\mu{}^a &= \partial_\mu \epsilon^a + \epsilon^c B_\mu{}^b f_{bc}{}^a \tag{\discolorlinks{\ref{transgauge}}.r}
\\ \notag
\delta(\xi)e_\mu{}^a &= \partial_\mu \xi^a + \epsilon^c e_\mu{}^b f_{bc}{}^a + \epsilon^c \omega_\mu{}^{ab} f_{[ab],c}{}^a  \\ \notag
&= \partial_\mu \xi^a + \lambda^{ab} e_\mu{}^b f_{b,[ab]}{}^a + \xi^c \omega_\mu{}^{ab} f_{[ab],c}{}^a  \\  \notag
&= \partial_\mu \xi^a + \lambda^{ab} e_\mu{}^b (\eta_{ba}\delta_b^a) + \xi^c \omega_\mu{}^{ab} (-\eta_{ca}\delta_b^a) \\  \notag
&= \partial_\mu \xi^a + \lambda^{ab} e_\mu{}^b (1) + \xi^c \omega_\mu{}^{ab} (-\eta_{cb}) \\ 
&= \partial_\mu \xi^a + \lambda^{ab} e_\mu{}^b  - \xi^b \omega_\mu{}^{ab} \tag{\discolorlinks{\ref{gtvielbein}}.r}
\end{align}

As an aside that will be useful for Sections \ref{sec:sold} and \ref{sec:soldeq}, lets break \discolorlinks{\ref{gtvielbein}}.r up into pieces \cite{roelsphysicsforum}. We will call the terms with the P-translations' $\xi^a$ parameter $\delta_P$ and the terms with the Lorentz transformations' $\lambda^{ab}$ parameter $\delta_M$. This is just a labeling/bookkeeping system. Then \discolorlinks{\ref{gtvielbein}}.r becomes 

\begin{align} \notag
\delta(\xi)e_\mu{}^a 
&= (\partial_\mu \xi^a - \xi^b \omega_\mu{}^{ab}) + (\lambda^{ab} e_\mu{}^b)  \\
&= \delta_P (\xi^a) e_\mu{}^a + \delta_M (\lambda^{ab}) e_\mu{}^a \label{split}
\end{align}

Onto the curvature.
Equation \discolorlinks{\ref{Re}} follows from \discolorlinks{\ref{gaugecurvature}} as follows, where we utilize \discolorlinks{\ref{antisymedit}} again

\begin{align} 
R_{\mu\nu}{}^a &= 2 \partial_{[\mu} B_{\nu]}{}^a + B_\nu{}^c B_\mu{}^b f_{bc}{}^a \tag{\discolorlinks{\ref{gaugecurvature}}.r}
\\ 
R_{\mu\nu}{}^a (e) &= 2 \partial_{[\mu} e_{\nu]}{}^a + e_\nu{}^c e_\mu{}^b f_{bc}{}^a + e_\nu{}^b \omega_\mu{}^{cd} f_{[cd],b}{}^a + \omega_\nu{}^{cd} e_\mu{}^b f_{b,[cd]}{}^a + \omega_\nu{}^{cd}\omega_\mu{}^{ba} f_{[ba],[cd]}{}^{[ef]} \label{gtcurv}
\end{align}

Note that the $e$ $e$ term's structure constant is zero so that term vanishes. Note also that the last term is not including in this calculation since we are calculating $R_{\mu\nu}{}^a(e)$ which, as the gauge field $e_\mu{}^a$ does, has only the one upper $a$ index. This structure constant with its double upper indices is not included in the curvature of this gauge field. So we continue with

\begin{align} \notag
R_{\mu\nu}{}^a (e) &= 2 \partial_{[\mu} e_{\nu]}{}^a + e_\nu{}^c e_\mu{}^b f_{bc}{}^a + e_\nu{}^b \omega_\mu{}^{cd} f_{[cd],b}{}^a + \omega_\nu{}^{cd} e_\mu{}^b f_{b,[cd]}{}^a + \omega_\nu{}^{cd}\omega_\mu{}^{ba} f_{[ba],[cd]}{}^{[ef]} \\ \notag
&= 2 \partial_{[\mu} e_{\nu]}{}^a + 0 + e_\nu{}^b \omega_\mu{}^{cd} (f_{[cd],b}{}^a) + \omega_\nu{}^{cd} e_\mu{}^b (f_{b,[cd]}{}^a) \\ \notag
&= 2 \partial_{[\mu} e_{\nu]}{}^a + e_\nu{}^b \omega_\mu{}^{cd} (-\eta_{bc}\delta_d^a) + \omega_\nu{}^{cd} e_\mu{}^b (\eta_{bc}\delta_d^a) \\ \notag
&= 2 \partial_{[\mu} e_{\nu]}{}^a - e_{\nu c} \omega_\mu{}^{ca} + \omega_\nu{}^{ca} e_{\mu c} \\ 
&= 2 (\partial_{[\mu} e_{\nu]}{}^a - \omega_{[\mu}{}^{ca}e_{\nu] c}) \tag{\discolorlinks{\ref{Re}}.r}
\end{align}

\pagebreak

\subsection{Soldering}
\label{sec:sold}
Take a moment to return to Carroll's quote on the situation we find ourselves in:

\begin{displayquote}
``In the language of noncoordinate bases, it is possible to compare the formalism of connections and curvature in Riemannian geometry to that of gauge theories in particle physics. In both situations, the fields of interest live in vector spaces that are assigned to each point in spacetime. In Riemannian geometry the vector spaces include the tangent space, the cotangent space, and the higher tensor spaces constructed from these. In gauge theories, on the other hand, we are concerned with `internal' vector spaces. The distinction is that the tangent space and its relatives are intimately associated with the manifold itself, and are naturally defined once the manifold is set up; the tangent space, for example, can be thought of as the space of directional derivatives at a point. In contrast, an internal vector space can be of any dimension we like, and has to be defined as an independent addition to the manifold. In math jargon, the union of the base manifold with the internal vector spaces (defined at each point) is a \textbf{fiber bundle}, and each copy of the vector space is called the `fiber' (in accord with our definition of the tangent bundle)." \cite{carrollbook}
\end{displayquote}

\begin{figure}[H]
\centering
\includegraphics[scale=0.7]{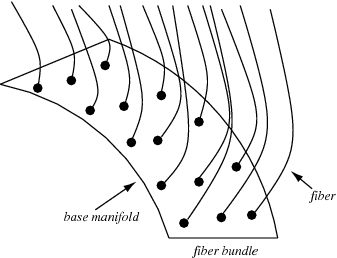}
\captionsetup{format=hang}
\caption{An illustration of a fiber bundle \cite{fiber}. Our base manifold is spacetime.}
\label{hairs}
\end{figure}

Up to this point, our purely gauge theoretic point of view has not explicitly given any indication that we are working with a spacetime symmetry. We considered symmetry transformations acting on fields only in the internal vector spaces/fibers.

This seemed feasible, until we extended to the Poincaré algebra. The issue that arises is with regard to the local spacetime translations, $P_a$, of the Poincaré algebra. Viewed in the purely gauge theoretic point of view, these $P_a$ generators are abstract and act solely in the internal vector spaces/fibers. 

However, in addition to the these internal translations, there exist ``physical" translations, usually called general coordinates transformations (GCTs), which act on the manifold of spacetime.

The fact that these GCTs act on the manifold and these local Poincaré P-translations act in the internal space becomes an issue for object that live in \textit{both} spaces. Notably, consider the indices of the gauge field $e_\mu{}^a$. 

The $\mu$ index of the object tells us that the object lives (at least partially) in the spacetime manifold. It follows that the object transforms under ``translation" according to the GCTs. But the object also has an $a$ index which tells us that it transforms under ``translation" in the internal space according to the local Poincaré P-translations.

In order for this gauging procedure to yield something that leads to the physical reality of general relativity, these two transformation ought to match up in some way. A priori, they do not.

Equation \discolorlinks{\ref{gctvector}} gives the transformation of the vielbein under GCT 

\begin{equation}
\delta_{gct} e_\mu{}^a = \xi^\nu \partial_\nu e_\mu{}^a +e_\nu{}^a \partial_\mu \xi^\nu 
\end{equation}

and equation \discolorlinks{\ref{split}} gives the purely P-translation part of the gauge theory transformation for the vielbein

\begin{equation}\delta_{P} e_\mu{}^a = \partial_\mu \xi^a - \xi^b \omega_\mu{}^{ab}  \tag{\discolorlinks{\ref{split}}.r}
\end{equation}

\begin{figure}[H]
\centering
\includegraphics[scale=0.5]{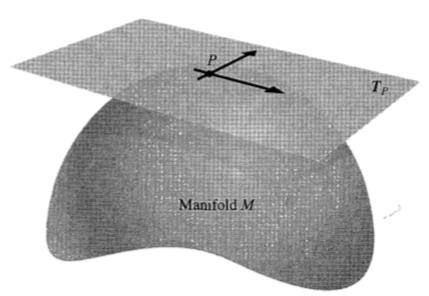}
\captionsetup{format=hang}
\caption{Soldering: attaching fibers to a base manifold so that all fibers are tangent and thus the internal vector spaces become tangent spaces, and the fiber bundle becomes a tangent bundle. \cite{carrollbook}}
\label{solder}
\end{figure}

How can we resolve this? What is we impose some geometric constrain what forces the local P-translations to line up with GCTs? Notabley, what if we attach the internal vector spaces/fibers to the spacetime manifold in such a way that the arbitrary fibers become tangent fibers? In this way, the transformation in the internal vector space would become what they are when they act specifically in the tangent space. This process of massaging the fibers to be attached to the base manifold in such a way that they produce a tangent bundle (see figure \ref{solder}) is known as ``soldering" \cite{solder}.

In some sense, this is not too wild an idea if one considers the operator representation of the P-translations, and the basis for tangent bundles. On one hand, page 64 and 65 of Carroll tell use that the basis elements for a tangent bundle are partial derivatives $\{\partial_\mu\}$ \cite{carrollbook}. On the other hand, our equation \discolorlinks{\ref{oppoin}}, page 81 of \cite{QFTGA}, and equation 1.54 of F-VP show us that P-translations act as operators like $\partial_\mu$ \cite{SUGRA}.

How can we do this? How can we impose some geometric constrain that forces the local P-translations to become GCTs, and the fibers to become tangent? The later requirement is suggestive. What is a tangent space? A \textit{flat} space where vectors live attached to points on the manifold. Who says the internal vectors spaces/fibers need to be ``flat?" No one, just look to the curvatures we calculated in a pure gauge theoretic context in Section \ref{naive}. The curvature of the P-translation is not zero, $R(e) = de+\omega e$, the space is not flat. What if we \textit{made} it flat? In other words, what if we set the curvature of the P-translations, R(e), equal to zero — flattening the a priori curvy fibers of Figure \ref{hairs} into tangent spaces like Figure \ref{solder}?

There exist at least four explanations for why the curvature of the P-translations ought to be set to zero.

\phantomsection
\label{marker1}
1) Replacing P-translations with GCTs (+ field-dependent LLTs, more on this below) is possible if the curvature of the P-translations R(e) is zero by what we will refer to as the ``soldering equation". This will be expounded on in Section \ref{sec:soldeq}.

\iffalse
, but the main result is 

\begin{equation}
    \delta_{gct} e - \delta_{gt} e = R(e)
\end{equation}

This implies $\delta_{gct} e = \delta_{gt} e$ if $R(e)=0$.
\fi
2) In gravity, the spin connections are not independent fields, they depend on the vielbein. If we insist that $R(e)=0$ (as well as use the vielbein in its full-fledged gravitational interpretation, i.e. let it be invertible) then $0=R(e)=de +\omega e$ can be solved for $\omega$. This encapsulates the difference between the local Poincaré gauge theory with GCTs and a typical Yang-Mills gauge theory — there is no notion of an invertible gauge field or a dependent gauge field in Yang-Mills.

Note: There exists another way to determine the spin connection in terms of the vielbein seemingly without using the curvature constraint. One can use ``anholonomy coefficients," as done in equation 2.11 of \cite{anholo} or exercise 7.11 on page 145 of \cite{SUGRA}. We will stick to the curvature constrain method here.

3) The curvature two-form R(e) is precisely that of the torsion two-form $T^a$ from equation \discolorlinks{\ref{torsion1}}. If we desire to work in a geometry without torsion, then $T^a=R(e)$ must be zero.

4) Somewhat a mix of \#3 and supergravity/symmetry, exercise 11.10 on page 224 of \cite{SUGRA} results in requiring $R(e)=0$.

\phantomsection
\label{marker2}

\pagebreak
\subsubsection{The soldering equation}
\label{sec:soldeq}

We follow Freedman-Van Proeyen in introducing ``covariant general coordinate translations" (CGCTs) \cite{SUGRA}.

CGCTs are defined by equation 11.61 in F-VP as 

\begin{equation}
\delta_{cgct} (\xi) = \delta_{gct}(\xi) - \delta(\xi^\mu B_\mu)
\end{equation}

where the second term can be thought of as field-dependent gauge theory transformations.

The necessity of these field-dependent gauge theory transformations will be made clear through the identity we denote the soldering equation below, but F-VP also motivate the definition by arguing that GCTs of scalar fields do not transform covariantly under internal symmetry. This is not entirely clear but we make one remark on it.

Consider the standard transformation of scalar fields given by equation 

\begin{equation}
\delta(\epsilon) \phi^i(x) = \epsilon^A(x) T_A \phi^i(x)
\end{equation}

The transformation of a scalar field under GCTs is given by letting the symmetry ($T_A$) be GCT, i.e. $\partial_\mu$, and letting the parameter ($\epsilon$) be $\xi$.

Then we have, as before in equation \discolorlinks{\ref{gctphi}}

\begin{equation}
\delta(\xi) \phi^i(x) = \xi^\mu(x) \partial_\mu \phi^i(x)
\end{equation}

If we combine GCTs with gauge transformations then we have 

\begin{align} \notag
\delta_{gct} \delta(\epsilon) \phi^i(x) &= \xi^\mu(x) \partial_\mu (\epsilon^a(x) T_a \phi^i (x)) \\ 
&= \xi^\mu(x) \partial_\mu (\epsilon^a(x)) T_a \phi^i (x) + \xi^\mu(x) \epsilon^a(x) \partial_\mu ( T_a \phi^i (x)) 
\end{align}

and it is clear that the first term is not covariant — it involves a derivative of the gauge parameter. It is not clear, after some not-included-here computations, how replacing the GCT in the above expression with a CGCT restores covariance. 

\pagebreak

\iffalse
F-VP then state on page 228,

We fix this by adding a field dependent gauge transformation and thus define

\begin{equation}
\delta_{cgct}(\xi) \phi^i = \xi^\mu \partial_\mu \phi^i(x)+(\xi^\mu A_\mu{}^A)t_{A}{}^i{}_j \phi^j"
\end{equation}

where $\phi^i$ and $\xi$ are still functions of spacetime, the $(x)$ has just been neglected for brevity.

\textbf{My confusion lies in ``but it has the undesirable property that it does not transform covariantly under internal symmetry."
What does this means?}

\vspace*{4cm}
\fi

Modeled after equation 11.65 on page 229 of \cite{SUGRA}, we call the following equation the soldering equation

\begin{align} \notag
\delta_{cgct}(\xi) B_\mu{}^a &= \delta_{gct}(\xi) B_\mu{}^a - \delta(\xi^\lambda B_\lambda) B_\mu{}^a \\
&= - \xi^\lambda R_{\mu\lambda}{}^a \label{soldereq}
\end{align}

\iffalse
\begin{equation}0= \delta_{gct}(\xi^\lambda) B_\mu{}^A + \xi^\lambda R_{\mu\lambda}{}^A - \sum_C \delta(\xi^\lambda B_\lambda{}^C)B_\mu{}^A \label{soldereq}
\end{equation}

where $A,C$ label the gauge fields so that $A,C=\{a, ab\}$ for the  P-transformation's gauge field ($e_\mu{}^a$) and LLT's gauge field ($\omega_\mu{}^{ab}$) respectively.
\fi

Another form of the soldering equation is equation 2.4 of \cite{bergshoeff} or equivalently equation B.10 of \cite{Roelthesis}.

%\begin{equation}\delta e_\mu{}^a = \partial_\mu \xi^a + \xi^c B_\mu{}^B f_{Bc}{}^a + \epsilon^C (B_\mu{}^B f_{BC}{}^a + \epsilon_\mu{}^b f_{bC}^a)\end{equation}

Before working on the soldering equation of the P-translations/GCTs, we will follow appendix B of \cite{Roelthesis} and verify the equation for U(1) gauge theory.

\iffalse
Let's rearrange equation \discolorlinks{\ref{soldereq}} first 

\begin{equation}
\delta_{gct}(\xi^\lambda) B_\mu{}^A - \sum_C \delta(\xi^\lambda B_\lambda{}^C)B_\mu{}^A = -\xi^\lambda R_{\mu\lambda}{}^A 
\tag{\discolorlinks{\ref{soldereq}}.r}
\end{equation}
\fi

Recall the structure constants of U(1) are zero. So equation \discolorlinks{\ref{transgauge}} tells us

\begin{align}
    \delta(\theta) A_\mu = \partial_\mu(\theta)
\end{align}

so then 

\begin{align} \notag
    \delta(\xi^\lambda A_\lambda) A_\mu &= \partial_\mu(\xi^\lambda A_\lambda) \\
    &= (\partial_\mu \xi^\lambda) A_\lambda + \xi^\lambda( \partial_\mu A_\lambda)
\end{align}

Equation \discolorlinks{\ref{gctvector}} tells us how a vector transforms under GCTs

\begin{align} \notag
    \delta_{gct}(\xi^\lambda ) A_\mu &= L_\xi A_\mu \\ 
    &=  \xi^\lambda( \partial_\lambda A_\mu)+ (\partial_\mu \xi^\lambda) A_\lambda 
\end{align}

Equation \discolorlinks{\ref{fmunuu1}} tells the curvature for U(1)

\begin{equation}
    R_{\mu\lambda} = \partial_\mu A_\lambda - \partial_\lambda A_\mu
\end{equation} 

Plugging all of this into \discolorlinks{\ref{soldereq}}, we get

\begin{align} \notag
\delta_{cgct}(\xi) B_\mu{}^a &= -\xi^\lambda R_{\mu\lambda}{}^a \\ \notag
\delta_{gct}(\xi) B_\mu{}^a - \delta(\xi^\lambda B_\lambda)B_\mu{}^a 
&= -\xi^\lambda R_{\mu\lambda}{}^a \\ \notag
\delta_{gct}(\xi) A_\mu - \delta(\xi^\lambda A_\lambda)A_\mu
&= -\xi^\lambda R_{\mu\lambda} \\ \notag
\xi^\lambda( \partial_\lambda A_\mu)+ (\partial_\mu \xi^\lambda) A_\lambda - \partial_\mu(\xi^\lambda A_\lambda)
&= -\xi^\lambda R_{\mu\lambda} \\ \notag
\xi^\lambda( \partial_\lambda A_\mu)+ (\partial_\mu \xi^\lambda) A_\lambda  - (\partial_\mu \xi^\lambda) A_\lambda - \xi^\lambda( \partial_\mu A_\lambda)
&= - \xi^\lambda (\partial_\mu A_\lambda - \partial_\lambda A_\mu) \\ \notag
\xi^\lambda( \partial_\lambda A_\mu) - \xi^\lambda( \partial_\mu A_\lambda)
&= - \xi^\lambda \partial_\mu A_\lambda + \xi^\lambda \partial_\lambda A_\mu \\ \notag
\xi^\lambda( \partial_\lambda A_\mu) - \xi^\lambda( \partial_\mu A_\lambda)
&= - \xi^\lambda \partial_\mu A_\lambda + \xi^\lambda \partial_\lambda A_\mu \\
- \xi^\lambda \partial_\mu A_\lambda +\xi^\lambda \partial_\lambda A_\mu 
&= - \xi^\lambda \partial_\mu A_\lambda + \xi^\lambda \partial_\lambda A_\mu
\end{align}

Now onto the Poincaré algebra.

\begin{align} \notag
\delta_{cgct}(\xi) e_\mu{}^a  &= \delta_{gct}(\xi) e_\mu{}^a - \delta(\xi^\lambda B_\lambda) e_\mu{}^a \\ \notag
&= L_\xi e_\mu{}^a - \partial_\mu(\xi^\lambda B_\lambda{}^a)  - (\xi^\lambda B_\lambda{}^c) B_\mu{}^b f_{bc}{}^a \\ \notag
&= L_\xi e_\mu{}^a - \partial_\mu(\xi^\lambda e_\lambda{}^a)  - (\xi^\lambda e_\lambda{}^c) e_\mu{}^b f_{bc}{}^a - (\xi^\lambda e_\lambda{}^c) \omega_\mu{}^{bd} f_{[bd],c}{}^a - (\xi^\lambda \omega_\lambda{}^{cd}) e_\mu{}^{b} f_{b,[cd]}{}^a \\ \notag
&= L_\xi e_\mu{}^a - \partial_\mu(\xi^\lambda e_\lambda{}^a)  - (\xi^\lambda e_\lambda{}^c) e_\mu{}^b (0) - (\xi^\lambda e_\lambda{}^c) \omega_\mu{}^{bd} (-\eta_{cb}\delta_d^a) - (\xi^\lambda \omega_\lambda{}^{cd}) e_\mu{}^{b} (\eta_{bc}\delta_d^a) \\ \notag
&= L_\xi e_\mu{}^a - \partial_\mu(\xi^\lambda e_\lambda{}^a) + (\xi^\lambda e_{\lambda b}) \omega_\mu{}^{ba} - (\xi^\lambda \omega_\lambda{}^{ca}) e_{\mu c} \\ \notag
&= \xi^\lambda \partial_\lambda e_\mu{}^a +e_\lambda{}^a \partial_\mu \xi^\lambda - \partial_\mu(\xi^\lambda e_\lambda{}^a) + \xi^\lambda  e_{\lambda b} \omega_\mu{}^{ba} - \xi^\lambda \omega_\lambda{}^{ca} e_{\mu c} \\ \notag
&= \xi^\lambda \partial_\lambda e_\mu{}^a +e_\lambda{}^a \partial_\mu \xi^\lambda - e_\lambda{}^a \partial_\mu \xi^\lambda - \xi^\lambda \partial_\mu e_\lambda{}^a  + \xi^\lambda \omega_\mu{}^{ba} e_{\lambda b}  - \xi^\lambda \omega_\lambda{}^{ca} e_{\mu c} \\ \notag
&= \xi^\lambda \partial_\lambda e_\mu{}^a + 0 -  \xi^\lambda\partial_\mu e_\lambda{}^a  +  \xi^\lambda\omega_\mu{}^{ca} e_{\lambda c}  - \xi^\lambda \omega_\lambda{}^{ca} e_{\mu c} \\ \notag
&= \xi^\lambda (\partial_\lambda e_\mu{}^a -  \partial_\mu e_\lambda{}^a  +  \omega_\mu{}^{ca} e_{\lambda c}  -  \omega_\lambda{}^{ca} e_{\mu c}) \\ \notag
&= -\xi^\lambda (-\partial_\lambda e_\mu{}^a +  \partial_\mu e_\lambda{}^a  -  \omega_\mu{}^{ca} e_{\lambda c}  +  \omega_\lambda{}^{ca} e_{\mu c}) \\ \notag
&= -\xi^\lambda (\partial_\mu e_\lambda{}^a  -\partial_\lambda e_\mu{}^a  -  \omega_\mu{}^{ca} e_{\lambda c}  +  \omega_\lambda{}^{ca} e_{\mu c}) \\ \notag
&= -\xi^\lambda (2 \partial_{[\mu} e_{\lambda]}{}^a - 2\omega_{[\mu}{}^{ca} e_{\lambda] c}) \\
&= - \xi^\lambda R_{\mu\lambda}{}^a (e) \label{soldcalc}
\end{align}

where going from the second equality to the third is where care need be taken —

$\partial_\mu(\xi^\lambda B_\lambda{}^a)$ goes to $\partial_\mu(\xi^\lambda e_\lambda{}^a)$ because the LHS of the equation has one upper $a$ index, i.e. we are looking at the cgct of $e_\mu{}^a$ only, not $\omega_\mu{}^{ab}$, and

the sum $(\xi^\lambda B_\lambda{}^c) B_\mu{}^b f_{bc}{}^a$ excludes the term  $(\xi^\lambda \omega_\lambda{}^{cd}) \omega_\mu{}^{ab} f_{[ab],[cd]}{}^{[ef]}$ because again, $[ef] \neq a$, and we have only one upper $a$ index on the LHS.

Going from the first to second equality uses equation \discolorlinks{\ref{transgauge}}.

In the same way we did in equation \discolorlinks{\ref{split}}, we will write out the 5th equality in \discolorlinks{\ref{soldcalc}} to show what the soldering equation does for us. Recall that the $\delta_P$/$\delta_M$ notation is just bookkeeping.

\begin{align} \notag
- \xi^\lambda R_{\mu\lambda}{}^a (e)    &= L_\xi e_\mu{}^a - (\partial_\mu(\xi^\lambda e_\lambda{}^a) + (\xi^\lambda e_{\lambda b}) \omega_\mu{}^{ba}) - (\xi^\lambda \omega_\lambda{}^{ca}) e_{\mu c} \\ \notag
&= \delta_{gct} e_\mu{}^a - [\partial_\mu(\xi^\lambda e_\lambda{}^a) + (\xi^\lambda e_{\lambda}{}^b) \omega_\mu{}^{ba}] - (\xi^\lambda \omega_\lambda{}^{ca}) e_{\mu c} \\ \notag
&= \delta_{gct} e_\mu{}^a - [\partial_\mu(\xi^\lambda e_\lambda{}^a) - (\xi^\lambda e_{\lambda}{}^b) \omega_\mu{}^{ab}] - (\xi^\lambda \omega_\lambda{}^{ca}) e_{\mu c} \\
&= \delta_{gct} e_\mu{}^a - [\delta_P(\xi^\lambda e_\lambda{}^b) e_\mu{}^a] - \delta_M(\xi^\lambda \omega_\lambda{}^{bc}) e_{\mu}{}^a \label{thisding}
\end{align}

As advertised in reason \#1 for why the curvature R(e) ought to be set to zero (on page \pageref{marker1}), if we set $R_{\mu\lambda}{}^a (e)=0$, equation \discolorlinks{\ref{thisding}} becomes 

\begin{align} \notag
0 &= \delta_{gct} e_\mu{}^a - [\delta_P(\xi^\lambda e_\lambda{}^b) e_\mu{}^a] - \delta_M(\xi^\lambda \omega_\lambda{}^{bc}) e_{\mu}{}^a \\
\delta_P(\xi^\lambda e_\lambda{}^b) e_\mu{}^a &= \delta_{gct} e_\mu{}^a - \delta_M(\xi^\lambda \omega_\lambda{}^{bc}) e_{\mu}{}^a
\end{align}

If we finally make the definitive statement that the gauge field $e_\mu{}^a$ is the vielbein, then we can use the $V^a= e_\mu{}^a V^\mu$ property of the vielbein to write 

\begin{align} \notag
\delta_P(\xi^\lambda e_\lambda{}^b) e_\mu{}^a &= \delta_{gct} e_\mu{}^a - \delta_M(\xi^\lambda \omega_\lambda{}^{bc}) e_{\mu}{}^a \\
\delta_P(\xi^b) e_\mu{}^a &= \delta_{gct} e_\mu{}^a - \delta_M(\xi^\lambda \omega_\lambda{}^{bc}) e_{\mu}{}^a
\end{align}

and we have a final expression showing that the local P-translations can be expressed in terms of GCTs and field dependent LLTs so long as the curvature of the P-translations $R(e)$ is set to zero. I.e. by imposing that curvature constraint, we have deformed the original Poincaré algebra to fit our purposes. These deformed algebras are called ``soft algebras" and are the algebraic structures used in supersymmetry. See Section 11.1.3 on page 219 of Freedman-Van Proeyen for more on this \cite{SUGRA}.

\pagebreak

\pagebreak
\subsection{Final steps to a theory of gravity}
\label{final steps}
 
\subsubsection{The metric} 
Now that we interpret our gauge field $e_\mu{}^a$ is a bona fide vielbein, we can introduce a metric through the definition of the vielbein

\begin{align} 
    g_{\mu\nu} &= e_{\mu}{}^a e_\nu{}^b \eta_{ab} \label{metricviel} \\
    g^{\mu\nu} &= e_{a}{}^\mu e_b{}^\nu \eta^{ab} 
\end{align}

\subsubsection{The Christoffel connection} 
\label{christcon}

Additionally, we can now introduce the familiar Christoffel connection in terms of our vielbein and spin connection. This connection is necessary if we want a covariant derivative in a curved background, i.e. if the the base manifold of our fiber bundle (spacetime) is curved. \cite{Thomasthesis}.

Using \discolorlinks{\ref{gaugcovder}}, with two different connections — the Christoffel connection and the spin connection — we can introduce two covariant (with respect to GCTs and LLTs respectively) derivatives, $\nabla_\mu$ and $D_\mu$ respectively.

\begin{align}
    \nabla_\mu V^\nu &= \partial_\mu V^\nu + \Gamma_{\mu\lambda}^\nu V^\lambda \label{covchrist}\\
    \nabla_\mu V_\nu &= \partial_\mu V_\nu - \Gamma_{\mu\nu}^\lambda V_\lambda\\
    D_\mu V^a &= \partial_\mu V^a - \omega_{\mu}{}^a{}_b V^b  \label{idkman}\\ \notag
    D_\mu V_a &= \partial_\mu V_a + \omega_\mu{}^b{}_a V_b
\end{align}

Note that the covariant (w.r.t LLTs) derivative $D_\mu$ is defined in a different way than is standard in \cite{SUGRA}, \cite{zee}, \cite{carrollbook}, where the upper index gets a plus spin connection just like in \discolorlinks{\ref{covchrist}}. We do this so that the LLT covariant derivative lines up with the gauge theory definition of covariant derivatives from \discolorlinks{\ref{gaugcovder}}. See equations 2.22 and 2.50 in \cite{Roelthesis}. 

The Christoffel symbol in terms of the vielbein and spin connection is derived by comparing \discolorlinks{\ref{covchrist}} with its equivalent formulation as a modification of \discolorlinks{\ref{idkman}}

\begin{equation}
    \nabla_\mu V^\nu = e^\nu{}_a D_\mu V^a
\end{equation}

as follows \cite{SUGRA}

\begin{align} \notag
    \nabla_\mu V^\nu &= e^\nu{}_a D_\mu V^a \\ \notag
    &= e^\nu{}_a D_\mu (e_\lambda{}^a V^\lambda) \\ \notag
    &= e^\nu{}_a \partial_\mu(e_\lambda{}^a V^\lambda) -e^\nu{}_a [\omega_\mu{}^a{}_b(e_\lambda{}^b V^\lambda)] \\ \notag
    &= e^\nu{}_a e_\lambda{}^a \partial_\mu(V^\lambda) + e^\nu{}_a \partial_\mu(e_\lambda{}^a) V^\lambda -e^\nu{}_a (\omega_\mu{}^a{}_b e_\lambda{}^b) V^\lambda \\ \notag
    &= \delta^\nu{}_\lambda{} \partial_\mu V^\lambda + e^\nu{}_a ( \partial_\mu e_\lambda{}^a  - \omega_\mu{}^a{}_b e_\lambda{}^b) V^\lambda \\
    &=  \partial_\mu V^\nu + e^\nu{}_a ( \partial_\mu e_\lambda{}^a  - \omega_\mu{}^a{}_b e_\lambda{}^b) V^\lambda \label{ref}
    \end{align}

thus, according to \discolorlinks{\ref{covchrist}},

\begin{align} \notag
    \Gamma^\nu_{\mu\lambda} &= e^\nu{}_a ( \partial_\mu e_\lambda{}^a  - \omega_\mu{}^a{}_b e_\lambda{}^b) \label{torsionthing} \\
&= e^\nu{}_a D_\mu e_\lambda{}^a
\end{align}

Recall however that the Christoffel connection is symmetric in its lower indices, so if an explicit expression for the Christoffel connection is written out, like \discolorlinks{\ref{torsionthing}} the indices on the other side need to be symmetrized,

\begin{equation}
    \Gamma^\nu_{(\mu\lambda)} = e^\nu{}_a ( \partial_{(\mu} e_{\lambda)}{}^a  - \omega_{(\mu}{}^a{}_b e_{\lambda)}{}^b)
\end{equation}

The Christoffel symbol in terms of the vielbein and spin connection can also be derived from the so-called ``vielbein postulate." The vielbein postulate can be derived from a similar relation to \discolorlinks{\ref{ref}}  (see page 603 of \cite{zee})

\begin{align} \notag
D_\mu V^a &= e_\nu{}^a \nabla_\mu V^\nu \\ \notag
D_\mu (e_\nu{}^a V^\nu)    
&= e_\nu{}^a (\partial_\mu V^\nu + \Gamma_{\mu\lambda}^\nu V^\lambda) \\ \notag
\partial_\mu (e_\nu{}^a V^\nu) - \omega_\mu{}^a{}_b e_\nu{}^b V^\nu    
&= e_\nu{}^a \partial_\mu V^\nu + e_\nu{}^a \Gamma_{\mu\lambda}^\nu V^\lambda \\ \notag
V^\nu \partial_\mu e_\nu{}^a  + e_\nu{}^a \partial_\mu V^\nu -\omega_\mu{}^a{}_b e_\nu{}^b V^\nu 
&= e_\nu{}^a \partial_\mu V^\nu + e_\lambda{}^a \Gamma_{\mu\nu}^\lambda V^\nu \\ \notag
V^\nu \partial_\mu e_\nu{}^a  + e_\nu{}^a \partial_\mu V^\nu -e_\nu{}^a \partial_\mu V^\nu-\omega_\mu{}^a{}_b e_\nu{}^b V^\nu -e_\lambda{}^a \Gamma_{\mu\nu}^\lambda V^\nu
&= 0 \\ \notag
V^\nu (\partial_\mu e_\nu{}^a  -\omega_\mu{}^a{}_b e_\nu{}^b  -e_\lambda{}^a \Gamma_{\mu\nu}^\lambda)
&= 0 \\ 
\partial_\mu e_\nu{}^a  -\omega_\mu{}^a{}_b e_\nu{}^b  -e_\lambda{}^a \Gamma_{\mu\nu}^\lambda
&= 0 \label{vielpostulate}
\end{align}

where the final line is often summarized as 

\begin{align}
\notag
\nabla_\mu e_\nu{}^a &= \partial_\mu e_\nu{}^a  -\omega_\mu{}^a{}_b e_\nu{}^b  -\Gamma_{\mu\nu}^\lambda e_\lambda{}^a  \\ \notag
&= D_\mu e_\nu{}^a - \Gamma_{\mu\nu}^\lambda e_\lambda{}^a\\
&=0  \label{vielpost}
\end{align}

The Christoffel connection can be obtained from the last line of \discolorlinks{\ref{vielpostulate}} by bringing the $-e_\lambda{}^a \Gamma_{\mu\nu}^\lambda$ term to the RHS and then hitting both sides with an inverse vielbein.

For completeness, the standard (full curved index) Riemann curvature of the Christoffel connection — defined here

\begin{equation}
    R^\rho{}_{\sigma\mu\nu}(\Gamma) = \partial_\mu \Gamma_{\nu\sigma}^\rho - \partial_\nu \Gamma_{\mu\sigma}^\rho + \Gamma_{\mu\lambda}^\rho \Gamma_{\nu\sigma}^\lambda - \Gamma_{\nu\lambda}^\rho \Gamma_{\mu\sigma}^\lambda \label{reimann}
\end{equation}

can be related to the curvature of the spin connection, equation \discolorlinks{\ref{Romega}}, with some vielbein manipulations 
as dictated by the vielbein postulate (our version with the minus sign — \discolorlinks{\ref{vielpost}})
as follows

\begin{equation}
    R^\rho{}_{\sigma\mu\nu}(\Gamma) = e^\rho{}_a e_{\sigma b}R^{ab}{}_{\mu\nu} (\omega)  = -R_{\mu\nu}{}^{ab} (\omega) e^\rho{}_a e_{\sigma b} \label{relate}
\end{equation}

\subsubsection{Torsion-free geometry} 
As promised in reason \# 3  (on page \pageref{marker2}) of why we ought to set the curvature R(e) to zero, it is clear now why $R(e)=0 \leftrightarrow$ no torsion. 
The torsion tensor is defined as \cite{carrollbook}

\begin{equation}
    T_{\mu\nu}{}^\lambda = \Gamma_{\mu\nu}^\lambda -\Gamma_{\nu\mu}^\lambda = 2 \Gamma_{[\mu\nu]}^\lambda \label{torsion}
\end{equation}

where it is clear that if $\Gamma_{[\mu\nu]}^\lambda=0$ then $T_{\mu\nu}{}^\lambda = 2\cdot 0 =0$ and there is no torsion. Antisymmetrizing \discolorlinks{\ref{torsionthing}}, in addition to $R(e)=0$, gives us precisely this

\begin{align} \notag
\Gamma^\nu_{[\mu\lambda]} &= e^\nu{}_a ( \partial_{[\mu} e_{\lambda]}{}^a  + \omega_{[\mu}{}^a{}_b e_{\lambda]}{}^a) \\ \notag
&= e^\nu{}_a (R_{\mu\lambda}{}^a(e)=0) \\
&= 0
\end{align}

\subsubsection{Spin connection as a dependent field} 
Additionally, as promised in reason \# 2 (page \pageref{marker1}), we can write the spin connection in terms of the vielbein (and its derivatives). This is in-line with the conventions of GR where the vielbein is the independent field, and the spin connection depends on it. 

Mimicking the process for finding the Christoffel connection in terms of the metric/its derivatives, see box 17.4 on page 205 of \cite{moore}, we can use the $R(e)=0$ condition and some non-trivial index gymnastics to solve for the spin connection in terms of the vielbein.
\begin{align} \notag
0&= 0+0+0 \\ \notag
0 &=   R_{\mu\nu}{}^a e_{\rho a} + R_{\rho\mu}{}^a e_{\nu a} -R_{\nu\rho}{}^a e_{\mu a} \\ \notag
R_{\nu\rho}{}^a e_{\mu a} &=   R_{\mu\nu}{}^a e_{\rho a} + R_{\rho\mu}{}^a e_{\nu a}  \\ \notag
(\partial_{[\nu} e_{\rho]}{}^a - \omega_{[\nu}{}^{ab} e_{\rho]}{}_b) e_{\mu a} &=   \begin{aligned}[t]
        &{} (\partial_{[\mu} e_{\nu]}{}^a - \omega_{[\mu}{}^{ab} e_{\nu]}{}_b) e_{\rho a}\\ &{} \quad+ (\partial_{[\rho} e_{\mu]}{}^a - \omega_{[\rho}{}^{ab} e_{\mu]}{}_b) e_{\nu a}
\end{aligned}  \\ \notag
\partial_{[\nu} e_{\rho]}{}^a e_{\mu a} - \omega_{[\nu}{}^{ab} e_{\rho]}{}_b e_{\mu a}&=  
\begin{aligned}[t]
         &{} \partial_{[\mu} e_{\nu]}{}^a e_{\rho a} - \omega_{[\mu}{}^{ab} e_{\nu]}{}_b e_{\rho a}\\ & {} \quad + \partial_{[\rho} e_{\mu]}{}^a e_{\nu a} - \omega_{[\rho}{}^{ab} e_{\mu]}{}_b e_{\nu a}
\end{aligned}   \\ \notag
\omega_{[\mu}{}^{ab} e_{\nu]}{}{}_b e_{\rho a} + \omega_{[\rho}{}^{ab} e_{\mu]}{}_b e_{\nu a} - \omega_{[\nu}{}^{ab} e_{\rho]}{}_b e_{\mu a} &=   \partial_{[\mu} e_{\nu]}{}^a e_{\rho a}  + \partial_{[\rho} e_{\mu]}{}^a e_{\nu a} - \partial_{[\nu} e_{\rho]}{}^a e_{\mu a} 
\\ \notag
\begin{aligned}[t]
    &{} \quad \frac{1}{2}
\omega_{\mu}{}^{ab} e_{\nu}{}{}_b e_{\rho a} - \frac{1}{2}\omega_{\nu}{}^{ab} e_{\mu}{}{}_b e_{\rho a}  \\
    +&{\frac{1}{2}} \omega_{\rho}{}^{ab} e_{\mu}{}_b e_{\nu a}  -\frac{1}{2}\omega_{\mu}{}^{ab} e_{\rho}{}_b e_{\nu a} \\
    - &{\frac{1}{2}}\omega_{\nu}{}^{ab} e_{\rho}{}_b e_{\mu a} +\frac{1}{2}\omega_{\rho}{}^{ab} e_{\nu}{}_b e_{\mu a}    
\end{aligned}&=   \partial_{[\mu} e_{\nu]}{}^a e_{\rho a}  + \partial_{[\rho} e_{\mu]}{}^a e_{\nu a} - \partial_{[\nu} e_{\rho]}{}^a e_{\mu a} 
\\[1ex] \notag
\begin{aligned}[t]
    &{} \quad \frac{1}{2}\omega_{\mu}{}^{ab} e_{\nu}{}{}_b e_{\rho a} -\frac{1}{2}\omega_{\mu}{}^{ba} e_{\rho}{}_a e_{\nu b} \\
    + &{\frac{1}{2}}\omega_{\rho}{}^{ab} e_{\nu}{}_b e_{\mu a} + \frac{1}{2} \omega_{\rho}{}^{ba} e_{\mu}{}_a e_{\nu b} \\
    - &{\frac{1}{2}}\omega_{\nu}{}^{ab} e_{\mu}{}{}_b e_{\rho a}  - \frac{1}{2}\omega_{\nu}{}^{ba} e_{\rho}{}_a e_{\mu b}
\end{aligned} &=   \partial_{[\mu} e_{\nu]}{}^a e_{\rho a}  + \partial_{[\rho} e_{\mu]}{}^a e_{\nu a} - \partial_{[\nu} e_{\rho]}{}^a e_{\mu a} 
\\[1ex] \notag
\begin{aligned}[t]
    &{} \quad \frac{1}{2}\omega_{\mu}{}^{ab} e_{\nu}{}{}_b e_{\rho a} +\frac{1}{2}\omega_{\mu}{}^{ab} e_{\rho}{}_a e_{\nu b} \\
+  &{\frac{1}{2}}\omega_{\rho}{}^{ab} e_{\nu}{}_b e_{\mu a}  - \frac{1}{2} \omega_{\rho}{}^{ab} e_{\mu}{}_a e_{\nu b} \\ -&{\frac{1}{2}}\omega_{\nu}{}^{ab} e_{\mu}{}{}_b e_{\rho a}  + \frac{1}{2}\omega_{\nu}{}^{ab} e_{\rho}{}_a e_{\mu b}
\end{aligned}
 &=   \partial_{[\mu} e_{\nu]}{}^a e_{\rho a}  + \partial_{[\rho} e_{\mu]}{}^a e_{\nu a} - \partial_{[\nu} e_{\rho]}{}^a e_{\mu a} 
\\\notag
\omega_{\mu}{}^{ab} e_{\nu}{}{}_b e_{\rho a}  + 0 + 0 &=   \partial_{[\mu} e_{\nu]}{}^a e_{\rho a}  + \partial_{[\rho} e_{\mu]}{}^c e_{\nu c} - \partial_{[\nu} e_{\rho]}{}^c e_{\mu c} 
\\ \notag 
\omega_{\mu}{}^{ab} e_{\nu}{}{}_b e_{\rho a} e^{\rho a} e^{\nu b}&=
\begin{aligned}[t]
    &{} \partial_{[\mu} e_{\nu]}{}^a e_{\rho a}e^{\rho a} e^{\nu b}  + \partial_{[\rho} e_{\mu]}{}^c e_{\nu c}e^{\rho a} e^{\nu b}\\ &{} \quad - \partial_{[\nu} e_{\rho]}{}^c e_{\mu c} e^{\rho a} e^{\nu b}
\end{aligned}
\\ \notag 
\omega_{\mu}{}^{ab} &=   \partial_{[\mu} e_{\nu]}{}^a  e^{\nu b}  + \partial_{[\rho} e_{\mu]}{}^c \delta_c^b e^{\rho a} - \partial_{[\nu} e_{\rho]}{}^c e_{\mu c} e^{\rho a} e^{\nu b}
\\ \notag 
\omega_{\mu}{}^{ab} &=   \partial_{[\mu} e_{\nu]}{}^a  e^{\nu b}  + \partial_{[\rho} e_{\mu]}{}^b e^{\rho a} - \partial_{[\nu} e_{\rho]}{}^c e_{\mu c} e^{\rho a} e^{\nu b}
\\ \notag 
\omega_{\mu}{}^{ab} &=   e^{\nu b}\partial_{[\mu} e_{\nu]}{}^a    + e^{\nu a}\partial_{[\nu} e_{\mu]}{}^b  -e_{\mu c}e^{\rho a} e^{\lambda b}  \partial_{[\lambda} e_{\rho]}{}^c
\\ \notag 
\omega_{\mu}{}^{ab} &=   2 e^{\nu [b}\partial_{[\mu} e_{\nu]}{}^{a]}    +e_{\mu c}e^{\rho a} e^{\lambda b}  \partial_{[\rho} e_{\lambda]}{}^c
\\ 
\omega_{\mu}{}^{ab} &=   -2 e^{\nu [a}\partial_{[\mu} e_{\nu]}{}^{b]}    +  e_{\mu c}e^{\rho a} e^{\lambda b} \partial_{[\rho} e_{\lambda]}{}^c \label{depomega}
\end{align}

\subsubsection{On-shell}

Our theory can be ``put on-shell", i.e. the equations of motions (the Einstein equation) can be imposed. In the vacuum, the Einstein equation reduces to the Ricci curvature vanishing, $R_{\mu\nu}=0$, and in our framework that (contracted Riemann curvature over the first and third indices) reads

\begin{align}
        e^\mu{}_a R_{\mu\nu}{}^{ab}(\omega) &= 0 \\
        g^{\rho \sigma} R_{\rho\mu\sigma\nu}(\Gamma) &=0 
\end{align}

\subsubsection{An action}

We can structure an action for the theory with the Ricci scalar \cite{Thomasthesis}. In the vielbein and standard formulations respectively we have 

\begin{align}
    R(\omega) &= e^\mu{}^a e^\mu{}_b R_{\mu\nu}{}^{ab}(\omega) \\
    R(\Gamma) &= g^{\mu\nu} g^{\rho \sigma} R_{\rho\mu\sigma\nu}(\Gamma)
\end{align}

Recall from page 596 of Zee, that the vielbein is ``the square root of the metric" \cite{zee}. Taking the determinant of both sides of equation \discolorlinks{\ref{metricviel}} and then the square root gives us 

\begin{equation}
    \text{det}(e_\mu{}^a) = e = \sqrt{-g} = \sqrt{-\text{det}(g_{\mu\nu})}
\end{equation}

So then our equivalent actions (in the vacuum) read

\begin{align}
    S&= \int d^4x e R(\omega) \label{tet}\\
    S&= \int d^4 x \sqrt{-g} R(\Gamma) \label{EH}
\end{align}

where \discolorlinks{\ref{tet}} is sometimes referred to as the tetradic Palatini action, and \discolorlinks{\ref{EH}} is the standard Einstein-Hilbert action.

\pagebreak

\subsection{Interlude}

This concludes the main objective of this work — gauging the Poincaré algebra to obtain GR. We hope that the reader finds the route taken to be helpful pedagogically in moving from point a — an undergraduate education in GR and classical field theory — to point b — having a working understanding of gauge theory as it pertains to pure Yang-Mills theory as well as to gravity.

From this point forward, we will be looking into non-relativistic gravity. In particular, we begin by looking for algebras suitable to construct NR gravitational theories through the gauging procdure we've followed above.

\iffalse
Let's connect this with the notation of the vielbein postulate from Carroll \cite{carrollbook}.

Carroll writes out the following in equations J.20 and J.22,

\begin{eqnarray} \notag
\Gamma^\nu_{\mu\lambda} &=& e^\nu{}_a \partial_\mu e_\lambda{}^a + e^\nu{}_a e_\lambda{}^b \omega_\mu{}^a{}_b \\
\nabla_\mu e_\nu{}^a &=& \partial_\mu e_\nu{}^a - \Gamma^\lambda_{\mu\nu}e_\lambda{}^a + \omega_\mu{}^a{}_b e_\nu{}^b = 0
\end{eqnarray}

In comparing the two, it seems that 

\begin{equation}D_\mu e_\nu{}^a= \partial_\mu e_\nu^a+ \omega_\mu{}^a{}_b e_\nu{}^b\end{equation}
\fi

\pagebreak
\section[Lie algebra expansions and contractions: Galilei, Bargmann, and Poincaré]{Lie algebra expansions and contractions:\\ Galilei, Bargmann, and Poincaré}
\label{sec:NRlas}

Before moving onto formal Lie algebra studies, we motivate why we want to know about the Bargmann algebra.
We recommend returning to this section immediately prior to beginning section \ref{gaugebarg} for a reminder.

Firstly, if we want to consider massive particles in our theory, which of course we would when working with gravity, we are going to need a massive representation of the Galilei algebra. As we will see in section \ref{sec:g2b} below, the Bargmann algebra is indeed sometimes called the massive or quantum Galilei algebra.

To see how this follows from a physical point of view as opposed to the mathematical one of section \ref{sec:g2b}, note that the Lagrangian of a non-relativistic particle 

\begin{equation}
    L = \frac{1}{2} M \dot{x}^i \dot{x}^i
\end{equation}

is not invariant under Galilean boosts (which we will talk more about in section \ref{galsym})

\begin{equation}
    \delta x^i = (v^i=\dot{x}^i)t
\end{equation}

Rather, the Lagrangian transforms as a total derivative,

\begin{equation}
    \delta L = \frac{d}{d t} (M x^i v^i)
\end{equation}

For more details on this involving Noether charges, Poisson brackets, and central charges, see page 10 of \cite{bergshoeff}.

\vspace{0.5cm}
Secondly (to be returned to after section \ref{metstruc} and equations \discolorlinks{\ref{tran1}} — \discolorlinks{\ref{tran5}}), in order to define an inverse spatial metric $\bar{h}_{\mu\nu}$ and inverse temporal metric $\bar{\tau}^\mu$ which (unlike the basic $h_{\mu\nu}$ and $\tau^\mu$ of section \ref{metstruc}) are invariant under Galilean boosts, it is required to introduce a vector field $m_\mu$ which transforms under Galilean boosts in such a way to cancel invariant terms in the transformations of $h_{\mu\nu}$ and $\tau^\mu$. 

For more details on this, see page 12 of \cite{2015berg} and page 11 of \cite{jelle}.

\vspace{0.5cm}
Lastly (to be returned to after section \ref{convconstrs}), in the gauging procedure to follow, building up Newton-Cartan gravity, we impose a conventional constraint on the gauge curvature of the Bargmann's auxiliary central charge gauge field $m_\mu$  in order to solve for the spin connection(s) in terms of the independent fields of the theory.

\pagebreak

\subsection{Galilei to Bargmann}
\label{sec:g2b}
Using the conventions of page 445 of \cite{zeeGT}, the Galilei algebra is expressed via the following commutation relations

\begin{align}
[J_i,J_j] &= i \epsilon_{ijk}J_k \label{from} \\
[J_i,P_j] &= i \epsilon_{ijk}P_k \\
[J_i,K_j] &= i \epsilon_{ijk}K_k \\
[K_i,H] &= i P_i \label{missing}\\
[P_i,H] = [J_i,H]=[P_i,P_j]=[K_i,P_j]=[K_i,K_j] &= 0 \label{to} \\
[K_i,P_j] &= 0 \label{central}
\end{align}

where H is the generator of temporal translations, $P_i$ are the generators of spatial translations, $K_i$ are the generators of Galilean boosts, and $J_i$ are the generators of spatial rotations.

The generators can be written as operators as follows

\begin{eqnarray}
    H &\equiv& i \frac{\partial}{ \partial t}  \label{eq:ham}\\ 
    P_i &\equiv& i \partial_i \label{eq:mom}\\ 
    K_i &\equiv& \frac{1}{i} t\cdot \partial_i \label{gal} \\
    J_i &\equiv& \frac{1}{i} (\vec{x}\times \vec{\nabla})_i \label{eq:j}
\end{eqnarray}

The Galilei algebra can be ``made quantum" by performing a Lie algebra extension.
If we want to ``make the algebra quantum", i.e. get physical information from it, we ought to identity a Hamiltonian operator and a momentum operator. Things have been named suggestively (page 510 \cite{zeeGT}).

Notice that equations \discolorlinks{\ref{eq:ham}} and \discolorlinks{\ref{eq:mom}} tell us that the dimensions of H and $P_i$ are $\frac{1}{T}$, and $\frac{1}{L}$ respectively. To identify these generators with the Hamiltonian and momentum operators, which have dimensions

\begin{align}
    [H] = [\text{energy}] &= \frac{ML^2}{T^2}  \\ 
    [P_i] = [\text{momentum}] &= \frac{ML}{T} \label{dimmom}
\end{align}

we can multiple the algebra generators by $\hbar$ which has dimensions $\frac{ML^2}{T}$. Thus we replace $H$ and $P_i$ in the algebra with $\hbar H$ and $\hbar P_i$ and now we have generators with the dimensionality of their namesakes.

What remains to be seen in this ``quantum Galilei" algebra is mass. Notice that none of thew generators \discolorlinks{\ref{eq:ham}} - \discolorlinks{\ref{eq:j}} have dimensions of mass or 1 over mass. Focusing on equation \discolorlinks{\ref{central}}, let's look into what two generators/Lie algebra elements commuting means.

The equation can be read as ``$[K_i,P_j]$ commutes with all other generators." Let $\mathfrak{g}$ denote an algebra, and let $x$ be its elements. The above sentence can be rewritten as  

\begin{eqnarray}
[[K_i,P_j]=0,x] = 0
\end{eqnarray}

since it is a property of Lie algebras that $[0,x] = 0$ $\forall x\in \mathfrak{g}$. This can be said in another way as well, $0 \in Z(\mathfrak{g})$, where $Z(\mathfrak{g})$ denotes the \textit{center} of the Lie algebra $\mathfrak{g}$. The center is an ideal (algebraic equivalent of a subset) of the algebra containing all elements that commute with all other elements of the algebra (in some sense the measure of an algebra's ``abelian-ness" since $\mathfrak{g}=$ abelian $\leftrightarrow Z(\mathfrak{g})=\mathfrak{g}$) \cite{LA}.

Notice then, that the statement ``$[K_i,P_j]$ commutes with all other generators" would still hold even if they didn't perfectly commute, but rather 

\begin{eqnarray}
[K_i,P_j]= i M \delta_{ij} \label{new}
\end{eqnarray}

for some $M \in Z(\mathfrak{g})$.

Now, \discolorlinks{\ref{gal}} tells us that the dimension of the Galilean boost operators is $\frac{T}{L}$. Combining this with \discolorlinks{\ref{dimmom}} it becomes clear that the dimension of the product (and Lie bracket) of $K_i$ and $P_j$ is mass! And so we can reasonably say $[M]=$ mass.

This process of extending an algebra by replacing a perfect commutation (a zero) with an element of the center of the algebra is called a ``central extension." This new algebra we have created (exactly equations \discolorlinks{\ref{from}} - \discolorlinks{\ref{to}}, but equation \discolorlinks{\ref{central}} becomes \discolorlinks{\ref{new}}) is called the Bargmann algebra \cite{bargmann}.

\pagebreak

\subsection{Poincaré to Galilei}
\label{p2g}

Here we use a GR metric signature of $(-+++)$ in agreement with  \cite{bergshoeff}, \cite{SUGRA}, \cite{bergshoeff2017}, and \cite{bergshoeff2020} (and in contrast to the QFT metric signature $(+- {} - {} -)$ of \cite{zeeGT}, \cite{boer}, and \cite{Poincare}; as well as in contrast to the Newtonian metric signature $(++++)$ of \cite{levyleblond}). We also neglect imaginary $i$'s in the commutation relations. $\mu,\nu$ are spacetime indices, and $i,j$ are space indices.

The Poincare algebra is given by equations \discolorlinks{\ref{poin1}}-\discolorlinks{\ref{poin3}}, reproduced here

\begin{align}
    [P_\mu,P_\nu] &=0 
    \tag{\discolorlinks{\ref{poin1}}.r} \\ 
    \notag
    [M_{\mu \nu },P_\rho] &= \eta_{\rho\nu}P_\mu -\eta_{\rho\mu }P_\nu \\  
    &= 2 \eta_{\rho[\nu}P_{\mu]} \tag{\discolorlinks{\ref{poin2}}.r}\\
    \notag
    [M_{\mu \nu},M_{\rho\sigma}] &= \eta_{\nu \rho}M_{\mu \sigma}+\eta_{\mu \sigma}M_{\nu \rho}-\eta_{\mu \rho}M_{\nu \sigma}-\eta_{\nu \sigma}M_{\mu \rho} \\ 
    &= 4\eta_{[\mu[\rho}M_{\sigma]\nu ]} \tag{\discolorlinks{\ref{poin3}}.r}
\end{align}

The Lorentz generators can be broken up into Lorentz boosts $\tilde{K_i}$ and spatial rotations $\tilde{J_{ij}}$ by the identification 

\begin{equation}
    M_{\mu\nu} \rightarrow (M_{0i}=\tilde{K_i}, \tilde{J_{ij}}) \label{breakup1}
\end{equation}

Similarly, the Poincaré translation generators can be broken up into temporal translations $\tilde{H}$ and spatial translations $\tilde{P_i}$ by the identification 

\begin{equation}
    P_{\mu} \rightarrow (P_{0}=\tilde{H}, \tilde{P_{i}}) \label{breakup2}
\end{equation}

With these decompositions, the algebra \discolorlinks{\ref{poin1}}.r - \discolorlinks{\ref{poin3}}.r can be rewritten as 

\begin{align}
[P_\mu,P_\nu] &= 0   & \xrightarrow{\hspace*{4cm}}  && 
    [\tilde{H},\tilde{H}] &= 0 \label{1}\\
    &&&& [\tilde{P_i},\tilde{H}] &= 0 \label{2}\\
    &&&& [\tilde{P_i},\tilde{P_j}] &= 0 \label{3}\\ 
[M_{\mu \nu },P_\rho] &= 2 \eta_{\rho[\nu}P_{\mu]}   & \xrightarrow{\hspace*{4cm}}  &&  
    [\tilde{K_i},\tilde{H}] &= \tilde{P_i} \label{4}\\
    &&&& [\tilde{K_i},\tilde{P_j}] &= \delta_{ij} \tilde{H} \label{5}\\
    &&&& [\tilde{J_{ij}},\tilde{P_k}] &= 2 \delta_{k[j}\tilde{P_{i]}} \label{6}\\
    &&&& [\tilde{J_{ij}},\tilde{H}] &= 0 \label{7}\\ 
[M_{\mu \nu},M_{\rho\sigma}] &= 4\eta_{[\mu[\rho}M_{\sigma]\nu ]}  & \xrightarrow{\hspace*{4cm}}  && 
    [\tilde{K_i},\tilde{K_j}] &= 0 \label{8}\\
    &&&& [\tilde{J_{ij}},\tilde{K_k}] &= 2 \delta_{k[j}\tilde{K}_{i]} \label{9}\\
    &&&& [\tilde{J_{ij}},\tilde{J_{kl}}] &= 4\delta_{[i[k}\tilde{J}_{l]j ]} \label{10}
\end{align}

Equation \discolorlinks{\ref{1}} can be disregarded since $H$ is a single element in the algebra, and by the alternativity property of Lie algebras, $[x,x]=0$  $\forall$  $x \in \mathfrak{g}$. So equation \discolorlinks{\ref{1}} adds no information.

To see how the Galilei algebra can be extracted from this via a Lie algebra (Inönü-Wigner \cite {iw}) contraction, we redefine the generators as follows

\begin{eqnarray}
\tilde{J_{ij}} &\rightarrow& J_{ij}  \label{lim1}\\
\tilde{H} &\rightarrow& H \label{lim2}\\
\tilde{K_{i}} &\rightarrow& c K_i \label{lim3}\\
\tilde{P_{i}} &\rightarrow& c P_i \label{lim4}
\end{eqnarray}

and then let $c \rightarrow \infty$ . This parameter $c$ is labeled as such to be suggestive. We call the Galilei algebra, the non-relativistic limit ($c \rightarrow \infty$) of the Poincaré algebra. The $c \rightarrow \infty$ limit can be understood by imagining yourself as a particle in a race with some photons. You are traveling at $v=0.99 c$ so you (obviously won't win the race but) don't feel totally outmatched. Suddenly your competitors all get some insane abilities and their speed goes from $c$ to infinity. Regardless of your remarkable speed, you can't compete with this, and at the moment they all shoot away from you, it feel like you're practically standing still. Hence your relativistic speed now feels woefully non-relativistic.

Given the redefinitions and limiting procedure of equations \discolorlinks{\ref{lim1}-\ref{lim4}}, equations \discolorlinks{\ref{1}-\ref{10}} become

\begin{align}
c[P_i,H] = 0  \rightarrow &[P_i,H]=c^{-1} \cdot 0 = 0  &   \hspace*{2.25cm}     && [P_i,H] &= 0 \tag{\discolorlinks{\ref{2}}.r}\\
c^2[P_i,P_j] =0 \rightarrow &[P_i,P_j] = c^{-2} \cdot 0 =0    &    \hspace*{2.25cm}   && [P_i,P_j] &= 0 \tag{\discolorlinks{\ref{3}}.r}\\
c [K_i,H] = c P_i  \rightarrow &[K_i,H] = P_i &   c\rightarrow \infty    \hspace{.5cm}    && [K_i,H] &= P_i \tag{\discolorlinks{\ref{4}}.r}\\
c^2 [K_i,P_j]=\delta_{ij}   H  \rightarrow &[K_i,P_j]= c^{-2}\delta_{ij}   H  &   \xrightarrow{\hspace*{2cm}}      && [K_i,P_i] &=0 \tag{\discolorlinks{\ref{5}}.r}\\
c[J_{ij},P_k] = c \cdot 2  \delta_{k[j}P_{i]} \rightarrow &[J_{ij},P_k] = 2 \delta_{k[j}P_{i]}    &     \hspace*{2.25cm}       && [J_{ij},P_k] &= 2 \delta_{k[j}P_{i]} \tag{\discolorlinks{\ref{6}}.r}\\
&[J_{ij},H]=0 &     c\rightarrow \infty   \hspace{.5cm}   && [J_{ij},H]&= 0 \tag{\discolorlinks{\ref{7}}.r}\\
c^{2}[K_i,K_j]=J_{ij} \rightarrow  &[K_i,K_j]=c^{-2}J_{ij}  &     \xrightarrow{\hspace*{2cm}}      && [K_i,K_j]&=0 \tag{\discolorlinks{\ref{8}}.r}\\
c[J_{ij},K_k]= c \cdot 2 \delta_{k[j}K_{i]} \rightarrow &[J_{ij},K_k]=  2 \delta_{k[j}K_{i]}  &     \hspace*{2.25cm}      && [J_{ij},K_k]&= 2 \delta_{k[j}K_{i]} \tag{\discolorlinks{\ref{9}}.r}\\
&[J_{ij},J_{kl}] = 4\delta_{[i[k}J_{l]j ]}    &     \hspace*{2.25cm}      && [J_{ij},J_{kl}] &= 4\delta_{[i[k}J_{l]j ]} \tag{\discolorlinks{\ref{10}}.r}
\end{align}

which is precisely the (4-dimensional, in agreement with \cite{cohomo}, and in contrast to 3-dimensional as given by equations \discolorlinks{\ref{from}-\ref{central}}) Galilei algebra.

Note that the limiting procedure of equations \discolorlinks{\ref{lim1}-\ref{lim4}} is not unique. See footnote 3 on page 5 of \cite{bergshoeff2020} for more details on why. For instance, while we will stick to the limit as stated above (as is presented in \cite{bergshoeff} and \cite{bergshoeff2020}), we could have have taken the limits presented in \cite{bergshoeff2017}, notably
\begin{eqnarray}
\tilde{J_{ij}} &\rightarrow& J_{ij} \\
\tilde{H} &\rightarrow& c^{-1} H \\
\tilde{K_{i}} &\rightarrow& c K_i \\
\tilde{P_{i}} &\rightarrow&  P_i
\end{eqnarray}
This is a totally valid limit and yields the Galilei algebra just the same.

\pagebreak

\subsection{Extended Poincaré to Bargmann}
\subsubsection{Extending Poincaré}

In Section \ref{sec:g2b}, in going from Galileli to Bargmann via a Lie algebra extension, we added a generator (M). Thus the Bargmann algebra is 11-dimensional, 10 from Galilei, and M.

Thus, it is not surprising that we cannot go directly from Poincaré (which is also 10-dimensional) to Bargmann. In other words, the following diagram does not commute 

\[
  \begin{tikzcd}[sep=huge]
    \large{\text{Poincaré}} \arrow[r,"\text{C}"] \arrow[rd, red, "\text{E} \hspace{1 em}\circ\hspace{1 em} \text{C}"'] & \large{\text{Galilei}} \arrow[d,"\text{E}"] \\
     & \large{\text{Bargmann}}
  \end{tikzcd}
\]

where the C's represent contractions and the E's represent extensions.

But consider what we added to Galilei to get Bargmann — just a single generator, M, which 
in turn generates a 1-dimensional algebra. Up to isomorphism there is only one 1-dimensional algebra, so we call ours $\mathfrak{u}$(1), the circle group U(1)'s Lie algebra, and we'll call its single generator $\tilde{M}$.

We then perform a Lie algebra extension of the Poincaré by ``tacking" on this $\mathfrak{u}$(1). This is a \textit{trivial} extension, and consists of simply a direct sum of the base Lie algebra we want to extend (Poincaré) and the other algebra we want to use to augment/extend your base algebra ($\mathfrak{u}$(1)). Thus our result is Poincaré $\oplus$  $\mathfrak{u}$(1). See \cite{LAext} as well as pages 191-193 of \cite{spainLA} for more on this.

Note that a direct sum Lie algebra $\mathfrak{s}=\mathfrak{p} \oplus \mathfrak{u}$ is defined with a bracket

\begin{equation}
    [(p_1,u_1),(p_2,u_2)]= ([p_1,p_2],[u_1,u_2])
\end{equation}
for elements $p_1,p_2$ $\in$ $\mathfrak{p}$ and $u_1,u_2$ $\in$ $\mathfrak{u}$. But our $\mathfrak{u}$ is $\mathfrak{u}$(1), which is 1-dimensional so there is no bracket between two elements. So the bracket of $\mathfrak{s}$ is just 

\begin{equation}
    [(p_1,u_1),(p_2,u_2)]= ([p_1,p_2],0) \label{itsjust}
\end{equation}

This makes it seem as though the algebra of Poincaré $\oplus$  $\mathfrak{u}$(1) is simply given by the algebra of Poincaré. This is not the case. The algebra is 11-dimensional now, not 10-dimensional. Moreover, even though the extra (central) generator $\tilde{M}$ commutes with everything in the algebra, it still exists. So one can think of the structure of the algebra Poincaré $\oplus$  $\mathfrak{u}$(1) as all the commutation relations of Poincaré plus a lonely generator $\tilde{M}$ just hanging out, not part of any commutator.

\pagebreak

\subsubsection{A new contraction}

We will now define the contraction of Poincaré $\oplus$  $\mathfrak{u}$(1) to the Bargmann as we did in Section \ref{p2g} such that the following diagram commutes 

\[
\begin{tikzcd}[sep=huge]
\large{\text{Poincaré}} \arrow[r, "\text{C}"] \arrow[d, "\text{E}"] & \large{\text{Galilei}} \arrow[d, "\text{E}"] \\
\large{\text{Poincaré}} \oplus \large{\mathfrak{u(1)}}\arrow[r, "\text{C '}"]& \large{\text{Bargmann}} \end{tikzcd}
\]

Motivated by breaking up ``relativistic" energy ($\tilde{H}$) into rest mass energy (M$c^2$) and non-relativistic kinetic energy (H) we add the following redefinition to equations \discolorlinks{\ref{lim1}-\ref{lim4}}, replacing \discolorlinks{\ref{lim2}} in particular with $\tilde{H} \rightarrow M c^2 + \frac{1}{2}H.$ And further, motivated by the equivalence of mass and energy in relativity, we define a mirrored redefinition for $\tilde{M}$ given by $\tilde{M} \rightarrow -M c^2 + \frac{1}{2}H.$ See equation 2.23 of \cite{bergshoeff2020}.

So all in all equations \discolorlinks{\ref{lim1}-\ref{lim4}} are replaced with the following limiting procedure

\begin{align}
\tilde{J_{ij}} &\rightarrow J_{ij} \label{newc1}\\
\tilde{H} &\rightarrow M c^2 + \frac{1}{2} H \label{newc2}\\
\tilde{K_{i}} &\rightarrow c K_i \label{newc3}\\
\tilde{P_{i}} &\rightarrow c P_i \label{newc4} \\
\tilde{M} &\rightarrow -M c^2 + \frac{1}{2}H \label{newc5}
\end{align}

It will turn out that we ought to define a bit more before taking the $c\rightarrow \infty$ limit. But to show what goes wrong, we will include the calculation of applying only \discolorlinks{\ref{newc1}-\ref{newc5}}
to Poincaré $\oplus$  $\mathfrak{u}$(1).

First, make the re-(d)efinitions prescribed by \discolorlinks{\ref{newc1}-\ref{newc5}}

\begin{align} 
[\tilde{P_i},\tilde{H}] &= 0  & \longrightarrow &&
[P_i,M] + \frac{1}{2} c^{-2}  [P_i,H] &= 0 \tag{\discolorlinks{\ref{2}}.d} \\
[\tilde{P_i},\tilde{P_j}] &= 0  & \longrightarrow && [P_i, P_j] &= 0 \tag{\discolorlinks{\ref{3}}.d} \\
[\tilde{K_i},\tilde{H}] &= \tilde{P_i} & \longrightarrow &&
[K_i,M]+\frac{1}{2}c^{-2}[K_i,H] &= c^{-2} P_i \tag{\discolorlinks{\ref{4}}.d} \\
[\tilde{K_i},\tilde{P_j}] &= \delta_{ij} \tilde{H} & \longrightarrow && [K_i,P_j] &= \delta_{ij} (M  + \frac{1}{2}c^{-2}H)
\tag{\discolorlinks{\ref{5}}.d} \\
[\tilde{J_{ij}},\tilde{P_k}] &= 2 \delta_{k[j}\tilde{P_{i]}} & \longrightarrow && [J_{ij},P_k] &= 2 \delta_{k[j}P_{i]} \tag{\discolorlinks{\ref{6}}.d} \\
[\tilde{J_{ij}},\tilde{H}] &= 0 & \longrightarrow &&[J_{ij},M] + \frac{1}{2}  c^{-2}[J_{ij},H] &= 0  \tag{\discolorlinks{\ref{7}}.d} \\
[\tilde{K_i},\tilde{K_j}] &= 0 & \longrightarrow && [K_i,K_j] &= 0 \tag{\discolorlinks{\ref{8}}.d} \\
[\tilde{J_{ij}},\tilde{K_k}] &= 2 \delta_{k[j}K_{i]} & \longrightarrow && [J_{ij},K_k] &= 2 \delta_{k[j}K_{i]} \tag{\discolorlinks{\ref{9}}.d} \\
[\tilde{J_{ij}},\tilde{J_{kl}}] &= 4\delta_{[i[k}\tilde{J}_{l]j ]} & \longrightarrow && [J_{ij},J_{kl}] &= 4\delta_{[i[k}J_{l]j ]}
\tag{\discolorlinks{\ref{10}}.d}
\end{align}

Factors of c only show up in equations \discolorlinks{\ref{2}}.d, \discolorlinks{\ref{4}}.d, \discolorlinks{\ref{5}}.d, and \discolorlinks{\ref{7}}.d. Taking the $c\rightarrow \infty$ ($\ell$)imit we have

\begin{align} 
[P_i,M] + \frac{1}{2}c^{-2} [P_i,H] &= 0  & \longrightarrow && [P_i,M] &= 0 \tag{\discolorlinks{\ref{2}}.$\ell$} \\
[K_i,M]+\frac{1}{2}c^{-2}[K_i,H] &= c^{-2} P_i & \longrightarrow && [K_i,M] &= 0 \tag{\discolorlinks{\ref{4}}.$\ell$} \\
[K_i,P_j] &= \delta_{ij} (M  + \frac{1}{2}c^{-2}H )& \longrightarrow && [K_i,P_j] &= \delta_{ij} M \tag{\discolorlinks{\ref{5}}.$\ell$} \\
[J_{ij},M] + \frac{1}{2}c^{-2}[J_{ij},H] &= 0 & \longrightarrow && [J_{ij},M] &= 0 \tag{\discolorlinks{\ref{7}}.$\ell$}
\end{align}

%\JL{One thing; you haven't used the fact that $\tilde M$ is central, that is $[anything, \tilde M]=0$. This would give additional constraints as far as I can see.}

All in all we are left with 

\begin{align}
    [P_i,M] &= 0 \tag{\discolorlinks{\ref{2}}.$\ell$} \\
    [P_i, P_j] &= 0 \tag{\discolorlinks{\ref{3}}.$\ell$} \\
    [K_i,M] &= 0 \tag{\discolorlinks{\ref{4}}.$\ell$} \\
    [K_i,P_j] &= \delta_{ij} M \tag{\discolorlinks{\ref{5}}.$\ell$} \\
    [J_{ij},P_k] &= 2 \delta_{k[j}P_{i]} \tag{\discolorlinks{\ref{6}}.$\ell$} \\
    [J_{ij},M] &= 0 \tag{\discolorlinks{\ref{7}}.$\ell$} \\
    [K_i,K_j] &= 0 \tag{\discolorlinks{\ref{8}}.$\ell$} \\
    [J_{ij},K_k] &= 2 \delta_{k[j}K_{i]} \tag{\discolorlinks{\ref{9}}.$\ell$} \\
    [J_{ij},J_{kl}] &= 4\delta_{[i[k}J_{l]j ]} \tag{\discolorlinks{\ref{10}}.$\ell$}
\end{align}

Notice why this is wrong. We have lost information about the algebra. Notably equation 4.4 of \cite{bergshoeff} (or our equation \discolorlinks{\ref{missing}} noting the different conventions in Section \ref{sec:g2b}), $[K_i,H]=P_i$.

Taking a look at equation \discolorlinks{\ref{4}}.d, $[K_i,M]+\frac{1}{2}c^{-2}[K_i,H]=c^{-2} P_i$ it seems we lost the information about $[K_i,H]$ by taking the limit to get to equation \discolorlinks{\ref{4}}.$\ell$, $[K_i,M] = 0$.

Maybe having two commutators on the left hand side of \discolorlinks{\ref{4}}.d is special? Two unknowns so we need to take the limit twice? Notably, what is we FIRST took the c to infinity limit of \discolorlinks{\ref{4}}.d to get $[K_i,M] =0$, then used $[K_i,M] =0$ as a fact, to write instead

\begin{align} \notag
[\tilde{K_i},\tilde{H}] = \tilde{P_i} \longrightarrow [cK_i,Mc^2+\frac{1}{2}H] &= cP_i\\ \notag
c^3[K_i,M]+\frac{1}{2}c[K_i,H] &= cP_i \\\notag
0+c[K_i,H] &= cP_i \\
[K_i,H] &= P_i \label{missinglink}
\end{align}

Using the result of the limit in taking this new limit is not concrete.

It turns out there is a way to more elegantly take this limit. What if, as opposed to only looking at what the relativistic generators are sent to, by \discolorlinks{\ref{newc1}}-\discolorlinks{\ref{newc5}}, we look at what the non-relativistic generators $J_{ij},K_i,P_i,H,$ and $M$ are sent to as well? Equations \discolorlinks{\ref{newc1}}-\discolorlinks{\ref{newc5}} are certainly invertible maps. Adding \discolorlinks{\ref{newc2}} and \discolorlinks{\ref{newc5}} gives us \discolorlinks{\ref{correctc2}}, subtracting \discolorlinks{\ref{newc5}} from \discolorlinks{\ref{newc2}} gives us \discolorlinks{\ref{correctc5}} and the remaining follow immediately

\begin{align}
J_{ij} &\rightarrow  \tilde{J_{ij}} \label{correctc1}\\
H &\rightarrow \tilde{M} + \tilde{H} \label{correctc2}\\
K_i &\rightarrow c^{-1} \tilde{K_{i}}  \label{correctc3}\\
P_i &\rightarrow c^{-1} \tilde{P_{i}} \label{correctc4} \\
M &\rightarrow \frac{1}{2} c^{-2} \tilde{H} - \frac{1}{2} c^{-2} \tilde{M} \label{correctc5}
\end{align}

Now the unique part. We want the commutation relations for the non-relativistic algebra right? Well instead of taking the limit of the relativistic algebra's commutation relations, let's go right to the NR commutators. We will \newline write all the possible commutators given the NR generators  $J_{ij},K_i,P_i,H,$ and $M$, \newline plug in the NR $\rightarrow$ Rel. redefinitions \discolorlinks{\ref{correctc1}}-\discolorlinks{\ref{correctc5}}, \newline use the commutators of the relativistic algebra \discolorlinks{\ref{2}}-\discolorlinks{\ref{10}}, 
\newline plug in the Rel. $\rightarrow$ NR definitions \discolorlinks{\ref{newc1}}-\discolorlinks{\ref{newc5}}, \newline and then take the limit.

Note that $M$ is in the center so we don't include that in the ``all the possible commutators." Between $J_{ij},K_i,P_i,H$ we will have 4 choose 2 (6) commutators between distinct elements, and then 3 commutators for each of $J_{ij},K_i,P_i$ with themselves. We will use \newline $\mapsto$ for the \discolorlinks{\ref{correctc1}}-\discolorlinks{\ref{correctc5}} (NR $\rightarrow$ Rel.) redefinitions, \newline$=$ for basic simplifications, \newline$\asymp$ for commutations according to the Poincaré algebra  \discolorlinks{\ref{2}}-\discolorlinks{\ref{10}}, \newline$\hookrightarrow$ for the \discolorlinks{\ref{newc1}}-\discolorlinks{\ref{newc5}} (Rel. $\rightarrow$ NR) redefinitions, and 
\newline $\Rrightarrow$ for the $c\rightarrow \infty$ limit. We have

\begin{align} \notag
    [K_i,H] &\mapsto [c^{-1} \tilde{K_i}, \tilde{H} + \tilde{M}] \\\notag
    &= c^{-1} [\tilde{K_i}, \tilde{H}] \\\notag
    &\asymp c^{-1} \tilde{P_i}\\\notag
    &\hookrightarrow c^{-1} c P_i  \label{barg1} \\
    &= P_i \\ \notag
    [J_{ij},H] &\mapsto [\tilde{J_{ij}},\tilde{H}+ \tilde{M}] \\ \notag
    &= [\tilde{J_{ij}},\tilde{H}] \\ 
    &\asymp 0 \\ \notag
    [P_i, H] &\mapsto c^{-1} [ \tilde{P_i}, \tilde{H}+ \tilde{M}]  \\ \notag
    &= c^{-1} [\tilde{P_i},\tilde{H}] \\ \notag
    &\asymp c^{-1} \cdot 0 \\
    &= 0 \\ \notag
    [K_i,P_i] &\mapsto c^{-2} [\tilde{K_i}, \tilde{P_j}] \\ \notag
    &\asymp c^{-2} \delta_{ij} \tilde{H} \\ \notag
    &\hookrightarrow c^{-2} \delta_{ij} (Mc^2 + \frac{1}{2} H) \\ \notag
    &= \delta_{ij} (M + \frac{1}{2} c^{-2} H) \\
    &\Rrightarrow \delta_{ij} M 
    \end{align}
    \begin{align}\notag
    [J_{ij},K_k] &\mapsto c^{-1} [\tilde{J_{ij}},\tilde{K_k}] \\ \notag
    &\asymp c^{-1} 2 \delta_{k[j}\tilde{K}_{i]} \\ \notag
    &\hookrightarrow c^{-1} c 2 \delta_{k[j}K_{i]} \\ 
    &= 2 \delta_{k[j}K_{i]} \\ \notag
    [J_{ij},P_k] &\mapsto c^{-1} [\tilde{J_{ij}},\tilde{P_k}] \\ \notag
    &\asymp c^{-1} 2 \delta_{k[j}\tilde{P}_{i]} \\ \notag
    &\hookrightarrow c^{-1} c 2 \delta_{k[j}P_{i]} \\ 
    &= 2 \delta_{k[j}P_{i]} \\ \notag
    [K_i,K_j] &\mapsto c^{-2} [\tilde{K_i},\tilde{K_j}] \\ \notag
    &\asymp c^{-2} 0\\ 
    &= 0 \\ \notag
    [P_i,P_j] &\mapsto c^{-2} [\tilde{P_i},\tilde{P_j}] \\ \notag
    &\asymp c^{-2} 0\\ 
    &= 0 \\ \notag
    [J_{ij},J_{kl}] &\mapsto [\tilde{J_{ij}},\tilde{J_{kl}}] \\ \notag
    &\asymp 4\delta_{[i[k}\tilde{J}_{l]j ]} \\
    &\hookrightarrow 4\delta_{[i[k}J_{l]j ]} \label{barg9}
\end{align}

Equations \discolorlinks{\ref{barg1}}- \discolorlinks{\ref{barg9}} are precisely the commutation relations of the Bargmann algebra. See equations 2.5.a - 2.5.d and 2.24 of \cite{bergshoeff2020}. The non-vanishing relations read

\begin{align}
[K_i,H] &= P_i \label{fbarg1}\\
[J_{ij},K_k] &= 2 \delta_{k[j}K_{i]} \label{fbarg2}\\
[J_{ij},P_k] &= 2 \delta_{k[j}P_{i]} \label{fbarg3}\\
[J_{ij},J_{kl}] &= 4\delta_{[i[k}J_{l]j ]} \label{fbarg4}\\
[K_i,P_i] &= \delta_{ij} M \label{fbarg5}
\end{align}

\pagebreak

\section{Newton-Cartan gravity and Bargmann algebra gauge theory}
\label{sec:nrgauge}

\subsection{Newtonian gravity}
\label{galsym}

In Newtonian gravity, the sacred spacetime of GR is decoupled into space \textit{and} time, which are now \textit{separately} invariant.

Moreover, there exists an absolute time in Newtonian gravity unlike in GR. This is most easily realized by considering ``slicing the bread." Two events in the unsliced loaf of spacetime cannot absolutely be said to come before or after one another because of the dependence on a coordinate system as is standard in GR. But in the sliced loaf of Newtonian space and time, all observers agree whether an event occurs before or after another since there is an absolute definition of time.

\begin{figure}[H]
\centering
\begin{subfigure}[b]{.5\textwidth}
  \centering
  \includegraphics[scale=0.15]{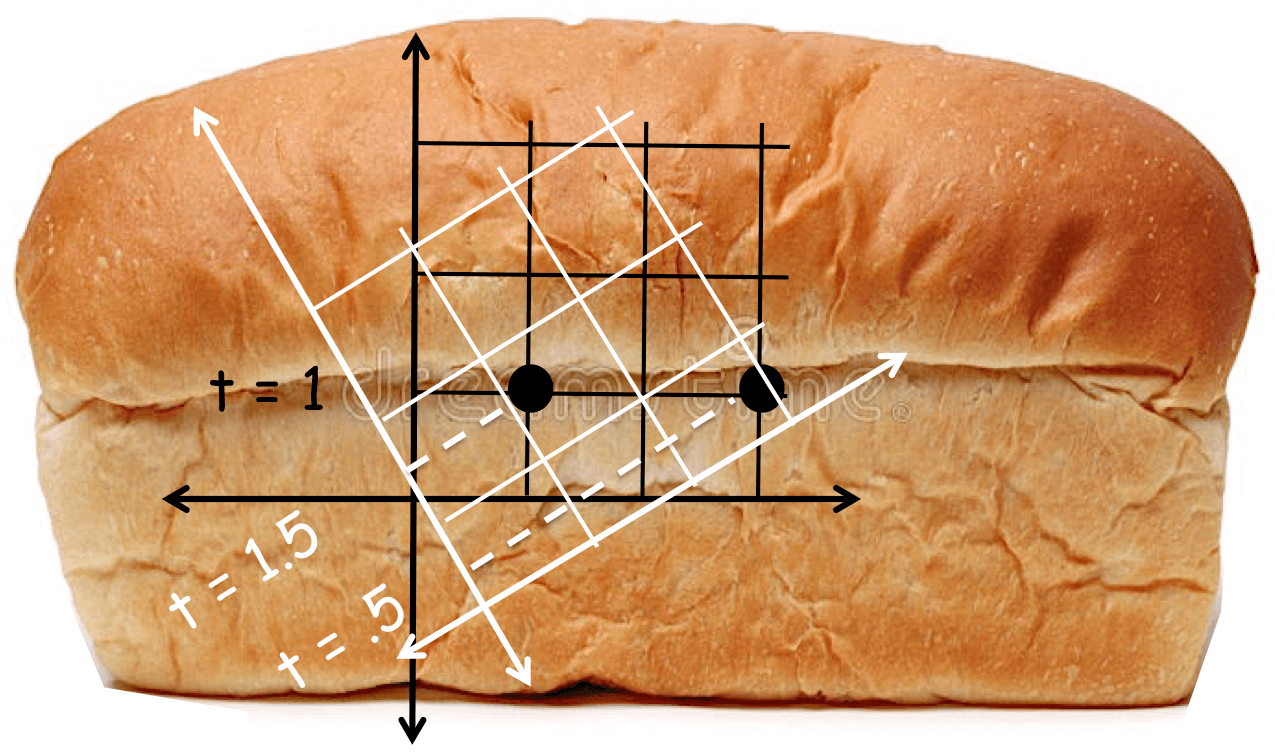}
\caption{Unsliced loaf of spacetime}
\end{subfigure}%
\begin{subfigure}[b]{.5\textwidth}
  \centering
  \includegraphics[scale=0.15]{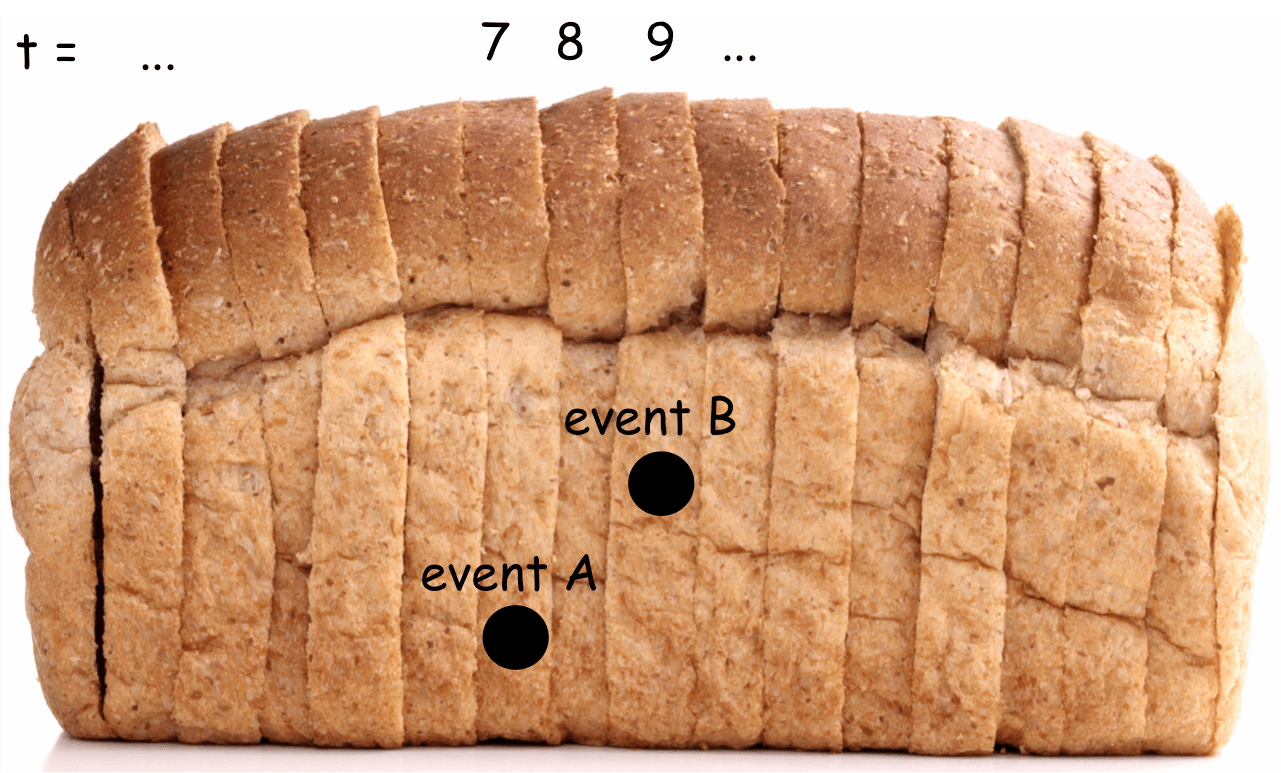}
\caption{Sliced loaf of space and time}
\end{subfigure}
\caption{Relative time versus absolute time in GR versus Newtonian gravity}
\label{fig:test}
\end{figure}

As opposed to the Poincaré symmetry group of GR, the Galilean group is the symmetry group for observers in a Newtonian setup. The group is characterized by the following coordinates transformations

\begin{align}
    (x^0 = t)' &= t + \xi^0 \label{coord1}\\
    x'{}^i &= R^i{}_j x^j + v^i t + \xi^i \label{coord2}
\end{align}

where $\xi^0$ are temporal translations, $\xi^i$ are spatial translations, $v^i$ are Galilean boosts, and $R^i{}_j$ are spatial rotations \cite{Roelthesis}. 

One can also consider the augmentation of \discolorlinks{\ref{coord1}} and \discolorlinks{\ref{coord2}} that includes any possible non-inertial reference frame. This can be done be letting $R^i{}_j$ and $v^i$ be time-dependent such that \discolorlinks{\ref{coord2}} becomes

\begin{equation}
    x'{}^i = R^i{}_j(t) x^j + \xi^i(t) \label{linegroup}
\end{equation}

This generalization of the Galilei group is called the Galilei line group \cite{grinnel}. 

The transformations \discolorlinks{\ref{coord1}} and \discolorlinks{\ref{coord2}} can be summarized with

\begin{equation}
    x'{}^\mu = G^\mu{}_\nu x^\nu + \xi^\mu
\end{equation}

where $G^\mu{}_\nu$ is the Galilean equivalent of the Lorentz matrix $\Lambda^\mu{}_\nu$ and encapsulates the Galilean boosts and spatial rotations as follows \cite{Roelthesis}

\renewcommand{\arraystretch}{1.3}
\begin{align}
    G^\mu{}_\nu &= \frac{\partial x ' {}^\mu}{\partial x^\nu} \\ \notag
&= \left( 
\begin{array}{c|ccc}
   \frac{\partial x ' {}^0}{\partial x^0}  & & \frac{\partial x ' {}^0}{\partial x^i} & \\
   \hline \\
   \frac{\partial x ' {}^i}{\partial x^0}  & & \frac{\partial x ' {}^i}{\partial x^j} & \\
   \\
\end{array}
\right) \\ \notag
&= \left( 
\begin{array}{c|ccc}
   1  & & 0 & \\
   \hline 
   \\
   v^i  & & R^i{}_j & \\
   \\
\end{array}
\right)
\end{align}

The zero in the 01, 02, and 03 positions of the matrix is indicative of the asymmetry of Galilean transformations. With Lorentz transformations, space rotates to time and time rotates to space, $x\leftrightarrow t$, but with Galilean transformations, space rotates to time but time does not rotate to space, $x\rightarrow t$ but $t \not\rightarrow x$.

The presence of an absolute time, and Galilean boosts taking space to time but not time to space, are two interconnected features of Newtonian gravity which lead to a final characteristic of the theory \cite{roelsNCphysforum}.

Newtonian gravity lacks a single non-degenerate metric like $g_{\mu\nu}$ in GR \cite{Roelthesis}. There are instead two separately invariant metrics
\begin{align} \tau_{\mu\nu} &= \text{diag} (1,0,0,0) \\
h^{\mu\nu} &= \text{diag} (0,1,1,1)
\end{align}
which together are degenerate, $\tau_{\mu\rho}h^{\rho\nu}=0.$

While this metric structure is more complicated that the $g_{\mu\nu}$ encountered in GR, having metric in hand nevertheless enables a geometric (built up from a metric) formulation of Newtonian gravity which mimicks the complete frame independence of GR. This is precisely what Cartan accomplished \cite{nc}.

\pagebreak
\subsubsection{Semi-geometric Newtonian gravity}
\label{subsec:semi}

What exactly does Cartan's theory need to do for it to ``succeed?" It turns out, even without the metric in hand, Newtonian gravity can be made geometric. The results of these manipulations (which produce a Christoffel connection, a Riemann curvature, a Ricci curvature, and Poisson's equation of gravity) are what Cartan's theory needs to reproduce. These quantities are given here before discussing how Cartan derived them from the metric(s) in a true frame-independent fashion through his \textit{Newton-Cartan} theory. We follow Misner-Thorne-Wheeler's page 289-290 \cite{mtw}, and Andringa's pages 19-22, 38/39, and 63 \cite{Roelthesis}. Since these derivations are done without metric in-hand, and just arise from comparing to the geodesic equation of GR, we will call what follows below the ``semi-geometric version" of Newtonian gravity and we will reserve the name ``geometric reformulation" of Newtonian gravity for Newton-Cartan theory.

The geodesic equation of GR for a particle moving on a curve parametrized by $\lambda$ is 

\begin{equation}
    \frac{d^2 x^\mu}{d \lambda^2} + \Gamma^\mu_{\nu\rho} \frac{d x^\nu}{d \lambda}\frac{d x^\rho}{d \lambda}=0 \label{geo}
\end{equation}

The equations of motion for Newtonian gravity, for a gravitational potential $\Phi(x)$, are 

\begin{equation}
    \frac{d^2 x^i}{d t^2} + \frac{\partial \Phi(x)}{\partial x^i}=0 \label{EOM}
\end{equation}

These can be rewritten by considering a particle moving on a curve parametrized by $\lambda$ in this potential $\Phi(x)$,

\begin{align}
    \frac{d^2 t}{d \lambda^2}&=0 \\\notag
    \frac{d^2 x^i}{d \lambda^2}+ \frac{\partial \Phi(x)}{\partial x^i}\bigg(\frac{d t}{d \lambda}\bigg)^2&=0 \\ 
    \frac{d^2 x^i}{d \lambda^2}+ \frac{\partial \Phi(x)}{\partial x^i} \frac{d x^0}{d \lambda}\frac{d x^0}{d \lambda} &=0
\end{align}

comparing this with \discolorlinks{\ref{geo}}, we see that the only non-zero coefficient of the Christoffel connection is

\begin{equation}
    \Gamma^i_{00} = \delta^{ij}\partial_j \Phi \label{Tcon}
\end{equation}

Using \discolorlinks{\ref{reimann}} with this connection, we have 

\begin{equation}
    R^i{}_{0j0} = \partial_j \Gamma^i_{00} = \delta^{ik}\partial_k \partial_j \Phi \label{riem}
\end{equation}

Poisson's equation for gravity (in natural $G=1$ units) reads

\begin{equation}
    \nabla^2 \Phi(x) = 4\pi \rho \label{poss}
\end{equation}

this is also called the source equation since $\rho$ is the density of the massive \textit{source} object.

Writing the left-hand side out ($\nabla^2 \Phi=\partial_i \partial_i \Phi$) and defining the Ricci tensor ($R^i{}_{0i0}=R_{00}$) gives us the geometric Poisson's equation for gravity

\begin{equation}
    R_{00} = 4\pi \rho \label{ricci}
\end{equation}

\textbf{Equations \discolorlinks{\ref{Tcon}}, \discolorlinks{\ref{riem}}, and \discolorlinks{\ref{ricci}} are what Newton-Cartan theory needs to reproduce solely from the metric structure.}

\vspace{0.5cm}
Returning briefly to our aside on the Galilei line group that we look at in \discolorlinks{\ref{linegroup}}, let us consider the equations of motion for some non-inertial frames \cite{Roelthesis}\cite{grinnel}. We will consider solely time-dependent rotations $R^i{}_j(t)$ such that
\begin{equation}
    x'{}^i = A^i{}_j(t) x^j \label{tdrot}
\end{equation}
and consider a free particle, i.e. $\Phi=0$ and so equation \discolorlinks{\ref{EOM}} is simply $\ddot{x}^i =0$.

After the time-dependent rotation transformation \discolorlinks{\ref{tdrot}}, $\ddot{x}^i =0$ becomes

\begin{equation}
    \ddot{x}^{'i} + 2 A_k{}^i\dot{A}^k{}_j \dot{x}^{'j} + A_k{}^i\ddot{A}^k{}_j x^j =0
\end{equation}

where the second and third terms account for the fictitious Coriolis and centrifugal forces respectively \cite{Roelthesis}\cite{fictious}.

\pagebreak
\subsection{Newton-Cartan gravity}

A helpful addition to ``slicing the bread," is to think of the limiting procedure of GR to NC gravity as follows:

\begin{displayquote}
``One may view the procedure geometrically as an opening of the light cones of spacetime, which finally become the space-like hypersurfaces of constant Newtonian time." \cite{datacourt}
\end{displayquote}

where the following graphic may provide additional clarity

\begin{figure}[H]
\centering
\includegraphics[width=15cm, keepaspectratio]{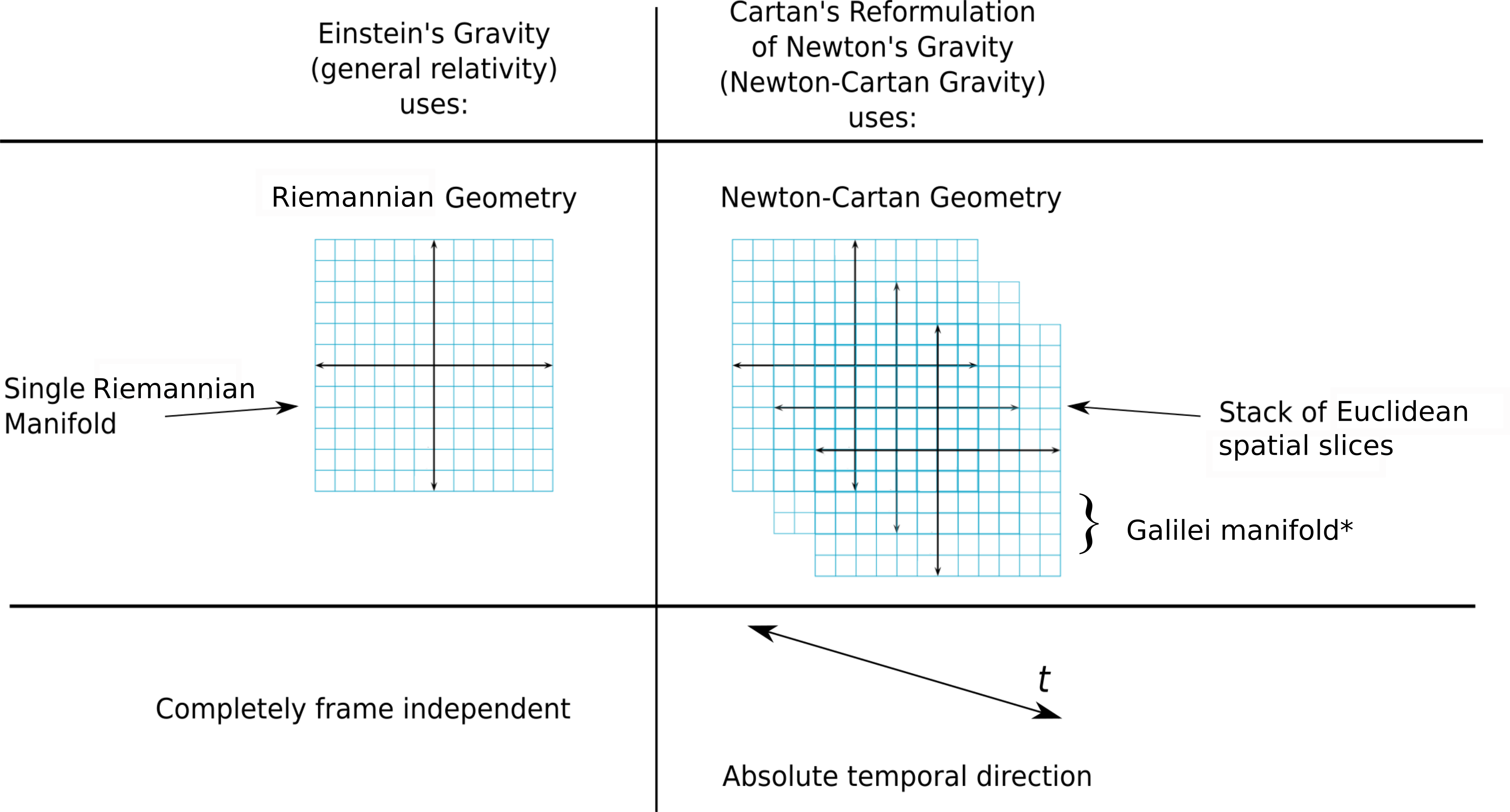}
\captionsetup{format=hang}
\caption{The manifold (which we will call a Galilei manifold and we will return to defining this manifold structure later in this chapter \cite{nccosmo} \cite{kuezle}) of Newton-Cartan geometry is foliated into constant time slices.}
\label{stack}
\end{figure}

This section will not be a comprehensive review of Newton-Cartan gravity, but rather a synopsis of the important ingredients to the theory that we will aim to recover through gauging. We follow section 4.3 of \cite{Roelthesis}, beginning on page 62, more or less verbatim.

At various points in this section we will make note that a particular calculation is non-trivial and requires attention to detail if the reader desires a rigorous account of the steps of logic in deriving Newton-Cartan.

The following references include many of the calculations that are skipped in this chapter in order to get to gauging the Bargmann:
\newline
\cite{dautcourtsOG} (as well as a more recent but shorter \cite{datacourt} from the same author),
\cite{nccosmo}, and 
\cite{topics}.

\pagebreak

\subsubsection{Metric and vielbein structure}
\label{metstruc}

We begin with the metric structure 
\begin{align} \tau_{\mu\nu} &= \text{diag} (1,0,0,0) \label{1x1}\\
h^{\mu\nu} &= \text{diag} (0,1,1,1)
\end{align}
which together are degenerate, 

\begin{equation}
    h^{\mu\nu}\tau_{\nu\rho}=0 \label{deg}
\end{equation}

Inverse temporal and spatial metrics $\tau^{\mu\nu}$ and $h_{\mu\nu}$ can be introduced which obey the following relations

\begin{align}
    h^{\mu\nu}h_{\nu\rho} + \tau^{\mu\nu} \tau_{\nu\rho} &= \delta^\mu_\rho  \label{met1}\\
    \tau^{\mu\nu}\tau_{\mu\nu} &= 1 \label{met2} \\
    h_{\mu\nu}\tau^{\nu\rho} &= 0 \label{met3}
\end{align}

We can introduce a vielbein structure for both metrics as follows: we decompose the full vielbein $e_\mu{}^a$ from GR into temporal and spatial vielbeine 
\begin{equation} e_\mu{}^a \rightarrow \{e_\mu{}^0=\tau_\mu, e_\mu{}^i\}
\end{equation}

In this way, both temporal and spatial metrics are composed of their respective vielbeine

\begin{align}
    h^{\mu\nu} &= e^\mu{}_{i} e^\nu{}_j \delta^{ij} \label{metspaceviel}\\
    \tau_{\mu\nu} &= \tau_\mu \tau_\nu
\end{align}

Inverse temporal and spatial vielbeine $\tau^\mu$ and $e^\mu{}_i$ can be introduced as follows

\begin{align}
\tau_\mu \tau^\mu &=1 \label{tinvert}\\
    e_\mu{}^{i} e^\mu{}_j &= \delta^i_j
\end{align}

The vielbein versions of the  \discolorlinks{\ref{deg}}, \discolorlinks{\ref{met1}}, and \discolorlinks{\ref{met3}} metric conditions now read 

\begin{align}
    e^\mu{}_i\tau_{\mu} &= 0 \tag{\discolorlinks{\ref{deg}}.v}\\
    e_\mu{}^{i} e^\nu{}_i + \tau_\mu \tau^\nu &= \delta^\nu_\mu  \tag{\discolorlinks{\ref{met1}}.v}\\
    e_\mu{}^i\tau^{\mu} &= 0 \tag{\discolorlinks{\ref{met3}}.v}
\end{align}

where we already have the vielbein version of \discolorlinks{\ref{met2}} from \discolorlinks{\ref{tinvert}}. 

\pagebreak

\subsubsection{Metric compatibilty, zero torsion, and a non-unique connection}
\label{subsec:torsionm}
As with GR, we introduce metric compatibility (on both metrics) as a first step towards a connection. Taking the covariant derivative with respect to some to be determined connection $\Gamma^\rho_{\mu\nu}$

\begin{align} 
    \nabla_\rho h^{\mu\nu} &= 0 \\
    \nabla_\rho \tau_{\mu\nu} &= 0 \label{oops}
\end{align}

where \discolorlinks{\ref{oops}} can be written using the temporal vielbein 

\begin{equation}
        \nabla_\rho \tau_{\mu}= 0 \label{ncmet}
\end{equation}

in order to facilitate the follow crucial discussion.

In Newton-Cartan gravity, one way to manually impose casuality is to require that the geometry is torsion-less (or equivalently that the connection is symmetric) \cite{ericNCtorsion}. As we will see, this results in the absolute time of Newtonian gravity from Figures \ref{fig:test} and \ref{stack}. Note however that there exist other methods of imposing causality that do not require one to impose a vanishing torsion. This is relevant in holographic applications of NC gravity with torsion \cite{ericNCtorsion}.

Zero-torsion, as per equation \discolorlinks{\ref{torsion}}, implies

\begin{equation}
    T_{\rho \mu}{}^\lambda = 0 \rightarrow \Gamma_{[\rho\mu]}{}^\lambda = 0 \label{notors}
\end{equation}

i.e. the connection is symmetric. Applying this to the (anti-symmetrized versions of the) metric compatibility equation \discolorlinks{\ref{ncmet}} gives us

\begin{align} \notag
    \nabla_{[\rho} \tau_{\mu]} &= 0 \\\notag
    \partial_{[\rho} \tau_{\mu]} - \Gamma_{[\rho\mu]}{}^\lambda \tau_\lambda &= 0 \\\notag
    \partial_{[\rho} \tau_{\mu]} - 0 &= 0 \label{vanishingcurl}\\
     \partial_{[\rho} \tau_{\mu]}&= 0 \\\notag
    \frac{1}{2}(\partial_{\rho} \tau_{\mu} - \partial_{\mu} \tau_{\rho})  &= 0 \\\notag
   \partial_{\rho} \tau_{\mu} &= \partial_{\mu} \tau_{\rho} 
\end{align}

The symmetry of mixed second derivatives enables us to solve this equation for $\tau_\mu$,

\begin{equation} 
\tau_\mu =\partial_\mu f(x)  \label{curl} 
\end{equation}

for some scalar function f. We will learn in the next section that this scalar f can be identified as the absolute Newtonian time t foliating the manifold.

The final interesting thing to note in this section is that since Newton-Cartan geometry is not defined by a single non-degenerate metric as in GR, the Galilei manfifold of Figure \ref{stack} is neither Riemannian nor psuedo-Riemannian. The implication of this is that the fundamental theorem of Riemannian geometry (for any Riemannian/psuedo-Riemannian manifold, there is a \textit{unique} torsion-less connection — the Christoffel/Levi-Civita connection) cannot be utilized. Thus, unlike in GR, there is \textit{no unique connection in Newton-Cartan theory}. See page 3 of \cite{nccosmo} for detailed discussion on this. Explicitly, this can be realized by noting that the metric compatibility conditions \discolorlinks{\ref{ncmet}} are still satisfied if our to-be-determined connection is shifted by some arbitrary anti-symmetric two-form $K_{\mu\nu}$ term (combined with the spatial metric and temporal vielbein)

\begin{equation}
    \Gamma^\rho_{\mu\nu} \rightarrow \Gamma^\rho_{\mu\nu} + h^{\rho\lambda} K_{\lambda(\mu} \tau_{\nu)} \label{nonuni}
\end{equation}

The rigour-minded reader ought to check this explicitly.

The equation in GR for the Christoffel symbols in terms of derivatives of the metric and inverse metric

\begin{equation}
    \Gamma^\lambda_{\mu\nu} = \frac{1}{2} g^{\lambda\sigma}[\partial_\mu g_{\nu\sigma}+\partial_\nu g_{\sigma\mu} - \partial_\sigma g_{\mu\nu}] \label{intermsof}
\end{equation}

Modeled after the GR expression \discolorlinks{\ref{intermsof}} and keeping the non-uniqueness of the connection \discolorlinks{\ref{nonuni}} in mind, we can write the most general possible connection 

\begin{equation}
    \Gamma^\rho_{\mu\nu} = \tau^\rho \partial_{(\mu}\tau_{\nu)} + \frac{1}{2} h^{\rho\sigma}(\partial_\nu h_{\sigma\mu} + \partial_\mu h_{\sigma\nu} -\partial_\sigma h_{\mu\nu}) + h^{\rho\lambda}K_{\lambda(\mu} \tau_{\nu)} \label{prop}
\end{equation}

The rigour-minded reader ought to look into box 17.4 of \cite{moore}, working through the derivation of \discolorlinks{\ref{intermsof}} to see where the minus on the third term comes from, and why the temporal version $\tau^\rho \partial_{(\mu}\tau_{\nu)}$ in \discolorlinks{\ref{prop}} does not require a third minus term, while the spatial version $\tfrac{1}{2} h^{\rho\sigma}(\partial_\nu h_{\sigma\mu} + \partial_\mu h_{\sigma\nu} -\partial_\sigma h_{\mu\nu})$ in \discolorlinks{\ref{prop}} does require it.

In order for this to become equation \discolorlinks{\ref{Tcon}}, a particular coordinate system need be adopted, curvature constraints need to be imposed, and coordinate transformations need be utilized.

\pagebreak

\subsubsection{Adapted coordinates, coordinates transformations, covariant Poisson's equation}

The so-called ``adapted coordinates" arise by letting the scalar function $f(x)$ from \discolorlinks{\ref{curl}} take the form, $f(x)=x^0=t$. Then \discolorlinks{\ref{curl}} becomes 

\begin{equation}
 \tau_\mu = \partial_\mu (t) = \delta_\mu^0 \label{adapt}
\end{equation}

Using \discolorlinks{\ref{adapt}}, equations \discolorlinks{\ref{met1}} - \discolorlinks{\ref{met3}}, and equations \discolorlinks{\ref{met1}}.v - \discolorlinks{\ref{met3}}.v,  it can be shown that 

\begin{align}
    \tau^\mu &= (1,\tau^i) \label{start}\\
    h^{\mu0} &= 0 \\
    h_{\mu0} &= -h_{\mu i} \tau^i
\end{align}

or equivalently, 

\begin{align}
    \tau_{\mu\nu} &= \begin{pmatrix} 1 & 0 \\ 0 & 0\end{pmatrix} \label{tellsus}\\
    \tau^{\mu\nu} &= \begin{pmatrix} 1 & \tau^i \\ \tau^i  & \tau^i \tau^j \end{pmatrix} \\
    h_{\mu \nu} &= \begin{pmatrix} h_{ij} \tau^i \tau^j & -h_{ij} \tau^j \\ h_{ij} \tau^i & h_{ij} \end{pmatrix} \\
    h^{\mu \nu} &= \begin{pmatrix} 0 & 0 \\ 0 & h^{ij}\end{pmatrix}  \label{end}
\end{align}

where \discolorlinks{\ref{start}} - \discolorlinks{\ref{end}} ought to be derived explicitly by the rigour-minded reader.

Additionally, the reader ought to check that the adapted coordinate are fixed up to the following transformation,

\begin{align}
    t' &= t + \xi^0 \label{cor1} \\
    x^{'i} &= x^j + f^i(x^j,t) \label{cor2}
\end{align}

for constant temporal translations from $\xi^0$ and arbitrary fucntions of space and time $f^i(x^j,t)$.

Equation \discolorlinks{\ref{prop}} broken up into components in the adapted coordinates reads

\begin{align}
    \Gamma^i_{00} &= h^{ij} (\partial_0 h_{j0} - \frac{1}{2} \partial_j h_{00} + K_{j0}) \hspace{0.6cm} = \hspace{0.2cm} h^{ij} \Phi_j \label{ad1}\\
    \Gamma^i_{0j} &= h^{ik} (\frac{1}{2}\partial_0 h_{jk} + \partial_{[j} h_{k]0} - \frac{1}{2} K_{jk}) \hspace{0.1cm} = \hspace{0.2cm} h^{ik} (\frac{1}{2}\partial_0 h_{jk} + \Omega_{jk}) \label{ad2}\\
    \Gamma^i_{jk} &= \frac{1}{2}h^{i\ell}(\partial_j h_{\ell k} + \partial_k h_{\ell j} - \partial_\ell h_{jk}) \label{ad3}\\
    \Gamma^0_{\mu\nu} &= 0 \label{ad4}
\end{align}

where $\Phi_i$ is a vector field we call the acceleration field, and $\Omega_{ij}$ is a pseudo-vector field we call the Coriolis field \cite{dautcourtsOG}. Recall $K_{\mu\nu}$ is anti-symmetric, thus so is $\Omega_{ij}$. That equations \discolorlinks{\ref{ad1}}- \discolorlinks{\ref{ad4}} follow from \discolorlinks{\ref{start}}-\discolorlinks{\ref{end}} needs to checked.

To proceed with turning \discolorlinks{\ref{ad1}}- \discolorlinks{\ref{ad4}} into \discolorlinks{\ref{Tcon}}, we make an ansatz for a covariant Ricci tensor. Since the metric $\tau_{\mu\nu}$, and equivalently its vielbein $\tau_\mu$, are more or less a 1x1 matrices of just the number 1, we make the following ansatz

\begin{equation}
    R_{\mu\nu} = 4 \pi \rho \tau_{\mu\nu} = 4 \pi \rho \tau_{\mu}\tau_{\nu}
\end{equation}

In the adapted coordinates, \discolorlinks{\ref{adapt}} tells us that this becomes 

\begin{align}
    R_{00} &= 4 \pi \rho \\
    R_{ij} &= 0 \label{flat}\\
    R_{i0} &= R_{0i} = 0
\end{align}

where \discolorlinks{\ref{flat}} confirms our picture of a Galilei manifold from Figure \discolorlinks{\ref{stack}} — that the stack of spatial slices in the manifold are flat. And so the adapted coordinates tell us that the spatial metric and its inverse are

\begin{equation}
    h^{ij} = \delta^{ij} \hspace{1cm} h_{ij} = \delta_{ij} \label{flat2}
\end{equation}

To be checked by the reader: equation \discolorlinks{\ref{flat2}} restricts the coordinate transformations of \discolorlinks{\ref{cor2}} to the more familiar form of 

\begin{equation}
    x'{}^i = R^i{}_j(t) x^j + \xi^i(t) \label{partline}
\end{equation}

mirroring the completely general transformations of the Galilei line group of \discolorlinks{\ref{linegroup}}

Finally, equation \discolorlinks{\ref{flat2}} also simplifies \discolorlinks{\ref{ad1}} and \discolorlinks{\ref{ad2}} 

\begin{align} \notag
\Gamma^i_{00} &= h^{ij} \Phi_j \\
                &= \delta^{ij} \Phi_j \label{simp1}\\ \notag
\Gamma^i_{0j} &=  h^{ik} (\frac{1}{2}\partial_0 h_{jk} + \Omega_{jk}) \\ \notag
            &= \delta^{ik} (\frac{1}{2}\partial_0 (\delta_{jk}) + \Omega_{jk}) \\ \notag
            &= \delta^{ik} (0 + \Omega_{jk}) \\ 
            &= \delta^{ik} \Omega_{jk} \label{simp2}
\end{align}

These can both be inverted, recalling that $\Omega_{ij}$ is anti-symmetric and, as per \discolorlinks{\ref{notors}}, $\Gamma^\lambda_{\mu\nu}$ is symmetric

\begin{align} 
\Gamma^i_{00} &= \delta^{ij} \Phi_j & \rightarrow && \Phi_j &= \delta_{ij} \Gamma^i_{00} \label{datrel}\\
\Gamma^i_{0j} &= \delta^{ik} \Omega_{jk} & \rightarrow && \Omega_{jk} &= \delta_{i[k}\Gamma^i_{j]0} \label{relation}
\end{align}

\pagebreak

\subsubsection{Trautman condition}
\label{trautsec}

The final two steps needed to recover the results of the semi-geometric picture of Newtonian gravity are to impose additional constraints on the Riemann curvature of a Galilei manifold. By doing so we narrow down the form of the Christoffel connection to recover a single non-vanishing part as in the semi-geometric picture.

The first constraint we impose is the so-called Trautman condition \cite{kuezle}

\begin{equation}
  R^{[\lambda}{}_{(\nu}{}^{\mu]}{}_{\rho)}=  h^{\sigma[\lambda} R^{\mu]}{}_{(\nu\rho)\sigma}=0 \label{traut}
\end{equation}

This can be written out explicitly in 4 terms, and one can use the following properties of the Riemann tensor to show this works just fine in standard GR — where one raises and lowers indices with the (inverse) metric tensor $g_{\mu\nu}$ ($g^{\mu\nu}$)

\begin{align}
    R_{\mu\nu\rho\sigma} &= R_{\rho\sigma\mu\nu} \label{riemsyms} \\
    R_{\mu\nu\rho\sigma} &= - R_{\mu\nu\sigma\rho} = - R_{\nu\mu\rho\sigma} \label{sym2}
\end{align}

For our purposes, we would like to express this Trautman condition in our adapted coordinate system to see what it tells us about the Christoffel connection/curvature.

The first step towards doing so is to express the curvature in \discolorlinks{\ref{traut}} in terms of the connection via \discolorlinks{\ref{reimann}} which is indeed what we know about given equations \discolorlinks{\ref{ad1}} - \discolorlinks{\ref{ad4}}. Skipping straight to $h^{ij}$ — because we know from \discolorlinks{\ref{end}} that $h^{ij}$ is the only non-zero component of $h^{\mu\nu}$ — and noting that Christoffel symbols are symmetric in their lower indices, we have

\begin{equation}
    h^{i[j}\partial_{(\rho} \Gamma^{\mu]}_{\nu) i} - h^{i[j}\partial_{i} \Gamma^{\mu]}_{\nu \rho} + h^{i[j} \Gamma^{\mu]}_{\lambda (\rho} \Gamma^\lambda_{\nu)i} - h^{i[j} \Gamma^{\mu]}_{\lambda i} \Gamma^\lambda_{\nu \rho}=0 \label{reco}
\end{equation}

Since $\Gamma^{i}_{00}$ and $\Gamma^{i}_{0j}$ are the non-trivial terms of \discolorlinks{\ref{ad1}} - \discolorlinks{\ref{ad4}}, we have two choices, $\nu=\rho=0$ or $\mu=k$,$\nu=0$. For $\nu=\rho=0$ we have

\begin{equation}
    h^{i[j}\partial_{(0} \Gamma^{\mu]}_{0) i} - h^{i[j}\partial_{i} \Gamma^{\mu]}_{00} + h^{i[j} \Gamma^{\mu]}_{\lambda (0} \Gamma^\lambda_{0)i} - h^{i[j} \Gamma^{\mu]}_{\lambda i} \Gamma^\lambda_{00}=0 \label{num1}
\end{equation}

Since we only have terms of the form \discolorlinks{\ref{ad1}} - \discolorlinks{\ref{ad4}}, we must have either $\mu$ or $\lambda$ as a spatial. Without loss of generality we choose $\mu=k$,

\begin{equation}
    h^{i[j}\partial_{(0} \Gamma^{k]}_{0) i} - h^{i[j}\partial_{i} \Gamma^{k]}_{00} + h^{i[j} \Gamma^{k]}_{\lambda (0} \Gamma^\lambda_{0)i} - h^{i[j} \Gamma^{k]}_{\lambda i} \Gamma^\lambda_{00} =0
\end{equation}

In the last two terms, again since we only have terms of the form \discolorlinks{\ref{ad1}} - \discolorlinks{\ref{ad4}}, $\lambda$ must be simultaneously temporal and spatial for the terms not to cancel, because the $\Gamma^\lambda_{00}$ parts need a spatial $\lambda$, but the $\Gamma^{k]}_{\lambda i}$ parts need a temporal $\lambda$. So whether we choose $\lambda=0$ or $\lambda=j$, the last two terms cancel either way since the other part of the term will be zero regardless of the choice.

Then plugging in \discolorlinks{\ref{simp1}} and \discolorlinks{\ref{simp2}} to the non-vanishing first two terms,

\begin{align} \notag
    h^{i[j}\partial_{0} (h^{k]j}\Omega_{ij}) - h^{i[j}\partial_{i} (h^{k]j}\Phi_j) &= 0 \\ \notag
    h^{i[j} h^{k]j} \partial_{0}\Omega_{ij} - h^{i[j} h^{k]j} \partial_{i}\Phi_j &= 0\\ \notag
      h^{i[j} h^{k]j} (\partial_{0}\Omega_{ij} -  \partial_{i}\Phi_j) &= 0\\ 
      \partial_{0}\Omega_{ij} -  \partial_{[i}\Phi_{j]} &= 0 \label{traut1}
\end{align}
where the anti-symmetry in the 2nd term of the last line comes from the anti-symmetry of $\Omega_{ij}$.

The same process but for $\mu=k$ and $\nu=0$ yields

\begin{align} \notag
    h^{i[j}h^{k]\ell}(\partial_j \Omega_{i\ell} - \partial_i \Omega_{j \ell}) &=0 \\
    \partial_{[j} \Omega_{i\ell]} &=0 \label{traut2}
\end{align}
where the anti-symmetry of $i$ and $\ell$ in the first term, and the anti-symmetry of $j$ and $\ell$ in the second term imply the full anti-symmetry expressed in the final line.

Finally, we can use the definitions of $\Omega_{ij}$ and $\Phi_i$ implicit in \discolorlinks{\ref{ad1}} and \discolorlinks{\ref{ad2}}

\begin{align}
    \Omega_{i\ell} &= \partial_{[i} h_{\ell]0} - \frac{1}{2} K_{i\ell} \\
    \Phi_k &= \partial_0 h_{k0} - \frac{1}{2}\partial_k h_{00} + K_{k0}
\end{align}

and plug them into \discolorlinks{\ref{traut1}} and \discolorlinks{\ref{traut2}} to obtain a constraint on the arbitrary (and anti-symmetric by virtue of $\Omega$ being anti-symmetric) two-form $K_{\mu\nu}$.

Noting that the symmetry of mixed second derivatives makes the term $\partial_{[j}\partial_i h_{\ell]}$ vanish, we have from \discolorlinks{\ref{traut2}}

\begin{equation}
    \partial_{[j} K_{i \ell]}=0 \label{dat1}
\end{equation}

Next, \discolorlinks{\ref{traut1}} gives (again noting mixed partials symmetry)

\begin{align} \notag
    \partial_0 \partial_{[i} h_{k]0} - \frac{1}{2} \partial_0 K_{ik} -\partial_0 \partial_{[i} h_{k]0} -\frac{1}{2} \partial_{[i} \partial_{k]} h_{00} + \partial_{[i}K_{k]0} &=0 \\
   - \frac{1}{2} \partial_0 K_{ik}  + \partial_{[i}K_{k]0} &=0
\end{align}

The second term here tells us that (since $K_{\mu\nu}$ is anti-symmetric) not only are $i$ and $k$ anti-symmetric, but so are $k$ and $0$. Thus

\begin{equation}
    \partial_{[i}K_{k0]} =0 \label{dat2}
\end{equation}

Putting \discolorlinks{\ref{dat1}} and \discolorlinks{\ref{dat2}} together we have generally

\begin{equation}
    \partial_{[\rho}K_{\mu\nu]} =0 \label{auxillary}
\end{equation}

Then we can mirror what we did in \discolorlinks{\ref{curl}} and use the symmetry of mixed partials to solve \discolorlinks{\ref{auxillary}} for $K_{\mu\nu}$

\begin{equation}
    K_{\mu\nu} = 2\partial_{[\mu}m_{\nu]} \label{tbd}
\end{equation}

for some vector field $m_\mu$, and where the 2 accommodates the necessary anti-symmetrization on the RHS.

This adapted coordinates-version of the Trautman condition is what we will reproduce when we gauge the Bargmann.

\pagebreak
\subsubsection{Ehlers conditions}

The second of the two steps needed to narrow down the form of the Christoffel connection is to impose the so-called Ehlers conditions.

As with the Trautman, the Ehlers conditions (all three are equivalent) are constraints on the Riemann curvature,

\begin{align}
    h^{\sigma[\lambda} R^{\mu]}{}_{\nu\rho\sigma}(\Gamma)&=0 \label{ehlers}\\ 
    h^{\rho\lambda} R^{\mu}{}_{\nu\rho\sigma}R^{\nu}{}_{\mu\lambda\xi}(\Gamma)&=0\\
    \tau_{[\lambda} R^{\mu}{}_{\nu]\rho\sigma}(\Gamma)&=0
\end{align}

As a sanity check, one can use the same symmetries of the Riemann tensor listed in \discolorlinks{\ref{riemsyms}} and \discolorlinks{\ref{sym2}} as well as the symmetry of the Ricci tensor $R_{\mu\nu}=R_{\nu\mu}$ to show that the Ehlers condition holds for the Riemann tensor standard in GR.

To use the Ehlers condition to learn about the form of the Christoffel connection, we follow the same step as in the Trautman case in section \ref{trautsec}. These steps go as follows:

\begin{enumerate}
    \item Express the curvature in the Ehlers condition in terms of the Christoffel connection using \discolorlinks{\ref{reimann}}, noting that  we know from \discolorlinks{\ref{end}} that $h^{ij}$ is the only non-zero component of $h^{\mu\nu}$. This leaves us with an Ehlers version of equation \discolorlinks{\ref{reco}}.
    \item Use the form of the various parts of the Christoffel connection in adapted coordinates (\discolorlinks{\ref{ad1}} - \discolorlinks{\ref{ad4}}) to reduce the resulting equation of the first step. This mimics the process of \discolorlinks{\ref{num1}}-\discolorlinks{\ref{traut2}}
\end{enumerate}

The result of taking these steps is the following constraint on the so-called Coriolis field $\Omega_{ij}$

\begin{equation}
    \partial_k \Omega_{ij} =0
\end{equation}

which can be read verbally as, ``the spatial derivative of that object is zero, and so it cannot depend on space. Thus it most be entirely time-dependent, or zero." 

For the moment we give the benefit of the doubt and let $\Omega_{ij}$ be time-dependent, $\Omega_{ij}(t)$. By the relation  \discolorlinks{\ref{relation}}, this tells us that equivalently, the $\Gamma^i_{j0}$ component of the connection is time-dependent.

We will show how to set it to zero in the next section and look at the implications of doing so.

\pagebreak
\subsubsection{Time-dependent rotations and their implications for the Ehlers and Trautman condition}

The time-dependent coordinate transformations of \discolorlinks{\ref{partline}} give us a way to set the time-dependent component of the connection ($\Gamma^i_{j0}$ from the previous section) to zero.

In particular, by using a particular for of \discolorlinks{\ref{partline}}, notably just a time-dependent rotation as opposed to including the spatial translations
\begin{equation}
    x'{}^i = R^i{}_j(t) x^j
\end{equation}
Using this time-dependent transformation, we are free to transform to a coordinate system in which the purely time-dependent object $\Gamma^i_{j0}=0$ . Of course via the relation \discolorlinks{\ref{relation}}, this tells us equivalently that
\begin{equation}
    \Omega_{ij}=0 \label{alter}
\end{equation}
Looking back to \discolorlinks{\ref{traut1}}, by imposing \discolorlinks{\ref{alter}} we alter the resulting restriction on the Trautman condition (in adapted coordinates). In other words, equation \discolorlinks{\ref{traut1}} reduces to 
\begin{equation}
    \partial_{[i}\Phi_{j]} = 0
\end{equation}
where we can once again use the symmetry of mixed 2nd partials to solve this for $\Phi_i$,
\begin{equation}
    \Phi_i = \partial_i \Phi  \hspace{1cm} \text{for some scalar} \hspace{0.3cm}\Phi \label{phi}
\end{equation}
Now via the relation \discolorlinks{\ref{datrel}}, we recover the final non-vanishing part of the Christoffel connection 

\begin{equation}
    \Gamma^i_{00} =\delta^{ij} \partial_j \Phi \label{res1}
\end{equation}

Using \discolorlinks{\ref{reimann}}, we have the only term of the Riemann tensor

\begin{equation}
   R^i{}_{0j0} = \partial_j \Gamma^i_{00} = \delta^{ik} \partial_k \partial_j \Phi \label{res2}
\end{equation}

and finally the first and third index of the Riemann tensor \discolorlinks{\ref{res2}} can be contracted to get the Ricci tensor, which using Poisson's equation \discolorlinks{\ref{poss}}, reads

\begin{equation}
    R^i{}_{0i0} = \partial_i \Gamma^i_{00} = \delta^{ij} \partial_i \partial_j \Phi = \nabla^2 \Phi = 4\pi \rho \label{res3}
\end{equation}

\textbf{The equations \discolorlinks{\ref{res1}}, \discolorlinks{\ref{res2}}, and \discolorlinks{\ref{res3}} match the calculations of the semi-geometric picture — \discolorlinks{\ref{Tcon}}, \discolorlinks{\ref{riem}}, and \discolorlinks{\ref{ricci}} respectively} — where we interpret the suggestively named $\Phi$ of equation \discolorlinks{\ref{phi}} above as the gravitational potential $\Phi$.

\pagebreak

\subsubsection{Key components of Newton-Cartan}

Here we summarize aspects of Newtonian gravity in the Newton-Cartan picture.

\begin{table}[H]
\centering 
\renewcommand{\arraystretch}{1.3}
\begin{tabular}{l  l}
\toprule
Degenerate metric(s) & $\begin{aligned} \MyLeftColumn{\tau_{\mu\nu}} &= \text{diag}(1,0,0,0) \\ 
\MyLeftColumn{h^{\mu\nu}} & {}= \text{diag} (0,1,1,1) \\
\MyLeftColumn{\tau_{\mu\rho}h^{\rho\nu}} & {}= 0 \end{aligned}$ \\ 
\midrule
Vielbeine & $\begin{aligned} \MyLeftColumn{e_\mu{}^a} &= \{e_\mu{}^0=\tau_\mu, e_\mu{}^i\} \\ 
\MyLeftColumn{\tau^{\mu}\tau_{\mu}} &= 1 \\ 
\MyLeftColumn{e_\mu{}^{i} e^\mu{}_j} &= \delta^i_j \\ 
\MyLeftColumn{h^{\mu\nu}} & {}= e^\mu{}_i e^\nu{}_j \delta^{ij} \\
\MyLeftColumn{\tau_{\mu\nu}} & {}= \tau_\mu \tau_\nu \\
\MyLeftColumn{e_\mu{}^{i} e^\nu{}_i} & {}= \delta^\nu_\mu -\tau_\mu \tau^\nu \\
\MyLeftColumn{e^\mu{}_i\tau_{\mu}} & {}= 0 \\
\MyLeftColumn{e_\mu{}^i\tau^{\mu}} & {}= 0
\end{aligned}$ \\ 
\midrule
Connection & $\MyLeftColumn{\Gamma^i_{00}} = \partial_i \Phi(x)$ \\  
\midrule
Riemann curvature & $\MyLeftColumn{R^i{}_{0j0}} = \partial_i \partial_j \Phi(x)$  \\  
\midrule
Equations of motion & $\MyLeftColumn{R_{00}} = 4 \pi \rho$ \\
\midrule
Covariant equations of motion & $\MyLeftColumn{R_{\mu\nu}} = 4 \pi \rho \tau_\mu \tau_\nu$ \\
\midrule
Trautman condition & $\MyLeftColumn{h^{\sigma[\lambda} R^{\mu]}{}_{(\nu\rho)\sigma}} = 0$ \\
\midrule
Ehlers conditions & $\begin{aligned} \MyLeftColumn{\tau_{[\lambda} R^{\mu}{}_{\nu]\rho\sigma}} &= 0 \\ 
\MyLeftColumn{h^{\rho\lambda} R^{\mu}{}_{\nu\rho\sigma}R^{\nu}{}_{\mu\lambda\xi}} &= 0 \\ 
\MyLeftColumn{h^{\sigma[\lambda} R^{\mu]}{}_{\nu\rho\sigma}} &= 0
\end{aligned}$ \\ 
\bottomrule
\end{tabular}
\end{table}

\iffalse
In the adapted coordinates \discolorlinks{\ref{adapt}}, we have

\begin{table}[H]
\centering 
\renewcommand{\arraystretch}{1.3}
\begin{tabular}{l  l}
\toprule
Adapted coordinates & $\MyLeftColumn{\tau_\mu} &= \partial_\mu (t) = \delta_\mu^0$ \\ 
\midrule
Covariant EoM & $\MyLeftColumn{R_{\mu\nu}} = 4 \pi \rho \tau_\mu \tau_\nu$ \\
\midrule
Flat spatial slices & $\begin{aligned}
         \MyLeftColumn{R_{ij}} &= 0 \\
         \MyLeftColumn{h^{ij}} &= \delta^{ij}
\end{aligned}$ \\
\bottomrule
\end{tabular}
\end{table}
\fi

\pagebreak
\subsection{Gauging the Bargmann}
\label{gaugebarg}

\subsubsection{Generators, gauge fields, and parameters}

Motivated by the breaking up of the Poincare generators as in equations \discolorlinks{\ref{breakup1}} and \discolorlinks{\ref{breakup2}}, we will break up the $e_\mu{}^a$ and $\omega_\mu{}^{ab}$ gauge fields that we used in gauging the Poincaré $P_a$/$M_{ab}$ transformations into ``temporal" and ``spatial" parts as follows

\begin{align}
e_\mu{}^a &\rightarrow (e_\mu{}^0 = \tau_\mu, e_\mu{}^i)    \label{g1}\\
\omega_\mu{}^{ab} &\rightarrow (\omega_\mu{}^{0i} = \omega_\mu{}^i, \omega_\mu{}^{ij})   \label{spinsplit}
\end{align}

where the ``  " are in reference to the fact that the $a,b$ indices are totally abstract internal vector space indices. Of course for the 4-dimensional Poincaré algebra $\mathfrak{iso}(4-1=3,1)$, we have $a,b=\{0,1,2,3\}$, but the identification of $a,b=0$ with ``temporal" is just to facilitate that (in hindsight) we know  we will be constructing a physcial theory from the algebra.

Further, we break up the the local parameters we associated to the Poincaré $P_a$/$M_{ab}$ transformations, $\xi^a(x)$ and $\lambda^{ab}(x)$ in a similar fashion

\begin{align}
\xi^a &\rightarrow (\xi^0 = \tau, \xi^i)    \label{g2}\\
\lambda^{ab} &\rightarrow (\lambda^{0i} = \lambda^i, \lambda^{ij})   
\end{align}

The Bargmann algebra (\discolorlinks{\ref{fbarg1}}-\discolorlinks{\ref{fbarg5}}) has 5 species of generators: H, $P_i$, $K_i$, $J_{ij}$, and M. In addition to the 4 gauge fields of \discolorlinks{\ref{g1}} and \discolorlinks{\ref{g2}}, we have a gauge field $m_\mu$ that we associate to the central generator $M$. We will called the local parameter of the M generator $\sigma(x)$. The rest of the generators H, $P_i$, $K_i$, and $J_{ij}$ have gauge fields and gauge parameters as follows:

\begin{center}
\large
\renewcommand{\arraystretch}{1.5}
\begin{tabular}{l l l}
\hline
Transformation \hspace{0.5cm} & Gauge field \hspace{1cm} & Local parameters\\ 
\hline
H & $\tau_\mu$ & $\tau(x)$ \\ 
\hline
$P_i$ & $e_\mu{}^i$ & $\xi^i(x)$ \\ 
\hline
$K_i$ & $\omega_\mu{}^i$ & $\lambda^i(x)$\\
\hline
$J_{ij}$ & $\omega_\mu{}^{ij}$ &  $\lambda^{ij}(x)$\\  
\hline
M & $m_\mu$ & $\sigma(x)$ \\
\hline
\end{tabular}
\end{center}

\pagebreak

\subsubsection{Gauge field transformations and curvatures}

Writing the transformations and curvatures for each gauge field follows how we derived equations \discolorlinks{\ref{gtvielbein}} and \discolorlinks{\ref{Re}} on pages \pageref{split} and \pageref{gtcurv}, but is slightly more complicated because of the 5 species of generators for Bargmann as opposed to the 2 of Poincaré. So we will introduce some notation that help keep track of things when working through the gauge theory calculations.

\begin{center}
\large
\renewcommand{\arraystretch}{1.5}
\begin{tabular}{l l l}
\hline
Transformation \hspace{0.5cm} & Gauge field \hspace{1cm} & Local parameters\\ 
\hline
H & $\tau_\mu\hspace{1ex}{}^{\_}$ & $\tau\hspace{1ex}{}^{\_}$ \\ 
\hline
$P_i$ & $e_\mu{}^i$ & $\xi^i$ \\ 
\hline
$K_i$ & $\omega_\mu{}^{\underline{i}}$ & $\lambda^{\underline{i}}$\\
\hline
$J_{ij}$ & $\omega_\mu{}^{ij}$ &  $\lambda^{ij}$\\  
\hline
M & $m_\mu^\square$ & $\sigma^\square$ \\
\hline
\end{tabular}
\end{center}

where these underscores, underlines, and square boxes are helpful in writing out the structure constants of the algebra. Note: admittedly, these structure constants (other than those explicitly based on those of the Poincaré) were constructed with the correct answers for the gauge field transformations/curvatures in-hand. And so, the rigour-minded reader ought to do either (or both) of the following: solve exercise 1.4 of Freedman-Van Proeyen or more generally learn how to turn any algebra's commutation relations into structure constants; or derive (if possible) an alternative expression for \discolorlinks{\ref{transgauge}} in terms of the the algebra commutation relations instead of the structure constants.

\begin{align} 
\large
\notag
&\text{Commutation relations} & &\text{Structure constants} \\
[K_i,H] &= P_i & f_{\underline{i}\hspace{0.25ex}\underline{\hspace{0.5ex}}}{}^j&=-\epsilon_{\underline{i}\hspace{0.25ex}\underline{\hspace{0.5ex}}}{}^j \\
[K_i,J_{jk}] &= 2 \delta_{i[j}K_{k]} & f_{\underline{i},[jk]}{}^{\underline{\ell}}&= 2\delta_{\underline{i}[j}\delta_{k]}{}^{\underline{\ell}} \\
[P_i,J_{jk}] &= 2 \delta_{i[j}P_{k]}  & f_{i,[jk]}{}^{\ell}&= 2\delta_{i[j}\delta_{k]}{}^{\ell} \\
[J_{ij},J_{k\ell}] &= 4\delta_{[i[k}J_{\ell]j ]}  & f_{[ij],[k\ell]}{}^{[mn]} &= 8 \delta_{[k[j}\delta_{i]}^{[m}\delta_{\ell]}^{n]}\\
[K_i,P_j] &= \delta_{ij} M  & f_{\underline{i}j}{}^\square &= -\epsilon_{\underline{i}j}{}^\square
\end{align}

Recall the antisymmetry rule of \discolorlinks{\ref{antisymedit}} as we proceed. We will derive two of the gauge field transformations and leave the rest to the reader. The derivations follow the procedure performed in deriving equations \discolorlinks{\ref{gtvielbein}} and \discolorlinks{\ref{Re}} on pages \pageref{split} and \pageref{gtcurv}.

\begin{align} \notag
\delta(\epsilon) B_\mu{}^a &= \partial_\mu \epsilon^a + \epsilon^c B_\mu{}^b f_{bc}{}^a  \\ \notag
\delta(\xi) e_\mu{}^i &= \partial_\mu \xi^i + \epsilon^k \tau_\mu{}^\_ f_{\underline{\hspace{0.7ex}} \hspace{0.25 ex} k}{}^i + \epsilon^k \epsilon_\mu{}^j f_{jk}{}^i + \epsilon^k \omega_\mu{}^{\underline{j}} f_{\underline{j}k}{}^i + \epsilon^k \omega_\mu{}^{ij} f_{[ij],k}{}^i 
\\ \notag
&= \partial_\mu \xi^i + \lambda^{\underline{k}} \tau_\mu{}^\_ f_{\underline{\hspace{0.7ex}} \hspace{0.25 ex} \underline{k}}{}^i + \lambda^{ij} \epsilon_\mu{}^j f_{j,[ij]}{}^i + \tau\hspace{0.5ex}^{\_} \omega_\mu{}^{\underline{j}} f_{\underline{j} \hspace{0.25 ex} \underline{\phantom{j}\hspace{0.1ex}} }{}^i + \xi^k \omega_\mu{}^{ij} f_{[ij],k}{}^i \\ \notag
&= \partial_\mu \xi^i + \lambda^{\underline{k}} \tau_\mu{}^\_ (\epsilon_{\underline{\hspace{0.7ex}} \hspace{0.25 ex} \underline{k}}{}^i) + \lambda^{ij} \epsilon_\mu{}^j (\delta_{ji}\delta_j^i) + \tau\hspace{0.5ex}^{\_} \omega_\mu{}^{\underline{j}} (-\epsilon_{\underline{j} \hspace{0.25 ex} \underline{\phantom{j}\hspace{0.1ex}}}{}^i) + \xi^k \omega_\mu{}^{ij} (-\delta_{ki} \delta_j{}^i) 
\\ \notag
&= \partial_\mu \xi^i + \lambda^{\underline{k}} \tau_\mu{}^\_ (\delta_{\underline{k}}{}^i) + \lambda^{ij} \epsilon_\mu{}^j (\delta_j{}^j = 1) + \tau\hspace{0.5ex}^{\_} \omega_\mu{}^{\underline{j}} (-\delta_{\underline{j}}{}^i) + \xi^k \omega_\mu{}^{ij} (-\delta_{kj}) \\ \notag
&= \partial_\mu \xi^i + \lambda^{i} \tau_\mu{}^\_  + \lambda^{ij} \epsilon_\mu{}^j  - \tau\hspace{0.5ex}^{\_} \omega_\mu{}^{i}  - \xi^j \omega_\mu{}^{ij}  \\ 
&= \partial_\mu \xi^i + \lambda^{i} \tau_\mu  + \lambda^{ij} \epsilon_\mu{}^j  - \tau \omega_\mu{}^{i}  - \xi^j \omega_\mu{}^{ij}  
\end{align}

\begin{align} \notag
\delta(\epsilon) B_\mu{}^a &= \partial_\mu \epsilon^a + \epsilon^c B_\mu{}^b f_{bc}{}^a 
\\ \notag
\delta(\sigma) m_\mu &= \partial_\mu \sigma^\square + \epsilon^k \epsilon_\mu{}^i f_{ik}{}^\square + \epsilon^k \omega_\mu{}^{\underline{j}} f_{\underline{j} k}{}^\square
\\ \notag
&= \partial_\mu \sigma^\square + \lambda^{\underline{j}} \epsilon_\mu{}^i f_{i\underline{j}}{}^\square + \xi^i \omega_\mu{}^{\underline{j}} f_{\underline{j} i}{}^\square
\\ \notag
&= \partial_\mu \sigma^\square + \lambda^{\underline{j}} \epsilon_\mu{}^i (\epsilon_{i\underline{j}}{}^\square) + \xi^i \omega_\mu{}^{\underline{j}} (-\epsilon_{\underline{j} i}{}^\square)
\\ \notag
&= \partial_\mu \sigma^\square + \lambda^{\underline{j}} \epsilon_\mu{}^i (\epsilon_{\underline{j}\square}{}^i) + \xi^i \omega_\mu{}^{\underline{j}} (-\epsilon_{\square \underline{j}}{}^i)
\\ \notag
&= \partial_\mu \sigma^\square + \lambda^{\underline{j}} \epsilon_\mu{}^i (\delta_{\underline{j}}{}^i) + \xi^i \omega_\mu{}^{\underline{j}} (-\delta_{\underline{j}}{}^i)
\\ 
&= \partial_\mu \sigma + \lambda^{i} \epsilon_\mu{}^i - \xi^i \omega_\mu{}^{i} 
\end{align}

All in all, the rest of the pure gauge theory calculations for the transformations and the curvatures (where we label the curvatures according to the more describe symmetry transformations as opposed to the gauge fields which would now be ambiguous) read

\begin{align}
    \delta(\tau) \tau_\mu &= \partial_\mu \tau \label{tran1} \\
    \delta(\xi) e_\mu{}^i &= \partial_\mu \xi^i + \lambda^{i} \tau_\mu  + \lambda^{ij} \epsilon_\mu{}^j  - \tau \omega_\mu{}^{i}  - \xi^j \omega_\mu{}^{ij} \\
    \delta(\lambda^i) \omega_\mu{}^i &= \partial_\mu \lambda^i - \lambda^j \omega_\mu{}^{ij} + \lambda^{ij} \omega_\mu{}^j \\
    \delta(\lambda^{ij}) \omega_\mu{}^{ij} &= \partial_\mu \lambda^{ij} + 2\lambda^{k[i} \omega_\mu{}^{j]k} \\
    \delta(\sigma) m_\mu &= \partial_\mu \sigma + \lambda^{i} \epsilon_\mu{}^i - \xi^i \omega_\mu{}^{i} \label{tran5}
\end{align}

\begin{align}
R_{\mu\nu}(H) &= 2 \partial_{[\mu} \tau_{\nu]} \label{Rh}\\
R_{\mu\nu}{}^i(P) &= 2(\partial_{[\mu} e_{\nu]}{}^i - \omega_{[\mu}{}^{ij}e_{\nu]}{}^j - \omega_{[\mu}{}^i \tau_{\nu]})\\
R_{\mu\nu}{}^i(K) &= 2(\partial_{[\mu} \omega_{\nu]}{}^i - \omega_{[\mu}{}^{ij}\omega_{\nu]}{}^j) \label{Rk}\\
R_{\mu\nu}{}^{ij}(J) &= 2(\partial_{[\mu} \omega_{\nu]}{}^{ij} - \omega_{[\mu}{}^{ki}\omega_{\nu]}{}^{jk}) \label{Rj}\\
R_{\mu\nu}(M) &= 2(\partial_{[\mu} m_{\nu]} - \omega_{[\mu}{}^i e_{\nu]}{}^i) \label{Rm}
\end{align}

\pagebreak

\subsubsection{Conventional constraints, geometric constraints, and Bianchi identities}
\label{convconstrs}

Just as we did with Poincaré and GR, we proceed to soldering/removing the local translations (where now we have two translation generators, temporal H and spatial P). 

One should notice that while we motivated the soldering in the Poincaré GR case with geometry, we called the constraint we imposed a ``conventional constraint" which we define as a constraint which allows us to derive an expression for the spin connection in terms of the vielbein — a hallmark of gravity. In the Bargmann Newton-Cartan case, we will have \textit{both} conventional and true/geometric constraints, which we define as a constraint we place on the gauge theory side that allows us to build up the geometry we need on the gravity side.

Notably, the constraint on the temporal translations we impose,
\begin{equation}
    R_{\mu\nu}(H) = 0 \label{geoconstr}
\end{equation}
is a geometric constraint because it enables us to reproduce a key component of Newton-Cartan geometry. To see how, we plug \discolorlinks{\ref{Rh}} into  \discolorlinks{\ref{geoconstr}} 
\begin{align}
    \notag
    R_{\mu\nu}(H) &= 0 \\ \notag
    2 \partial_{[\mu} \tau_{\nu]} &= 0 \\
    \partial_{[\mu} \tau_{\nu]} &= 0
\end{align}
This is precisely the zero torsion condition of equation \discolorlinks{\ref{vanishingcurl}} which enabled us to impose causality and reproduce the absolute time of Newtonian gravity. Thus, this constraint of \discolorlinks{\ref{geoconstr}} is indeed geometric in the sense that it reproduces from gauge theory one of the features of the geometry we wish to work with.

It is clear that to remove the local (spatial and temporal) translations, we ought to impose 
\begin{equation}
    R_{\mu\nu}{}^i(P) = 0 \label{Rp0}
\end{equation}

as well as equation \discolorlinks{\ref{geoconstr}}. However that is not that only additional vanishing curvature constraint we need to impose. Note that as per \discolorlinks{\ref{spinsplit}}, we now have two spin connections. Thus we have two unknowns, which we would like to solve for in terms of the independent fields. We need two conventional constraints to solve for these two unknowns.

To see what other conventional constraint we could possibly impose, we browse the gauge transformations of equations \discolorlinks{\ref{tran1}} — \discolorlinks{\ref{tran5}} and look for any terms indicative of the particular gauge field transforming with P or H translations (i.e. a term containing a parameter $\xi^i$ or $\tau$). In this way it is apparent that not only do the $\tau_\mu$ and $e_\mu{}^i$ gauge fields transform under P and H translations, but $m_\mu$ does as well. Thus we also impose 
\begin{equation}
    R_{\mu\nu}(M) = 0 \label{Rm0}
\end{equation}

For convincing that the constraints \discolorlinks{\ref{Rp0}} and \discolorlinks{\ref{Rm0}} are indeed conventional, i.e. enable us to solve for both spin connections $\omega_\mu{}^i$ and $\omega_\mu{}^{ij}$, see page 70 of \cite{Roelthesis} for instructions on the derivation. Because our previous derivation of the dependent spin connection from a conventional constraint (\discolorlinks{\ref{depomega}}) was so long, we will simply state the answers and leave the calculation to those readers with access to a chalkboard.

From \discolorlinks{\ref{Rp0}}, one obtains

\begin{equation}
    \omega_\mu{}^{ij} = - 2 e^{\nu[i} \partial_{[\mu}e_{\nu]}{}^{j]} + e_{\mu k} e^{\nu i} e^{\rho j} \partial_{[\nu} e_{\rho]}{}^k - \tau_\mu e^{\rho[i} \omega_\rho{}^{j]} \label{depspin1}
\end{equation}

Using \discolorlinks{\ref{depspin1}} with \discolorlinks{\ref{Rp0}}, as well as \discolorlinks{\ref{Rm0}}, one obtains

\begin{equation}
    \omega_\mu{}^i = e^{\nu i} \partial_{[\mu} m_{\nu]} + e_{\mu j} e^{\nu i} \tau^\rho  \partial_{[\nu} e_{\rho]}{}^j + \tau_\mu \tau^\nu e^{\rho i} \partial_{[\nu}m_{\rho]} + \tau^\nu \partial_{[\mu} e_{\nu]}{}^i \label{depspin2}
\end{equation}

So our two spin connections $\omega_\mu{}^{ij}$ and $\omega_\mu{}i$ are dependant gauge fields, written above in terms of the 3 other independent gauge fields, $\tau_\mu$, $e_\mu{}^i$, and $m_\mu$.

\vspace{1cm}
Before moving on to recovering NC gravity, a final set of constraints on the gauge curvatures exists (as opposed to being imposed manually like those above) via the (differential/second) Bianchi identity. In GR, this Bianchi identity reads

\begin{equation}
    \nabla_{[\lambda} R_{\mu\nu]\rho\sigma}(\Gamma) = 0
\end{equation}

where $\nabla_\lambda$ is the covariant (with respect to manifold GCTs) derivative of GR ($\nabla_\lambda=\partial_\lambda - \Gamma$) and the curvature $R_{\mu\nu\rho\sigma}$ is constructed from the Christoffel connection $\Gamma$ of GR.

We can also use this Bianchi identity in gauge theory 
\begin{equation}
    D_{[\lambda} R_{\mu\nu]}(A) = 0
\end{equation}
where $D_\lambda$ is the covariant (with respect to fiber LLTs) derivative of gauge theory ($D_\lambda=\partial_\lambda - \omega_\mu{}^{ij}$) and the curvature $R_{\mu\nu}$ is constructed from each respective gauge field $A$ of the algebra. Thus for the Bargmann's 5 species of generators, we have 5 Bianchi identities. They are, from page 141 of \cite{Roelthesis}

\begin{align}
    D_{[\lambda}R_{\mu\nu]}(H) &= 0 \label{bRh}\\
    D_{[\lambda}R_{\mu\nu]}{}^i(P) = 0 \rightarrow D_{[\lambda}R_{\mu\nu]}{}^i(P) &= -R_{[\lambda\mu}{}^{ij}(J)e_{\nu]j} -R_{[\lambda\mu}{}^{i}(K)\tau_{\nu]}+ R_{[\lambda\mu}(H)\omega_{\nu]}{}^i \label{pRHS}\\
    D_{[\lambda}R_{\mu\nu]}{}^{i}(K) = 0 \rightarrow D_{[\lambda}R_{\mu\nu]}{}^{i}(K) &= -R_{[\lambda\mu}{}^{ij}(J)\omega_{\nu]j} \label{kRHS}\\
    D_{[\lambda}R_{\mu\nu]}{}^{ij}(J) &= 0 \\
    D_{[\lambda}R_{\mu\nu]}(M) = 0 \rightarrow D_{[\lambda}R_{\mu\nu]}(M)&= R_{[\lambda\mu}{}^{i}(P)\omega_{\nu]i} -R_{[\lambda\mu}{}^{i}(K)e_{\nu]i} \label{mRHS}
\end{align}

where \discolorlinks{\ref{pRHS}}, \discolorlinks{\ref{kRHS}}, and \discolorlinks{\ref{mRHS}} are expressed as such by taking the terms that cancel to zero, and grouping them in terms of the other curvatures of the theory on the other side of the equation. Doing so gives us additional structure by relating different curvatures. Notably, even after imposing the constraints \discolorlinks{\ref{geoconstr}} — \discolorlinks{\ref{Rm0}} above, we are still left with the following relations via \discolorlinks{\ref{pRHS}} and \discolorlinks{\ref{mRHS}}

\begin{align}
    R_{[\lambda\mu}{}^{ij}(J)e_{\nu]j} &= -R_{[\lambda\mu}{}^{i}(K)\tau_{\nu]} \label{bianchi1}\\
    R_{[\lambda\mu}{}^{i}(K)e_{\nu]i} &= 0 \label{bianchi2}
\end{align}

The derivations of these Bianchi identities \discolorlinks{\ref{bRh}} — \discolorlinks{\ref{mRHS}} are massive and are more suited to those readers with a chalkboard. We will very briefly start one of the derivations (\discolorlinks{\ref{kRHS}}) here to give the reader a feel for what is required. 

First, note that all curvatures \discolorlinks{\ref{Rh}} — \discolorlinks{\ref{Rm}} are anti-symmetric, so 
\begin{align}
    D_{[\lambda}R_{\mu\nu]} = \frac{1}{6}(D_\lambda R_{\mu\nu}+D_\mu R_{\nu\lambda}+D_\nu R_{\lambda\mu}-D_\lambda R_{\nu\mu}-D_\mu R_{\lambda\nu}-D_\nu R_{\mu\lambda})
\end{align}
becomes
\begin{align}
    D_{[\lambda}R_{\mu\nu]} = \frac{1}{3}(D_\lambda R_{\mu\nu}+D_\mu R_{\nu\lambda}+D_\nu R_{\lambda\mu}) \label{supsimp1}
\end{align}

for any of the gauge curvatures.

Secondly, note that by the symmetry of mixed second derivatives ($\partial_\mu \partial_\nu= \partial_\nu \partial_\mu$), when the partial derivative of the covariant derivative hits the first term (derivative of the local parameter) of any of the curvatures, it vanishes. For example
\begin{align}
    D_{[\lambda}R_{\mu\nu]}(H) &= \partial_\lambda(\partial_\mu \tau_\nu-\partial_\nu \tau_\mu) + \partial_\mu(\partial_\nu \tau_\lambda-\partial_\lambda \tau_\nu) + \partial_\nu (\partial_\lambda \tau_\mu-\partial_\mu \tau_\lambda) \label{supsimp2} \\ \notag
    &{} \quad -\omega_\lambda{}^{ij}(2 \partial_{[\mu} \tau_{\nu]}) -\omega_\mu{}^{ij}(2 \partial_{[\nu} \tau_{\lambda]}) -\omega_\nu{}^{ij}(2 \partial_{[\lambda} \tau_{\mu]}) 
\end{align}
where all the first 6 terms cancel.

\pagebreak
Onto \discolorlinks{\ref{kRHS}}. Expanding and writing out the RHS (only up to the first term of the anti-symmetrization) explicitly,

\begin{align}
    -R_{[\lambda\mu}{}^{ij}(J)\omega_{\nu]}{}^j &= -R_{\lambda\mu}{}^{ij}(J)\omega_\nu{}^j - \dots \\ \notag
    &= {\color{red}- (\partial_\lambda \omega_\mu{}^i{}_j)\omega_\nu{}^j} {\color{orange}+ (\partial_\mu \omega_\lambda{}^i{}_j)\omega_\nu{}^j} {\color{green}+ (\omega_{\lambda k}{}^i\omega_{\mu j}{}^k)\omega_\nu{}^j} {\color{blue}- (\omega_{\mu k}{}^i\omega_{\lambda j}{}^k)\omega_\nu{}^j} - \dots
\end{align}

And now expanding and writing out the LHS (up to the second term of the anti-symmetrization) explicitly, while keeping \discolorlinks{\ref{supsimp1}} and \discolorlinks{\ref{supsimp2}} in mind

\begin{align}
    D_{[\lambda} R_{\mu\nu]}{}^i (K) &= D_{\lambda} R_{\mu\nu}{}^i +D_\mu R_{\nu\lambda]}{}^i + \dots \\ \notag
    &= [\{\partial_\lambda R_{\mu\nu}{}^i)\} - \{\omega_\lambda{}^i{}_j R_{\mu\nu}{}^j\}] + [\{\partial_\mu R_{\nu\lambda}{}^i\} - \{\omega_\mu{}^i{}_j R_{\nu\lambda}{}^j\}] + \dots \\ \notag
    &=  [\{-(\partial_\lambda \omega_\mu{}^i{}_j)\omega_\nu{}^j + (\partial_\lambda \omega_\nu{}^i{}_j)\omega_\mu{}^j -(\partial_\lambda \omega_\nu{}^j)\omega_\mu{}^i{}_j + (\partial_\lambda \omega_\mu{}^j)\omega_\nu{}^i{}_j \} \\ \notag
    &{} \hspace{0.7 cm} 
    \{ -(\partial_\mu \omega_\nu{}^j)\omega_\lambda{}^i{}_j + (\partial_\nu \omega_\mu{}^j)\omega_\lambda{}^i{}_j +\underline{\omega_\mu{}^j{}_k \omega_\nu{}^k \omega_\lambda{}^i{}_j} - \omega_\nu{}^j{}_k \omega_\mu{}^k \omega_\lambda{}^i{}_j\} ]\\[1ex] \notag
    &{} \hspace{0.6 cm} [ \{-(\partial_\mu \omega_\nu{}^i{}_j)\omega_\lambda{}^j + (\partial_\mu \omega_\lambda{}^i{}_j)\omega_\nu{}^j -(\partial_\mu \omega_\lambda{}^j)\omega_\nu{}^i{}_j + (\partial_\mu \omega_\nu{}^j)\omega_\lambda{}^i{}_j \}\\ \notag
    &{} \hspace{0.7 cm} \{-(\partial_\nu \omega_\lambda{}^j)\omega_\mu{}^i{}_j + (\partial_\lambda \omega_\nu{}^j)\omega_\mu{}^i{}_j +\omega_\nu{}^j{}_k \omega_\lambda{}^k \omega_\mu{}^i{}_j - \underline{\omega_\lambda{}^j{}_k \omega_\nu{}^k \omega_\mu{}^i{}_j}\}]
    \\[2ex] \notag
    &=  [\{{\color{red}-(\partial_\lambda \omega_\mu{}^i{}_j)\omega_\nu{}^j} + (\partial_\lambda \omega_\nu{}^i{}_j)\omega_\mu{}^j \mathcolorbox{SkyBlue}{-(\partial_\lambda \omega_\nu{}^j)\omega_\mu{}^i{}_j} + (\partial_\lambda \omega_\mu{}^j)\omega_\nu{}^i{}_j \} \\ \notag
    &{} \hspace{0.7 cm} 
    \{ \mathcolorbox{pink}{-(\partial_\mu \omega_\nu{}^j)\omega_\lambda{}^i{}_j} + (\partial_\nu \omega_\mu{}^j)\omega_\lambda{}^i{}_j {\color{green}+\underline{\omega_{\mu j}{}^k \omega_\nu{}^j \omega_{\lambda k}{}^i}} - \omega_\nu{}^j{}_k \omega_\mu{}^k \omega_\lambda{}^i{}_j\} ]\\[1ex] \notag
    &{} \hspace{0.6 cm} [ \{-(\partial_\mu \omega_\nu{}^i{}_j)\omega_\lambda{}^j {\color{orange}+ (\partial_\mu \omega_\lambda{}^i{}_j)\omega_\nu{}^j} -(\partial_\mu \omega_\lambda{}^j)\omega_\nu{}^i{}_j \mathcolorbox{pink}{+ (\partial_\mu \omega_\nu{}^j)\omega_\lambda{}^i{}_j} \}\\ \notag
    &{} \hspace{0.7 cm} \{-(\partial_\nu \omega_\lambda{}^j)\omega_\mu{}^i{}_j \mathcolorbox{SkyBlue}{+ (\partial_\lambda \omega_\nu{}^j)\omega_\mu{}^i{}_j} +\omega_\nu{}^j{}_k \omega_\lambda{}^k \omega_\mu{}^i{}_j {\color{blue}- \underline{\omega_{\lambda j}{}^k \omega_\nu{}^j \omega_{\mu k}{}^i}}\}]
\end{align}

where with colored text we've shown the matching terms on each side of the equation, and with highlighted text we've shown the remaining terms on the LHS that cancel. 

\pagebreak
\subsubsection{Vielbein postulates and a Christoffel connection}

As we did in section \ref{christcon}, we now use our metric/vielbein structure to define a Christoffel connection. As opposed to before, now with our two vielbeine, we have two vielbein postulates, a temporal and spatial one.

Recall from equation \discolorlinks{\ref{vielpost}} that the vielbein postulate reads

\begin{equation}
    \nabla_\mu e_\nu{}^a = \partial_\mu e_\nu{}^a  -\omega_\mu{}^a{}_b e_\nu{}^b  -\Gamma_{\mu\nu}^\lambda e_\lambda{}^a =0 \tag{\discolorlinks{\ref{vielpost}}.r}
\end{equation}

Note the upper index $a$ on the LHS. Since we are working with a spatial vielbein with an upper $i$, and we have two options for a spin connection and two options for a vielbein, the term $\omega_\mu{}^a{}_b e_\nu{}^b$ in the spatial vielbein postulate will indeed be a sum. Notably, we can construct two such terms that have a single upper $i$ index by using both spin connections and vielbeine,

\begin{equation}
    \nabla_\mu e_\nu{}^i = \partial_\mu e_\nu{}^i  -\omega_\mu{}^i{}_j e_\nu{}^j - \omega_\mu{}^i \tau_\nu  -\Gamma_{\mu\nu}^\rho e_\rho{}^i =0 \label{spacevielpost}
\end{equation}

The second vielbein postulate for the temporal vielbein is even simpler. The temporal vielbein has no upper indices so the $\nabla_\mu \tau_\nu$ doesn't call for any spin connection terms,

\begin{equation}
    \nabla_\mu \tau_\nu = \partial_\mu \tau_\nu  -\Gamma_{\mu\nu}^\rho \tau_\rho =0 \label{timevielpost}
\end{equation}

As we did in equation \discolorlinks{\ref{prop}}, we add the two components of the Christoffel connection which we get now from \discolorlinks{\ref{spacevielpost}} and \discolorlinks{\ref{timevielpost}} (by bringing the term with the Christoffel symbol to the RHS and then hitting both sides with the respective inverse vielbein) to get

\begin{equation}
    \Gamma^\rho_{\mu\nu} = \tau^\rho \partial_{(\mu}\tau_{\nu)} + e^\rho{}_i (\partial_{(\mu} e_{\nu)}{}^i -\omega_{(\mu}{}^{ij} e_{\nu)j} - \omega_{(\mu}{}^i \tau_{\nu)}) \label{gaugecon}
\end{equation}

where the symmetry of $\mu$,$\nu$ in the first parenthesis follows from the \discolorlinks{\ref{geoconstr}} constraint, and the symmetry of $\mu$,$\nu$ in the second parenthesis follows from the \discolorlinks{\ref{Rp0}} constraint. 

\pagebreak
\subsubsection{Trautman condition}

At this stage, we compare our generic expression of \discolorlinks{\ref{gaugecon}} to the generic expression of \discolorlinks{\ref{prop}}. In order to do so, it is clear we need to express the spatial metric and inverse spatial metric of \discolorlinks{\ref{prop}} in terms of the spatial vielbein via \discolorlinks{\ref{metspaceviel}} and the equivalent expression for the inverse spatial metric

\begin{equation}
    h_{\mu\nu} = e_\mu{}^i e_\nu{}^j \delta_{ij}
\end{equation}

The next step would be to express the spin connections in \discolorlinks{\ref{gaugecon}} as given by the lengthy \discolorlinks{\ref{depspin1}} and \discolorlinks{\ref{depspin2}}. It is not immediately clear that this is the case, and so the rigour-minded reader ought to verify thus, but it is claimed in equation 4.30 of \cite{bergshoeff} that in order for both expressions for the connection (\discolorlinks{\ref{prop}} and \discolorlinks{\ref{gaugecon}}) to match the following needs to hold

\begin{equation}
    K_{\mu\nu} = 2 \omega_{[\mu}{}^i e_{\nu]i} \label{k1}
\end{equation}

Here the second conventional constraint, \discolorlinks{\ref{Rm0}}, comes into play again. Looking at \discolorlinks{\ref{Rm}}, it is clear that if the \discolorlinks{\ref{Rm0}} constraint is imposed, then $2 \omega_{[\mu}{}^i e_{\nu]i}$ must equal $2\partial_{[\mu} m_{\nu]}$. So then \discolorlinks{\ref{k1}} becomes

\begin{equation}
    K_{\mu\nu} = 2\partial_{[\mu} m_{\nu]} \label{k2}
\end{equation}

which is precisely the adapted coordinates version of the Trautman condition for Newton-Cartan gravity that we found in \discolorlinks{\ref{tbd}}.

As a second check that we satisfy the Trautman condition with our gauging construct, we relate the curvature of the Christoffel connection to the non vanishing curvatures of our gauge connections — just as we did in equation \discolorlinks{\ref{relate}}

\begin{equation}
    R^\rho{}_{\sigma\mu\nu} (\Gamma) = -e^\rho{}_i \tau_\sigma R_{\mu\nu}{}^i(K) - e^\rho{}_i e_{\sigma j} R_{\mu\nu}{}^{ij}(J) \label{manfib}
\end{equation}

It is claimed below equation 4.32 in \cite{bergshoeff} that imposing the Trautman condition on the LHS of this is equivalent to imposing the \discolorlinks{\ref{bianchi1}} Bianchi identity on the RHS. This is not immediately clear.

Firstly, using the process discussed surrounding equation \discolorlinks{\ref{riemsyms}}, imposing the Trautman condition on the LHS of \discolorlinks{\ref{manfib}} seems to yield 

\begin{equation}
    R^{\rho}_{\sigma\mu\nu}  + R^{\rho}_{\mu\nu\sigma}+ R^{\rho}_{\mu\sigma\nu}+ R^{\rho}_{\sigma\nu\mu} =0
\end{equation}

and if we use \discolorlinks{\ref{manfib}}, we can expand this to

\begin{align} \notag
    \tau_\sigma R_{\mu\nu}{}^i(K) &+ e_{\sigma j} R_{\mu\nu}{}^{ij}(J) + \tau_\mu R_{\nu\sigma}{}^i(K) + e_{\mu j} R_{\nu\sigma}{}^{ij}(J)   \\
  +  \tau_\mu R_{\sigma\nu}{}^i(K) &+ e_{\mu j} R_{\sigma\nu}{}^{ij}(J) + \tau_\sigma R_{\nu\mu}{}^i(K) + e_{\sigma j} R_{\nu\mu}{}^{ij}(J) =0 \label{thoughts}
\end{align}

which does not seem to clearly state the same thing as the Bianchi identity \discolorlinks{\ref{bianchi1}} on the RHS of \discolorlinks{\ref{manfib}}, where we use the fact that all the gauge curvatures are anti-symmetric

\begin{equation}
    \tau_\nu R_{\sigma\mu}{}^i(K) + e_{\nu j} R_{\sigma\mu}{}^{ij}(J) +   \tau_\sigma R_{\mu\nu}{}^i(K) + e_{\sigma j} R_{\mu\nu}{}^{ij}(J) +    \tau_\mu R_{\nu\sigma}{}^i(K) + e_{\mu j} R_{\nu\sigma}{}^{ij}(J) =0 
\end{equation}

\pagebreak
\subsubsection{Ehlers conditions}
The final constraint we need apply to recover the gravitational theory is the so-called Ehlers conditions of \discolorlinks{\ref{ehlers}} (recall that all three Ehlers conditions are equivalent so we look at \discolorlinks{\ref{ehlers}} without loss of generality).

Using the symmetries of the curvature tensor, and (raising) lowering indices with the (inverse) spatial metric, \discolorlinks{\ref{ehlers}} becomes
\begin{align} \notag
h^{\sigma\lambda} R^{\mu}{}_{\nu\rho\sigma} -h^{\sigma\mu} R^{\lambda}{}_{\nu\rho\sigma}&=0 \\\notag
R^{\mu}{}_{\nu\rho}{}^\sigma - R^{\lambda}{}_{\nu\rho}{}^\mu&=0 \\\notag
h_{\mu\phi}h_{\lambda\sigma}(R^{\mu}{}_{\nu\rho}{}^\sigma - R^{\lambda}{}_{\nu\rho}{}^\mu)&=0 \\\notag
R_{\phi\nu\rho\sigma} - R_{\sigma\nu\rho\phi}&=0 \\\notag
R_{\phi\nu\rho\sigma} + R_{\phi\rho\sigma\nu}&=0 \\\notag
h^{\mu\phi}(R_{\phi\nu\rho\sigma} + R_{\phi\rho\sigma\nu})&=0 \\
R^\mu{}_{\nu\rho\sigma}(\Gamma) + R^\mu{}_{\rho\sigma\nu}(\Gamma) &=0
\end{align}
Using \discolorlinks{\ref{manfib}} this becomes
\begin{equation}
 e^\mu{}_i [ \tau_\nu R_{\rho\sigma}(K) + e_{\nu j} R_{\rho\sigma}{}^{ij}(J)+ \tau_\rho R_{\sigma\nu}(K) + e_{\rho j} R_{\sigma\nu}{}^{ij}(J) ] =0 \label{claim}
\end{equation}
By using the expressions for the gauge curvatures \discolorlinks{\ref{Rk}} and \discolorlinks{\ref{Rj}} (and possibly expressing the spin connections within the curvatures using \discolorlinks{\ref{depspin1}} and \discolorlinks{\ref{depspin2}}) it is claimed in equation 4.33 of \cite{bergshoeff} that \discolorlinks{\ref{claim}} implies 

\begin{equation}
    R_{\mu\nu}{}^{ij}(J) =0 \label{Rj0}
\end{equation}

presumably via the following equivalent equation (given the form of the curvature \discolorlinks{\ref{Rj}})

\begin{equation}
    \partial_{[\mu}\omega_{\nu]}{}^{ij} = \omega_{[\mu}{}^{ki}\omega_{\nu]}{}^{jk} 
\end{equation}

\vspace{0.5cm}
Instead of trying to show that one of the Ehlers conditions imply the $R(J)=0$ constraint (as we try above), a more elegant way to go about this could be to go in the reverse direction — i.e. start by imposing the $R(J)=0$ constraint on the gauge theory side, and then somehow use the translation between the curvatures (\discolorlinks{\ref{manfib}}) to show that $R(J)=0$ satisfies one of the Ehlers conditions. 

Imposing $R(J)=0$ reduces the Bianchi identities \discolorlinks{\ref{bianchi1}} and \discolorlinks{\ref{bianchi2}} to
\begin{align}
    R_{[\lambda\mu}{}^{i}(K)\tau_{\nu]} &= 0 \label{bianstep2a}\\
    R_{[\lambda\mu}{}^{i}(K)e_{\nu]i} &= 0 \label{bianstep2b}
\end{align}
and reduces \discolorlinks{\ref{manfib}} to
\begin{equation}
    R^\rho{}_{\sigma\mu\nu} (\Gamma) = -e^\rho{}_i \tau_\sigma R_{\mu\nu}{}^i(K) \label{manfib2}
\end{equation}

Equations \discolorlinks{\ref{bianstep2a}}, \discolorlinks{\ref{bianstep2b}}, and \discolorlinks{\ref{manfib2}} seem our master equations at our disposal to achieve our objective, $R(J)=0 \rightarrow$ Ehlers.

One idea on how to do this is to (since both Bianchi identities are zero) set the LHS's of each Bianchi identity equal to one another

\begin{equation}
    R_{[\lambda\mu}{}^{i}(K)\tau_{\nu]} = R_{[\lambda\mu}{}^{i}(K)e_{\nu]i} \label{equalbianchi}
\end{equation}
and then insert an inverted \discolorlinks{\ref{manfib2}}
\begin{equation}
    R_{\rho\sigma}{}^i(K) = - \tau^\nu e_\mu{}^i R^\mu{}_{\nu\rho\sigma}(\Gamma)
\end{equation}
into \discolorlinks{\ref{equalbianchi}}
\begin{align}
    R_{[\rho\sigma}{}^{i}(K)\tau_{\lambda]} &= R_{[\rho\sigma}{}^{i}(K)e_{\lambda]i} \\\notag
    - \tau^\nu e_\mu{}^i R^\mu{}_{\nu[\rho\sigma}(\Gamma)\tau_{\lambda]} &= - \tau^\nu e_\mu{}^i R^\mu{}_{\nu[\rho\sigma}(\Gamma)e_{\lambda]i} \\ \notag
    R^\mu{}_{\nu[\rho\sigma}(\Gamma)\tau_{\lambda]} &= R^\mu{}_{\nu[\rho\sigma}(\Gamma)e_{\lambda]i}
\end{align}

This however does not appear to yield anything fruitful for satisfying the Ehlers conditions.

\vspace{0.5cm}
Upon further consideration and discussion, it appears relating the constraint $R(J)=0$ directly to the Ehlers condition is not necessary. After all, our objective is to gauge Bargmann and recover Newton-Cartan, not recover the Ehlers condition. Because the Ehlers conditions were imposed in the Newton-Cartan picture does not necessarily imply we must satisfy Ehlers from the gauge side to produce Newton-Cartan, we are going about it from an entirely different again. If we can show that the $R(J)=0$ constraint enables us to reproduce Newton-Cartan, that is motivation enough to impose it.

This is of course not totally satisfactory. After all, how would one know to impose the $R(J)=0$ constraint if they started from scratch? From this perspective, we urge the rigour-minded reader to return to the calculations earlier in this sections to find a way to motivate the $R(J)=0$ constraint, be it from the Ehlers conditions or otherwise. Consider the discussion on page 81 of \cite{Roelthesis} for some possible inspiration.

\pagebreak
\subsubsection{Recovering a Riemann tensor and equation of motion}

Imposing $R(J)=0$ reduces the Bianchi identities \discolorlinks{\ref{bianchi1}} and \discolorlinks{\ref{bianchi2}} to \discolorlinks{\ref{bianstep2a}} and \discolorlinks{\ref{bianstep2b}} respectively, as well as reduces the curvature translation equation \discolorlinks{\ref{manfib}} to
\discolorlinks{\ref{manfib2}}.

Contracting all the spacetime indices in \discolorlinks{\ref{bianstep2a}} 

\begin{align} \notag
R_{[\mu\nu}{}^{i}(K)\tau_{\rho]} &= 0 \\\notag
e^\mu{}_i e^\nu{}_j    R_{[\mu\nu}{}^{i}\tau_{\rho]}  \tau^\rho &= 0 \\ \notag e^\mu{}_i e^\nu{}_j (R_{\mu\nu}{}^k\tau_{\rho}+R_{\nu\rho}{}^k\tau_{\mu}+R_{\rho\mu}{}^k\tau_{\nu})  \tau^\rho &= 0 \\\notag
e^\mu{}_i e^\nu{}_j R_{\mu\nu}{}^k(\tau_{\rho} \tau^\rho)+e^\nu{}_jR_{\nu\rho}{}^k(e^\mu{}_i\tau_{\mu})\tau^\rho+e^\mu{}_iR_{\rho\mu}{}^k(e^\nu{}_j\tau_{\nu})  \tau^\rho &= 0 \\ \notag
e^\mu{}_i e^\nu{}_j R_{\mu\nu}{}^k(1)+e^\nu{}_jR_{\nu\rho}{}^k(0)\tau^\rho+e^\mu{}_iR_{\rho\mu}{}^k(0)  \tau^\rho &= 0 \\
e^\mu{}_i e^\nu{}_j R_{\mu\nu}{}^k(K) &= 0
\end{align}

where we have used:
\begin{enumerate}
    \item the anti-symmetry of the gauge curvatures makes the a priori 6 terms of $R_{[\mu\nu}{}^{i}(K)\tau_{\rho]}$ only 3 and the equation is equal to zero so we do not bother writing the 1/3 factor,
\item $\tau_{\rho}  \tau^\rho=1$, and 
\item $e^\mu{}_i\tau_\mu=0$.
\end{enumerate}

All in all, the contracted Bianchi identities will read

\begin{align}
    e^\mu{}_i e^\nu{}_j    R_{\mu\nu}{}^{k}(K) &= 0 \label{cont1} \\
    \tau^\mu e^{\nu [i} R_{\mu\nu}{}^{j]}(K) &= 0 \label{cont2}
\end{align}

Mimicking the calculation of \discolorlinks{\ref{bianstep2a}} $\rightarrow$ \discolorlinks{\ref{cont1}} in the case of \discolorlinks{\ref{bianstep2b}} $\rightarrow$ \discolorlinks{\ref{cont2}} is not so straightforward. We urge the rigour-minded reader to work this out explicitly.

The interpretation of \discolorlinks{\ref{cont1}} is that the lower indices of this gauge curvature (and via \discolorlinks{\ref{manfib2}} the lower indices of the Riemann tensor) will be only temporal.

The interpretation of the \discolorlinks{\ref{cont2}} is that, while the $(ij)$ anti-symmetric version of that object vanishes, the $(ij)$ symmetric version may not be. Let's check. Starting with \discolorlinks{\ref{manfib2}}

\begin{align} \notag
    R^\rho{}_{\sigma\mu\nu} (\Gamma) &= -e^\rho{}_i \tau_\sigma R_{\mu\nu}{}^i(K) \\ \notag
    -e_\rho{}^i \tau^\sigma R^\rho{}_{\sigma\mu\nu} (\Gamma) &=  R_{\mu\nu}{}^i(K) \\\notag
    -  R^i{}_{0\mu\nu} (\Gamma) &=  R_{\mu\nu}{}^i(K) \\ \notag
    -  R^j{}_{0\mu\nu} (\Gamma) &=  R_{\mu\nu}{}^j(K) \\ \notag
    -  \tau^\mu e^{\nu(i} R^j{}_{0\mu\nu} (\Gamma) &=  \tau^\mu e^{\nu(i}R_{\mu\nu}{}^{j)}(K) \\ \notag
    -  R^{(j}{}_{00}{}^{i)} (\Gamma) &=   \\ \notag
    -  \delta^{k(j} \delta^{i)\ell} R_{k00\ell} (\Gamma) &=   \\ \notag
    -  \delta^{k(j} \delta^{i)\ell} R_{0\ell k0} (\Gamma) &=   \\ \notag
      \delta^{k(j} \delta^{i)\ell} R_{\ell0k0} (\Gamma) &=   \\ 
      \delta^{k(j} R^{i)}_{0k0} (\Gamma) &=   \label{result1}
\end{align}

which one can write as $R^{i}_{0j0}$ so long as the $(ij)$ symmetry is mentally kept track of.

This final non-vanishing part of the Riemann tensor is exactly that of Newton-Cartan gravity. Moreover, we can contract the first and third indices to form a Ricci tensor as follows,

\begin{equation}
     \tau^\mu e^{\nu}{}_i R_{\mu\nu}{}^{i}(K) = R^{i}_{0i0}(\Gamma) = R_{00}(\Gamma) \label{result2}
\end{equation}

One notices here that we have only obtained the non-zero parts of the curvature objects. This indeed reproduces Newton-Cartan, but not Newtonian gravity, where the objects are expressed in terms of the Newtonian potential $\Phi$. This has to do with the the completely arbitrary frame independence of Newton-Cartan compared to Newton. Newtonian gravity does not consider any arbitrary frame, but rather only ``earth-based" / constant acceleration frames. Thus, Newtonian gravity is a particular gauge-fixed version of Newton-Cartan. 

And so, in order to express the non-vanishing curvature, as well as the non-vanished part of the $\Gamma$-connection that forms it, in terms of a Newtonian potential $\Phi$, one would need to mimic that gauge fixing to bring the theory to only constant acceleration frames. This is discussed comprehensively in section 5.2 of \cite{Roelthesis} (equivalently section 2 of \cite{stringnc}) where more details on the $R(J)=0$ constraint of \discolorlinks{\ref{Rj0}} are elucidated. As a preview, the result of this gauge-fixed procedure yields (equation 5.22 of \cite{Roelthesis})

\begin{align}
    \Gamma^i_{00} &= \partial^i \Phi \\
    \Phi &= m_0 -\frac{1}{2}\delta_{ij} \tau^i \tau^j + \partial_0 m
\end{align}

.

\pagebreak

\section{Conclusion}
\label{sec:conclusion}

In this work we have shown how Einstein's general relativity (GR) can be viewed as a gauge theory of the Poincaré algebra. We have extended this to show how Cartan's geometric formulation of Newtonian gravity (Newton-Cartan gravity) can be viewed as a gauge theory of the Bargmann algebra. In doing so, we touched on the following auxiliary topics: the extension of Yang-Mills to a more generic formalism of gauge theory, the fiber bundle picture of gauge theory,  the vielbein formalism of GR, Lie algebra procedures such as central extensions and İnönü-Wigner contractions, and the hallmarks of Newtonian gravity which differentiate it from GR.

\vspace{1cm}
The main achievement of this work is a systematic review of how to view gravity as a gauge theory. The pivotal steps in that procedure are summarized here:

\begin{enumerate}
    \item Impose conventional (and possibly geometric) constants (as well) on the gauge curvatures which —
    \begin{enumerate}
        \item remove the local gauge translations from the algebra, replacing them with GCTs + other symmetries of the algebra
        \item enables solving for the spin connection(s) in terms of other fields
        \item (enable reproducing of a key feature of the geometry one's gravity theory lives in)
    \end{enumerate}
    \item Impose vielbein(e) postulate(s) which enable solving for the Christoffel $\Gamma$ connection
    \item Impose additional constraints to solve further conditions of the gravity theory (Trautman and Ehlers for example)
    \item Use the Bianchi gauge identities to narrow down non-vanishing parts of the Riemann tensor
\end{enumerate}

\pagebreak
\section{Further directions}
\label{sec:fd}

\iffalse
The present chapter is motivated by simultaneously looking forward and looking behind.

Looking forward: my future research efforts will at least partially be determined by whom I work under during my PhD.   

Looking behind: the amalgam of gauge theory, gravity, geometry, and algebra in this work has been extremely fulfilling for me.

As so, I have made an effort to look into research going on at (and around) Stony Brook University (or that I envision researchers there being interested in) that relates in one way or another to the aforementioned aspects of this work that have captivated me.

\pagebreak
\fi
\subsection{Condensed matter}

\subsubsection{Newton-Cartan and the quantum Hall effect}

Recall in Section \ref{subsec:torsionm}, we imposed zero torsion as one route to build a notion of causality. We mentioned there that this zero torsion constant is not the only way to impose causality, and moreover imposing causality in that fashion would make the bulk gravity impossible to be made dual to a boundary CFT for holographic applications \cite{ericNCtorsion}. In the holographic context, Newton-Cartan (NC) gravity with torsion is utilized, and other methods of imposing causality are explored.

On top of the NC gravity with torsion useful in holography, an even less restricted form of NC \textit{geometry} has been explored in the condensed matter (CM) community \cite{son} \cite{abanovgromov}. We emphasize geometry here as opposed to gravity to point out that, for what it is used for in this CM work, there is no need for a notion of causality in the gravitational theory. The role of ``gravity" in these works is solely as a background to which one couples the field theory describing a particular CM system of interest. In such a set up, one investigates the properties of the CM system of interest by studying how the field theory responds to the background. In addition to the papers above, the following — which include the determination of Hall viscosity by measuring an electromagnetic response — give more insight into these procedures \cite{wiegmann} \cite{hoyos}.

\subsubsection{Fractons and gauging algebras}

A current hot topic bringing together the CM and HEP-th communities is the study of systems know as fractons. These mobility-restricted charges have been shown to touch on the following areas: topological phases, quantum information, gravity, and QFT dualities \cite{fractons} \cite{seibergshao}. Here we will focus on studies of fractons that relate to our present work.

In particular, studying fractons from a gauge theory perspective, as initiated in \cite{pretko} \cite{slagle} \cite{idk}, may offer an application of the gauging procedure we outline in this present work.

In the following work \cite{gromov}, a symmetry algebra (taking the place of the Poincaré or Bargmann for example comparing to our construction) coined ``the multipole algebra" is constructed from the polynomial shift symmetries studied here \cite{kevin}.

The so-called multipole algebra of \cite{gromov} is then ``gauged" to construct an effective field theory that the author uses to study these fracton systems. Studying how this gauging procedure (symmetry algebra $\rightarrow$ effective field theory) of the CM community compares to that outlined in this work (symmetry algebra $\rightarrow$ gravitational theory) could be very interesting.

\pagebreak
\subsection{Non-relativistic supergravity and supersymetry}

\subsubsection{Non-relativistic supergravity}
Supergravity (SUGRA) can be obtained by gauging a supersymmetry (SUSY) algebra \cite{sugragauge}. In this picture of SUGRA, one could easily imagine constructing a non-relativistic (NR) SUGRA if one had an NR SUSY algebra to gauge. 

With this motivation in mind, SUSY extensions of the NR algebras (like the Galilei or Bargmann we have considered in this work) ought to be considered \cite{clarklove} \cite{quimtownsend}.

And indeed, gauging a SUSY extension of the Bargmann yields a Newton–Cartan SUGRA theory \cite{roelericSUGRA}.
The applications of NR SUGRA are explained in the next section.

\subsubsection{Non-relativistic supersymmetric quantum field theories in curved backgrounds}

Studying SUSY QFTs on curved backgrounds via Festuccia-Seiberg/Pestun's localization techniques \cite{festuccia} \cite{pestun} has enabled the uncovering of many new and interesting aspects of QFT, strongly-coupled systems, as well as precision holography \cite{res1} \cite{res2} \cite{res3} \cite{res4} \cite{res5} \cite{res6}.

There is no reason to assume those same localization techniques would not yield interesting results when applied to NR SUSY QFTs on curved backgrounds. Studying NR SUSY theories on curved manifolds is indeed the first step towards investigating what NR results can be garnered from Pestun/Festuccia-Seiberg's localization techniques \cite{johannesericSUSY}.

A further step in this direction could be taking the limit of an NR SUSY theory coupled to NR SUGRA, mimicking Festuccia-Seiberg's procedure in \cite{festuccia}. Doing so would necessitate finding an \textit{off-shell} formulation of NR SUGRA, unlike the on-shell NR SUGRA of \cite{roelericSUGRA}. Work towards this end is done in \cite{Thomasthesis} \cite{zoj1} \cite{zoj2}.

\pagebreak

\end{document}